\documentclass[aps,prl,twocolumn,floats,final,superscriptaddress,citeautoscript,longbibliography]{revtex4-1}
\usepackage{amssymb,bm}
\usepackage{graphicx}
\usepackage{amsmath}
\usepackage{braket}
\usepackage{bbm}
\usepackage[colorlinks=true,linkcolor=blue,citecolor=blue,urlcolor=blue]{hyperref}
\usepackage{xcolor,soul}

\newcommand{\Tr}{\operatorname{Tr}}

\usepackage{changes}

\begin{document}

%%%%%% title
\title{Emergent Universal Quench Dynamics in Randomly Interacting Spin Models}

%%%%%% authors
\author{Yuchen Li}
\thanks{These authors contribute equally to this work}
\affiliation{CAS Key Laboratory of Microscale Magnetic Resonance and School of Physical Sciences, University of Science and Technology of China, Hefei, Anhui 230026, China}

\author{Tian-Gang Zhou}
\thanks{These authors contribute equally to this work}
\affiliation{Institute for Advanced Study, Tsinghua University, Beijing 100084, China}

\author{Ze Wu}
\thanks{These authors contribute equally to this work}
\affiliation{CAS Key Laboratory of Microscale Magnetic Resonance and School of Physical Sciences, University of Science and Technology of China, Hefei, Anhui 230026, China}
\affiliation{CAS Center for Excellence in Quantum Information and Quantum Physics, University of Science and Technology of China, Hefei 230026, China}

\author{Pai Peng}
\affiliation{Department of Electrical and Computer Engineering, Princeton University, Princeton, NJ 08544, USA}

\author{Shengyu Zhang}
\affiliation{CAS Key Laboratory of Microscale Magnetic Resonance and School of Physical Sciences, University of Science and Technology of China, Hefei, Anhui 230026, China}
\affiliation{CAS Center for Excellence in Quantum Information and Quantum Physics, University of Science and Technology of China, Hefei 230026, China}
\affiliation{Hefei National Laboratory, Hefei 230088, China}

\author{Riqiang Fu}
\affiliation{National High Magnetic Field Laboratory, 1800 East Paul Dirac Drive, Tallahassee, FL 32310, USA}

\author{Ren Zhang}
\affiliation{School of Physics, Xi'an Jiaotong University, Xi'an, Shaanxi 710049, China}
\affiliation{Hefei National Laboratory, Hefei 230088, China}

\author{Wei Zheng}
\affiliation{Hefei National Research Center for Physical Sciences at the Microscale and School of Physical Sciences, University of Science and Technology of China, Hefei 230026, China}
\affiliation{CAS Center for Excellence in Quantum Information and Quantum Physics, University of Science and Technology of China, Hefei 230026, China}
\affiliation{Hefei National Laboratory, Hefei 230088, China}

\author{Pengfei Zhang}
\email{pengfeizhang.physics@gmail.com}
\affiliation{Department of Physics, Fudan University, Shanghai 200438, China}
\affiliation{Shanghai Qi Zhi Institute, AI Tower, Xuhui District, Shanghai 200232, China}
%\affiliation{Hefei National Laboratory, Hefei 230088, China}

\author{Hui Zhai}
\email{hzhai@tsinghua.edu.cn}
\affiliation{Institute for Advanced Study, Tsinghua University, Beijing 100084, China}
\affiliation{Hefei National Laboratory, Hefei 230088, China}

\author{Xinhua Peng}
\email{xhpeng@ustc.edu.cn}
\affiliation{CAS Key Laboratory of Microscale Magnetic Resonance and School of Physical Sciences, University of Science and Technology of China, Hefei, Anhui 230026, China}
\affiliation{CAS Center for Excellence in Quantum Information and Quantum Physics, University of Science and Technology of China, Hefei 230026, China}
\affiliation{Hefei National Laboratory, Hefei 230088, China}

\author{Jiangfeng Du}
\affiliation{CAS Key Laboratory of Microscale Magnetic Resonance and School of Physical Sciences, University of Science and Technology of China, Hefei, Anhui 230026, China}
\affiliation{CAS Center for Excellence in Quantum Information and Quantum Physics, University of Science and Technology of China, Hefei 230026, China}
\affiliation{Hefei National Laboratory, Hefei 230088, China}
\affiliation{Institute of Quantum Sensing and School of Physics, Zhejiang University, Hangzhou 310027, China}

%%%%%% Abstract part.
\begin{abstract}
Universality often emerges in low-energy equilibrium physics of quantum many-body systems, despite their microscopic complexity and variety. Recently, there has been a growing interest in studying far-from-equilibrium dynamics of quantum many-body systems. Such dynamics usually involves highly excited states beyond the traditional low-energy theory description. Whether universal behaviors can also emerge in such non-equilibrium dynamics is a central issue at the frontier of quantum dynamics. Here we report the experimental observation of universal dynamics by monitoring the spin depolarization process in a solid-state NMR system described by an ensemble of randomly interacting spins. The spin depolarization can be related to temporal spin-spin correlation functions at high temperatures. We discover a remarkable phenomenon that these correlation functions obey a universal functional form. This experimental fact helps us identify the dominant interacting processes in the spin depolarization dynamics that lead to this universality. Our observation demonstrates the existence of universality even in non-equilibrium dynamics at high temperatures, thereby complementing the well-established universality in low-energy physics. 
\end{abstract}

% make title
\date{\today}
\maketitle

\vspace{10pt}

The notion of \textit{universality} refers to simple rules and a small number of parameters that can universally describe a physical phenomenon across various systems, despite their complicated and distinct microscopic details. Numerous examples have demonstrated that universal behaviors can occur in different sub-fields of physics. For examples, in atomic physics, a single parameter, the $s$-wave scattering length, governs the low-energy scattering between two atoms \cite{Cheng,Pethick}. In other words, regardless of the specific atomic species with different interatomic Van der Waals potentials, their low-energy interaction properties tend to be identical as long as their $s$-wave scattering lengths are the same. Similarly, in condensed matter physics, systems within the quantum critical regime exhibit identical low-energy properties if they belong to the same universality class, even though their microscopic Hamiltonians can be vastly different \cite{Schadev}.

However, most known examples of universal behaviors occur in low-energy physics. In contrast, far-from-equilibrium quantum dynamics always involve highly excited states. In particular, we often study a type of quench dynamics where we start with an initial state at high temperature and follow its unitary evolution governed by a quantum many-body Hamiltonian, such as in cold atoms \cite{ColdAtom_KPZ2022,Rydberg_Floquet2021}, Ions \cite{Ion_OTOC2017,Ion_hydrodynamics2022}, NV centers \cite{NV_hydrodynamics2021,NV_LukinThermalization2023} and NMR systems \cite{NMR_UniversalDecay2008,NMR_UniversalDecay2011,NMR_UniversalDecay2012,NMR_transport2011,NMR_localization2015,NMR_OTOC2017,NMR_TimeCrystal2018,NMR_MBL2018,NMR_prethermal2019,NMR_Losch2020,NMR_prethermal2021,NMR_hydrodynamics2023}.  
Such dynamics can be attributed to temporal correlation functions at infinite temperatures \cite{Fine2004,Fine2005,ColdAtom_KPZ2022,Zhai,theo2017,theo_KPZ2019,theo_Transport2019_1,theo_Transport2019_2,theo_Moore2020,theo_EnergyTransport2023,NMR_UniversalDecay2008,NMR_UniversalDecay2011,NMR_UniversalDecay2012,NMR_transport2011,NMR_OTOC2017,NMR_prethermal2019,NMR_Losch2020,NMR_prethermal2021,NMR_hydrodynamics2023,NV_hydrodynamics2021,NV_LukinThermalization2023,Ion_OTOC2017,Ion_hydrodynamics2022,Rydberg_Floquet2021}. Discovering universality in such dynamics complements established universality on low-energy equilibrium physics. So far, such examples are still rare. A recent experiment in a cold atom system has revealed universal Kardar-Parisi-Zhang scaling for such quench dynamics in an integrable spin chain \cite{ColdAtom_KPZ2022}.
In contrast, we study spin models with random and all-to-all interactions using a solid-state NMR system. We reveal a couple of universal parameters in this system that can capture the main features of the quench dynamics, including both spin depolarization dynamics and multiple quantum coherence.

\begin{figure*}
    \begin{center}  
    \includegraphics[scale=1]{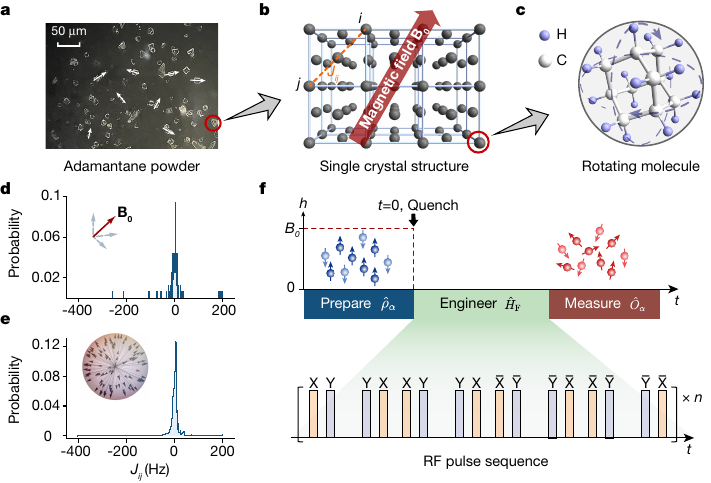}
    \caption{\textbf{The experimental protocol.} \textbf{a}, Microscopic picture of powder sample of adamantane (C$_{10}$H$_{16}$). The granules, whose sizes are of the order of micrometers, exhibit random orientations. \textbf{b}, In one granule, the adamantane molecules form a face-centered cubic lattice. The orientation of the static magnetic field $\mathbf{B}_0$ relative to the lattice principal axes determines the values of the secular dipolar coupling strength $J_{ij}$. \textbf{c}, Each molecule undergoes rapid rotation around its lattice site due to thermal motion, and the lattice site effectively serves as a time-averaged position for all the nuclear spins within the same molecule. ${^1}$H carries nuclear spin-$1/2$ and ${^{12}}$C carries no spin. \textbf{d,e}, The probability distribution of the intermolecular coupling $J_{ij}$. \textbf{d} shows the distribution for a given orientation of $\mathbf{B}_0$, and \textbf{e} shows the distribution averaged over $10^5$ random orientations denoted by arrows in the spherical surface. Up to the 13th neighbor couplings are incorporated in the calculation \cite{code}. \textbf{f}, The experimental protocol of the quench dynamics. Firstly we prepare a polarized initial density matrix $\hat{\rho}_\alpha\propto \mathbbm{1}+\epsilon\sum_{ia}\hat{S}_{ia}^\alpha$, $\alpha=\hat{x},\ \hat{y}$ or $\hat{z}$. Then, the state evolves under the anisotropic random spin models engineered by the RF pulse sequence illustrated below and also used in Refs.~\cite{Suter1987,NMR_MBL2018,NMR_prethermal2019,NMR_Losch2020,NMR_prethermal2021,NMR_hydrodynamics2023}, after which we measure the magnetization $\hat{O}_\alpha=\sum_{ia}\hat{S}_{ia}^\alpha$.}
    \label{Fig1}
    \end{center}
\end{figure*}

To be concrete, let us consider an initial density matrix as $\hat{\rho}\propto \mathbbm{1}+\epsilon\hat{O}$,
where $\epsilon$ is a small parameter and $\hat{O}$ is a traceless operator as a perturbation to the infinite temperature ensemble. This density matrix undergoes time evolution governed by a quantum many-body Hamiltonian $\hat{H}$, given by $\hat{\rho}(t)=\text{e}^{-\text{i}\hat{H}t/\hbar}\hat{\rho}~\text{e}^{\text{i}\hat{H}t/\hbar}$. Then, by measuring the expectation value of operator $\hat{O}$, we can access the auto-correlation function as $\langle\hat{O}(t)\rangle=\text{Tr}[\hat{O}\hat{\rho}(t)]\propto \mathcal{C}(t)$,
where $\mathcal{C}(t)=\frac{1}{c_O} \text{Tr}[\hat{O}(t)\hat{O}(0)]$ is the auto-correlation function with normalization constant $c_O$ such that $\mathcal{C}(0)=1$. This auto-correlation function is defined at infinite temperatures because it equally incorporates contributions from all eigenstates, thereby reflecting the properties of the many-body Hamiltonian.
During the Heisenberg evolution, the operator complexity of $\hat{O}(t)$ continuously increases \cite{complexity1,complexity2}, resulting in decaying of $\mathcal{C}(t)$. Therefore, the universality observed in $\mathcal{C}(t)$ ultimately stems from the universal behavior in the complexity theory of operator growth \cite{Zhai}.     

Our experiment is conducted on a powder sample of adamantane (C$_{10}$H$_{16}$)~\cite{NMR_localization2015,NMR_Losch2020,Pastawski2009,Claudia2014,Suter2010}. Each adamantane molecule contains sixteen Hydrogen atoms (${}^1$H), and each ${}^1$H carries nuclear spin $S=1/2$. There are approximately $10^{9}$ to $10^{12}$ molecules contained in a single granule of the powder, which has a size on the order of micrometers (Fig.~\ref{Fig1}a). These spins interact with each other through magnetic dipolar interactions. Moreover, the sample is placed in a uniformed magnetic field of $B_0=9.4$ T along the $\hat{z}$ direction. Therefore, the Hamiltonian reads 

\begin{align}
&\hat{H}=-\hbar\gamma_\text{H} B_0\sum\limits_{ia}\hat{S}^z_{ia}+\nonumber\\
&\sum\limits_{(i,a)< (j,b)}\frac{\mu_0\hbar^2\gamma^2_\text{H}}{4\pi r^3_{ia,jb}}\Bigg[\hat{\bm{S}}_{ia}\cdot\hat{\bm{S}}_{jb}-\frac{3(\hat{\bm{S}}_{ia}\cdot{\bf r}_{ia,jb})(\hat{\bm{S}}_{jb}\cdot{\bf r}_{ia,jb})}{r^2_{ia,jb}}\Bigg]
, \label{Hstep1}
\end{align}
where $\hat{\bm{S}}_{ia}=(\hat{S}_{ia}^x,\hat{S}_{ia}^y,\hat{S}_{ia}^z)$ are spin operators for each ${}^1$H. $i$ and $j$ label molecules positioned on a face-centered cubic lattice (Fig.~\ref{Fig1}b). The indices $a,b=1,\dots,16$ label the spin-$1/2$ within each molecule. The constraint $(i,a)< (j,b)$ is defined as $a<b$ when $i=j$ and otherwise $i<j$. $\mu_0$ is the vacuum magnetic permeability and $\gamma_\text{H}$ is the proton's gyromagnetic ratio. ${\bf r}_{ia,jb}={\bf R}_{ij}+{\bf l}_a-{\bf l}_b$ and $r_{ia,jb}=|{\bf r}_{ia,jb}|$, where ${\bf R}_{ij}$ denotes the displacement between centers of two molecules. ${\bf l}_a$ and ${\bf l}_b$ are the vectors from the center of the molecule to each nuclear spin carrier ${}^1$H. $\gamma_{\text{H}}B_0$ represents the strength of the Zeeman splitting resulting from the external magnetic field.

\begin{figure*}
    \begin{center}
    \includegraphics[scale=1]{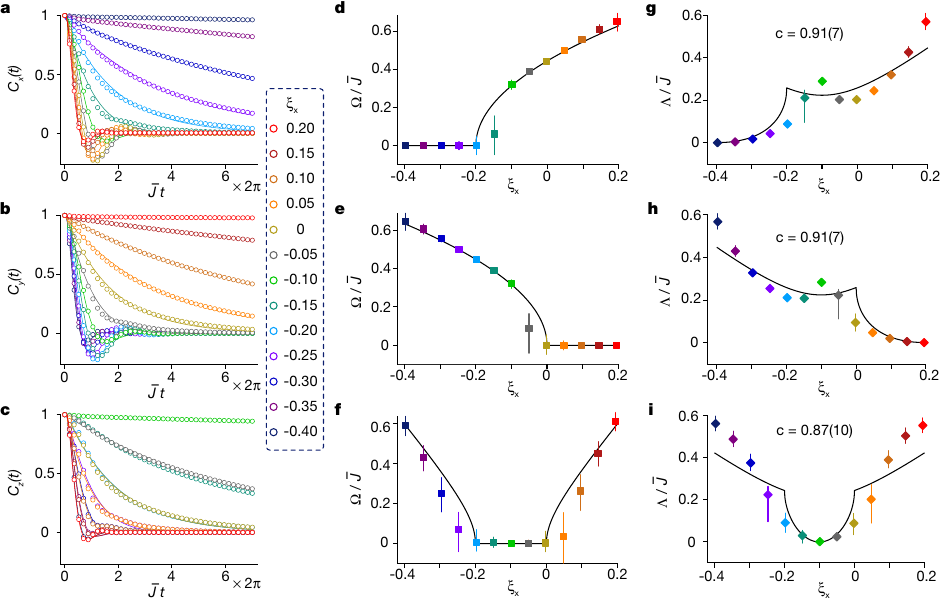}
    \caption{{\textbf{Dynamical evolution of spin polarization during the quench dynamics.}} \textbf{a-c}, Experimental measurements of $\mathcal{C}_\alpha(t)\propto\langle\hat{O}_\alpha\rangle$, with $\hat{O}_\alpha=\sum_{ia}\hat{S}_{ia}^\alpha$. The data is normalized by its value at $t=0$, with error bars ($\sim 10^{-4}$) incorporated within the markers of the data points. The initial state density matrix is prepared as $\hat{\rho}_\alpha\propto \mathbbm{1}+\epsilon\hat{O}_\alpha$, respectively. We have $\alpha=\hat{x}$ for (\textbf{a}), $\alpha=\hat{y}$ for (\textbf{b}), and $\alpha=\hat{z}$ for (\textbf{c}). Here we fix $\xi_z=0.2$ and $\sum_\alpha\xi_\alpha=0$ in the Hamiltonian Eq.~\eqref{Heisenberg}. Different colors in the figures denote different values of $\xi_x$, varying from $-0.4$ to $0.2$. The solid lines represent the fittings using a general function $A\cos(\Omega t+\Phi)\exp(-\Lambda t)$, from which both $\Omega$ and $\Lambda$ are obtained. \textbf{d-f}, The oscillation frequencies $\Omega/\bar{J}$ extracted from (\textbf{a-c}) are plotted as a function of $\xi_x$ for $\alpha=\hat{x}$ 
    (\textbf{d}), $\hat{y}$ (\textbf{e}) and $\hat{z}$ (\textbf{f}). The solid line denotes zero when $W_\alpha<0$ and fits $c\sqrt{W_\alpha}$ for $W_\alpha>0$. \textbf{g-i}, $\Lambda/\bar{J}$ extracted from (\textbf{a-c}) is plotted as a function of $\xi_x$ for $\alpha=\hat{x}$ (\textbf{g}), $\hat{y}$ (\textbf{h}) and $\hat{z}$ (\textbf{i}). The solid line fits $c\sqrt{\Gamma}$ when $W_\alpha>0$ and fits $c(\sqrt{\Gamma}-\sqrt{-W_\alpha})$ for $W_\alpha<0$. The constant $c$ is obtained from simultaneous fitting of both $\Omega/\bar{J}$ and $\Lambda/\bar{J}$, with the $95\%$ confidence interval in the parenthesis. The error bars of data points in (\textbf{d-i}) include both the $95\%$ confidence intervals estimated from the fitting residuals, and the fluctuation due to varying the fitting range (from $15$ to $41$ points) \cite{code}. This strategy reduces the fitting error caused by the ambiguity of the fitting range, thereby enhancing the reliability of the fitting results.}
    \label{polarization}
    \end{center}
\end{figure*}

At room temperature, each molecule undergoes rapid rotation around its center due to thermal motion, with a characteristic timescale of $10^{-11}$ s \cite{Resing1969} (Fig.~\ref{Fig1}c). This timescale is much faster than the timescale of dipolar interaction, which is approximately $10^{-3}$ s. By averaging the Hamiltonian over the solid angles of ${\bf l}_a$ and ${\bf l}_b$, and to the leading-order approximation, the Hamiltonian Eq.~\eqref{Hstep1} becomes \cite{Resing1969,McCall1960,Smith1961}

\begin{align}
\hat{H}=\sum\limits_{i< j,ab}&\frac{\mu_0\hbar^2\gamma^2_\text{H}}{4\pi R^3_{ij}}\left[\hat{\bm{S}}_{ia}\cdot\hat{\bm{S}}_{jb}-\frac{3(\hat{\bm{S}}_{ia}\cdot{\bf R}_{ij})(\hat{\bm{S}}_{jb}\cdot{\bf R}_{ij})}{R^2_{ij}}\right]\nonumber\\
&-\hbar\gamma_\text{H} B_0\sum\limits_{ia}\hat{S}^z_{ia}. \label{Hstep2}
\end{align}
That is to say, a nuclear spin in one molecule interacts identically with any other nuclear spin in another molecule. 

Furthermore, the presence of an external magnetic field causes all spins to rotate along the $\hat{z}$ direction, with a characteristic timescale of $10^{-9}$ s. This rapid motion can be effectively eliminated by applying a unitary transformation $\exp(-\text{i}\gamma_{\text{H}}B_0\sum\limits_{ia}\hat{S}^z_{ia}t)$. After taking the secular approximation \cite{Abragam1961}, we obtain the Hamiltonian
\begin{equation}
\hat{H}=\hbar\sum\limits_{i<j,ab}J_{ij}(-\hat{S}^x_{ia}\hat{S}^x_{jb}-\hat{S}^y_{ia}\hat{S}^y_{jb}+2\hat{S}^z_{ia}\hat{S}^z_{jb}), \label{dipole}
\end{equation}
where $J_{ij}\equiv(\mu_0/4\pi)(\hbar\gamma^2_\text{H}/2R^3_{ij})(1-3\cos^2\theta_{ij})$. $\theta_{ij}$ represents the angle between ${\bf R}_{ij}$ and the $\hat{z}$ direction. Now, the randomness arises because the molecules occupy lattice sites, and in a powder sample, the orientations between the lattice axes and the $\hat{z}$ direction are random. Fig.~\ref{Fig1}d-e present the probability distributions of $J_{ij}$ calculated from the lattice structure \cite{code}, supporting the notion that $J_{ij}$ can be regarded as random variables, with a mean and variance satisfying $\overline{J_{ij}}=0$ and $\overline{ J^2_{ij}}=4J^2/N$. $N=N_\text{m} N_\text{a}$ is the total number of spins, in which $N_\text{m}$ is the number of molecules and $N_\text{a} = 16$ represents the number of ${}^1$H in each molecule. We calibrate $J$ within the range of $2\pi [1432,1502]$ Hz \cite{supple}, with an average value $\bar{J}=(2\pi)\ 1460$ Hz, which is used later in the notation of the dimensionless time scale $\bar{J}t$. 

Next, by periodically applying a radio-frequency pulse sequence as shown in Fig.~\ref{Fig1}f~\cite{Suter1987,NMR_MBL2018,NMR_prethermal2019,NMR_Losch2020,NMR_prethermal2021,NMR_hydrodynamics2023}, the Hamiltonian in Eq.~\eqref{dipole} can be further engineered into a more general form according to the average Hamiltonian theory \cite{Haeberlen1968}
\begin{equation}
\hat{H}=\hbar\sum\limits_{i<j,ab}J_{ij}(\xi_x\hat{S}^x_{ia}\hat{S}^x_{jb}+\xi_y\hat{S}^y_{ia}\hat{S}^y_{jb}+\xi_z\hat{S}^z_{ia}\hat{S}^z_{jb})+\dots \label{Heisenberg}
\end{equation}   
Here $\xi_{\alpha}$ ($\alpha=\hat{x},\ \hat{y},\ \hat{z}$) represents three anisotropic parameters that are subjected to a constraint $\sum_\alpha\xi_\alpha=0$, which is inherited from the Hamiltonian \eqref{dipole} and conserved under global rotations. Note that the measurement is summed over random crystalline orientations, which facilitates Eq.~\eqref{Heisenberg} a random spin model in the sense of ensemble average. The various configurations of $(\xi_x,\ \xi_y,\ \xi_z)$ can be achieved by manipulating the pulse intervals (Methods and Supplementary Information \cite{supple}). The Floquet-engineered random spin model \eqref{Heisenberg} is non-integrable and generically prethermalizes an initial state with finite energy to quasi equilibrium, which could be characterized by a canonical ensemble $\hat{\rho}_\mathrm{pre}=\mathrm{e}^{-\beta\hat{H}}/\mathcal{Z}$, with $\mathcal{Z}$ the partition function and $\beta$ determined by the initial-state energy \cite{Deutsch2018,Mori2016,Abanin2017,NMR_prethermal2019,NMR_prethermal2021}. The Hamiltonian information is thus inherited by the prethermal state $\hat{\rho}_\mathrm{pre}$, which can be learned via state tomography (Methods and Supplementary Information \cite{supple}). The deviations from the target Hamiltonian configurations are calibrated to be within $3\%$. The term $\dots$ in Eq.~\eqref{Heisenberg} represents residual terms other than $\hat{S}^\alpha_{ia}\hat{S}^\alpha_{jb}$, and the total weight of these terms has been calibrated to be less than $20\%$.

\begin{figure*}[t]
    \begin{center}    \includegraphics[width=0.8\textwidth]{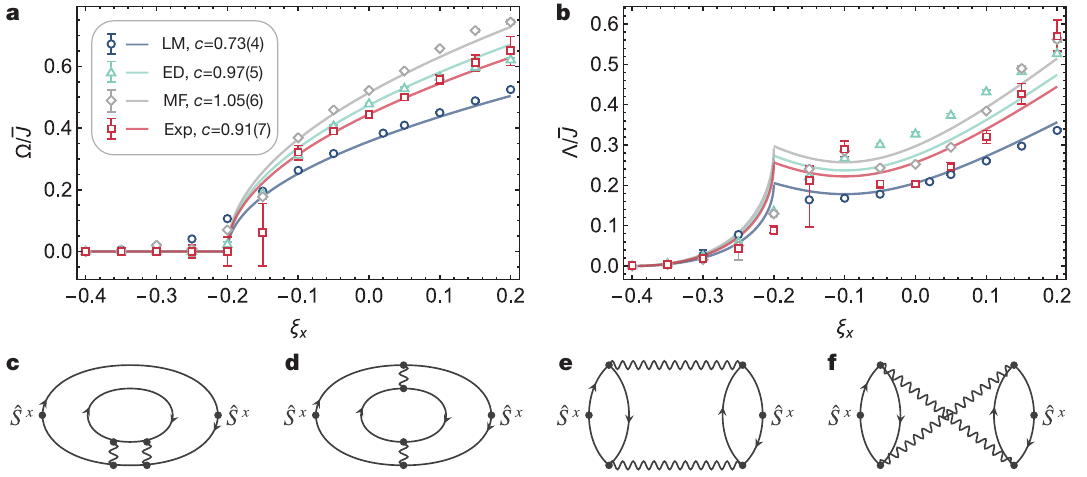}
    \caption{\textbf{Theoretical results and diagrammatic analysis.} \textbf{a},\textbf{b}, We present the frequency (\textbf{a}) and decay rate (\textbf{b}) obtained respectively from the large-$M$ expansion (LM), exact diagonalization (ED), mean-field theory (MF), and the experimental data (Exp), which are represented by different colors \cite{code}. The solid lines in (\textbf{a}),(\textbf{b}) are simultaneous fittings of the frequency $\Omega/\bar{J}$ and the decay rate $\Lambda/\bar{J}$ using their theoretical piecewise functions determined from Eq.~\eqref{eqn:universalform}: $(\Omega,\Lambda)/\bar{J}=c(\sqrt{W_x},\sqrt{\Gamma})$ when $W_x>0$, whereas $(\Omega,\Lambda)/\bar{J}=c(0,\sqrt{\Gamma}-\sqrt{-W_x})$ when $W_x<0$. The uncertainties of the constants $c$ in the parentheses denote the $95\%$ confidence intervals. \textbf{c-f}, These four diagrams represent contributions from (\textbf{c}) $\sum_{i,a} \Tr \left[ \hat{H}^2   \hat{S}_{ia}^x \hat{S}_{ia}^x \right]$, (\textbf{d}) $\sum_{i,a} \Tr \left[ \hat{H} \hat{S}_{ia}^x \hat{H} \hat{S}_{ia}^x \right]$, (\textbf{e}) $\sum_{(i,a)\neq (j,b)} \Tr \left[ \hat{H} \hat{S}_{ia}^x \hat{H} \hat{S}_{jb}^x \right]$ and (\textbf{f}) $\sum_{(i,a)\neq (j,b)} \Tr \left[\hat{H}^2 \hat{S}_{ia}^x \hat{S}_{jb}^x \right]$. In each diagram, the dots represent spin operators, and two of them are labeled as $\hat{S}^x$ since they are the corresponding operators in the two-point correlator. The loops represent the trace of spin operators, while the arrows indicate the order of spin operator contractions. The wavy lines and their associated dots denote the vertices of random spin interactions. } 
    \label{diagram}
    \end{center}
\end{figure*}

In this experiment, we consider three different initial density matrices, denoted as $\hat{\rho}_\alpha\propto \mathbbm{1}+\epsilon\hat{O}_\alpha$. Here, $\hat{O}_\alpha=\sum_{ia}\hat{S}^\alpha_{ia}$ ($\alpha=\hat{x}$, $\hat{y}$, or $\hat{z}$) represents the total spin along different directions, and $\epsilon\approx 6.4\times 10^{-5}$. We evolve the initial density matrix under the Hamiltonian \eqref{Heisenberg}. Subsequently, we measure $\langle\hat{O}_\alpha(t)\rangle$. As discussed earlier, the result corresponds to the normalized auto-correlation function $\mathcal{C}_\alpha(t)$. The most remarkable finding of this experiment is the discovery of a universal functional form for $\mathcal{C}_\alpha(t)$. Specifically, for $\alpha=\hat{x}$, we introduce two quantities, namely, $W_x$ and $\Gamma$, which are quadratic polynomials of the microscopic parameters $\xi_\alpha$ proposed in Ref.~\cite{Zhou}:

\begin{align}
&W_x\equiv-\xi^2_x+\xi^2_y-4\xi_y\xi_z+\xi^2_z, \label{W} \\
&\Gamma\equiv\xi^2_x+\xi^2_y+\xi^2_z. \label{Gamma}
\end{align} 

Using these two polynomials, we can introduce two characteristic energy scales $\hbar\omega_x \equiv c\hbar \sqrt{|W_x|}J$ and $\hbar\lambda \equiv c\hbar\sqrt{\Gamma}J$. Here $c$ is an $o(1)$ constant.  
For $\alpha=\hat{y}$ or $\alpha=\hat{z}$, we can introduce $W_y$ and $W_z$ through permutation as $W_y=-\xi^2_y+\xi^2_z-4\xi_z\xi_x+\xi^2_x$ and $W_z=-\xi^2_z+\xi^2_x-4\xi_x\xi_y+\xi^2_y$. $\hbar\omega_y$ and $\hbar\omega_z$ are then defined correspondingly. We find that $\mathcal{C}_\alpha(t)$ can be well described by
\begin{equation}\label{eqn:universalform}
\left\{ \begin{aligned}
&a\cos(\omega_\alpha t+\phi)\mathrm{e}^{-\lambda t},  \   \  \text{if} \  \ W_\alpha>0, \\
&a\cosh(\omega_\alpha t+\phi)\mathrm{e}^{-\lambda t}\approx a'\mathrm{e}^{-(\lambda-\omega_\alpha)t}  \   \  \text{if} \  \ W_\alpha<0, \end{aligned}
\right.
\end{equation} 
where $a$, $a'$ and $\phi$ are non-universal constants. This functional form is motivated by the quasinormal
mode analysis for non-equilibrium dynamics. Quasinormal modes are collective modes with complex frequencies $\omega_a-i\lambda_a$, which govern the dynamically oscillatory and decaying response in strongly interacting systems \cite{normalmode1,normalmode2}. Here, $a$ labels different modes. In the long-time limit, we can retain only the mode with the smallest $\lambda_a$, resulting in the functional form proposed in Eq.~\eqref{eqn:universalform} (see Methods and Supplementary Information \cite{supple} for detailed derivation). Notably, our results firstly reveal the universal scaling functions between $(\omega_\alpha, \lambda)$ and the microscopic parameters in the Hamiltonian, which has not been accomplished before to our best knowledge. As a consequence, this framework easily enables the establishment of a precise criterion for determining the presence of oscillatory or monotonic decay in spin relaxation dynamics. By offering a quantitative understanding, this advancement marks a significant step forward in alignment with prior research~\cite{Fine2004, Fine2005, NMR_UniversalDecay2008, NMR_UniversalDecay2011, NMR_UniversalDecay2012}. Despite of the effectiveness and simplicity of Eq.~(7), it is observed that it leads to larger deviations from the experimental data around the transition point $W_\alpha=0$, where the multi-mode dynamics become more significant.

Eq.~\eqref{eqn:universalform} is demonstrated by the experimental data presented in Fig.~\ref{polarization}. We polarize the system initially in three different directions $\alpha=\hat{x},\ \hat{y},\ \hat{z}$ respectively and then measure $\langle\hat{O}_\alpha\rangle$ for $\xi_z=0.2$ and $\xi_x \in [-0.4,0.2]$. The spin depolarization dynamics of $\langle\hat{O}_\alpha\rangle$ are depicted in Fig.~\ref{polarization}a-c. We fit these curves using the function $A\cos(\Omega t+\Phi)e^{-\Lambda t}$ and obtain $\Omega$ and $\Lambda$ for each case. Importantly, this approach does not assume the existence of an oscillating-to-monotonic transition. Comparing to Eq.~\eqref{eqn:universalform}, we predict $(\Omega,\Lambda)=(\omega_\alpha,\lambda)$ when $W_\alpha>0$, whereas $(\Omega,\Lambda)=(0,\lambda-\omega_\alpha)$ when $W_\alpha<0$.

Given the constraint $\sum_\alpha\xi_\alpha=0$, we have $W_x=-6\xi_y\xi_z$, $W_y=-6\xi_z\xi_x$, and $W_z=-6\xi_x\xi_y$. In this experiment, we fix $\xi_z=0.2$. Thus, when $\alpha=\hat{x}$, we find $W_x=1.2(0.2+\xi_x)>0$ for $\xi_x>-0.2$. As shown in Fig.~\ref{polarization}a,d, the auto-correlation function oscillates when $\xi_x>-0.2$, and the frequency $\Omega/\bar{J}$ scales as $c\sqrt{W_x}$, where $\bar{J}$ is the average value of $J$ determined by experiment. Fig.~\ref{polarization}g also demonstrates that $\Lambda/\bar{J}$ scales with $c\sqrt{\Gamma}$ for $\xi_x>-0.2$ and scales with $c(\sqrt{\Gamma}-\sqrt{-W_x})$ for $\xi_x<-0.2$. From the fitting, we obtain $c=0.91(7)$. Similarly, for $\alpha=\hat{y}$, we have $W_y=-1.2 \xi_x>0$ for $\xi_x<0$, where the frequency $\Omega/\bar{J}$ fits $c\sqrt{W_y}$ and $\Lambda/\bar{J}$ fits $c\sqrt{\Gamma}$. For $\xi_x<0$, the frequency is zero, and $\Lambda/\bar{J}$ fits $c(\sqrt{\Gamma}-\sqrt{-W_y})$. The fittings result is $c=0.91(7)$, as shown in Fig.~\ref{polarization}b,e,h. For $\alpha=\hat{z}$, $W_z=6\xi_x(0.2+\xi_x)>0$ for $\xi_x>0$ or $\xi_x<-0.2$, where the frequency fits $c\sqrt{W_z}$ and $\Lambda/\bar{J}$ also fits $c\sqrt{\Gamma}$. For $-0.2<\xi_x<0$, the frequency is zero, and $\Lambda/\bar{J}$ fits $c(\sqrt{\Gamma}-\sqrt{-W_z})$. The fittings yield $c=0.87(10)$, as shown in Fig.~\ref{polarization}c,f,i. The constants $c$ obtained from the three fittings are consistent with each other within the error bars.

As a self-consistent check, we note that when $\Gamma=-W_x$, $\Lambda=0$, and our ansatz shows that $\mathcal{C}_x(t)$ does not decay at all. It is easy to observe that $\Gamma=-W_x$ implies $\xi_y=\xi_z$, and the system restores spin rotational symmetry along $\hat{x}$. Therefore, $\hat{O}_x$ commutes with the Hamiltonian, and the total spin along $\hat{x}$ should not evolve in time. Similar conditions hold for $\alpha=\hat{y}$ or $\hat{z}$. This observation is also consistent with our experimental findings, indicating that our system remains coherent and the decoherence effect is negligible within the experimental timescale. Furthermore, these scaling behaviors have been confirmed by exact diagonalization calculations and approximation methods such as large-$M$ expansion and mean-field theory (see the Methods for further details) \cite{Zhou,code}. Each theoretical approach has its own advantages and disadvantages: The exact-diagonalization method captures the exact non-equilibrium quantum dynamics for $\operatorname{SU}(2)$ spin, but only with a small system size $N$. The semi-classical method captures the non-equilibrium spin dynamics through the Landau–Lifshitz equation of the non-equilibrium dynamics, with intermediate system size $N$. The large-$M$ expansion captures the leading order contribution to non-equilibrium dynamics in large system size $N$, which is rigorous at large-$M$ for $\operatorname{SU}(2)\times \operatorname{SU}(M)$ spin. It is remarkable that the same combinations of the anisotropic parameters in Hamiltonian enter the non-equilibrium dynamics, leading to the universal polynomial scaling on oscillation frequency and decay rate. In Fig.~\ref{diagram}a,b, we compare $\Omega$ and $\Lambda$ obtained by these three theoretical methods with the experimental data and find good agreements. All the theory results also obey the universal function form shown in Eqs.~\eqref{W}-\eqref{eqn:universalform}, with a slightly different value $c$. 

Below, we will discuss some physical intuitions as to why the quantities $W_\alpha$ and $\Gamma$ emerge as universal parameters in the quench dynamics. In low-energy physics, universality arises when a specific set of diagrams becomes the most relevant one and dominates the physical process under consideration, for instance, near a symmetry-breaking phase transition point in the Landau paradigm \cite{RMP}. In our case, we argue that the same reasoning applies to the emergence of universality in the quench dynamics, albeit with a focus on the infinite temperature auto-correlation function $\mathcal{C}_\alpha(t)$. 

Without loss of generality, we consider $\alpha=\hat{x}$ and the correlation function $\mathcal{C}_x(t)=\frac{1}{c_O}\sum_{ij,ab}\langle\hat{S}^x_{ia}(t)\hat{S}^x_{jb}(0)\rangle$. The experimental result suggests that dominant contributions to $\langle\hat{S}^x_{ia}(t)\hat{S}^x_{jb}(0)\rangle$ contain a few interaction channels, which can be identified by examining these terms $\langle\hat{H}\hat{S}^x_{ia}\hat{H}\hat{S}^x_{jb}\rangle$ and $\langle\hat{H}^2\hat{S}^x_{ia}\hat{S}^x_{jb}\rangle$ \cite{supple}. This argument can be justified using the large-$M$ and mean-field theory \cite{Zhou}, which are two of the most popular approximation schemes for studying spin models. The large-$M$ expansion has been particularly successful in studying a randomly interacting spin model known as the Sachdev-Ye (SY) model \cite{SY}, which was later extended to the celebrated Sachdev-Ye-Kitaev model (SYK) \cite{SYK1,SYK2,SYK3}. These two terms can be illustrated by the Feynman diagrams shown in Fig.~\ref{diagram}c-f, with contributions from $i=j$ given by Fig.~\ref{diagram}c,d, and contributions from $i\neq j$ given by Fig.~\ref{diagram}e,f. Lengthy but straightforward calculations demonstrate that the contribution from diagram Fig.~\ref{diagram}c is exactly proportional to $\Gamma$ as defined in Eq.~\eqref{Gamma}, while the contributions from diagrams Fig.~\ref{diagram}d-f can be combined into $W_x$ as defined in Eq.~\eqref{W} \cite{supple}. The fact that the experimental data can be well captured by these parameters reveals the underlying physics behind the dynamics, indicating that this non-equilibrium process is dominated by the interaction processes shown in Fig.~\ref{diagram}.

This universal behavior can also be applied to similar models realized in other physical systems. As a concrete example, a similar random spin model has been realized by Rydberg atoms excited in an ultracold atomic gas, and a non-monotonic dependence of the relaxation dynamics on the anisotropic parameter ratio has been observed \cite{Rydberg_Floquet2021}. This dependence also aligns with the dependency of the decay rate on the anisotropic parameters presented in this work.

\begin{figure}
    \begin{center}
    \includegraphics[width=0.95\linewidth]{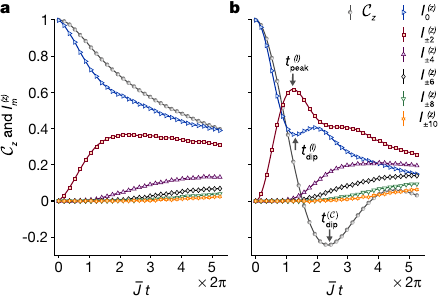}
    \caption{\textbf{The multiple quantum coherences for randomly interacting spin models.} MQC intensities of different order in the eigenbasis of $\hat{O}_z$, denoted as $I_m^{(z)}$, as a function of the evolution time $\bar{J}t$. The values of $(\xi_x,\xi_y,\xi_z)$ are $(-0.125,-0.025,0.15)$ for \textbf{a} and $(-0.1,0.1,0)$ for \textbf{b}. At each time of $\bar{J}t$, the data is normalized by the value $F(\phi,t)|_{\phi=0}=\sum_m I_m^{(z)}(t)$, in order to ensure that $\sum_m I_m^{(z)}(t)=1$. The different curves represent various values of the coherence order $m$. The error bars ($\sim 10^{-4}$) are incorporated within the markers of the data points. The solid lines denote the cubic spline interpolations. In \textbf{b}, we extract that $I_0^{(z)}(t)$ reaches its initial trough at $\bar{J}t_{\mathrm{dip}}^{(I)}/2\pi\approx1.29$, and $I_{\pm 2}^{(z)}(t)$ reaches its initial peak at $\bar{J}t_{\mathrm{peak}}^{(I)}/2\pi\approx1.22$, while $\mathcal{C}_z(t)$ reaches its initial trough at $\bar{J}t_{\mathrm{dip}}^{(\mathcal{C})}/2\pi\approx 2.37$.} 
    \label{MQC}
    \end{center}
\end{figure}

For a given direction $\alpha$, whether $W_\alpha>0$ or $W_\alpha<0$ not only distinguishes two types of quench dynamics for the two-point correlator but also marks the difference in higher-order correlators. We now investigate the higher-order correlation by studying the multiple quantum coherences (MQCs) \cite{Ernst1976,Ernst1977,Drobny1978,Bodenhausen1980,Pines1983,Pines1985}. The experimental protocol, such as described in Refs.~\cite{Pines1983,Pines1985}, can be used to extract the MQC spectrum by utilizing its relation with the out-of-time-order (OTO) correlator $F(\phi, t)=\text{Tr}[\text{e}^{-\text{i}\hat{O}_\alpha \phi} \hat{O}_\alpha(t) \text{e}^{\text{i}\hat{O}_\alpha \phi} \hat{O}_\alpha(t)]$ \cite{NMR_OTOC2017,Ion_OTOC2017,Garttner2018,OTOC1,OTOC2,OTOC3,Maldacena2016,Hosur2016,Landsman2019}. $F(\phi, t)$ can be expanded as $F(\phi,t)=\sum_m I_m^{(\alpha)}(t)e^{-\text{i}m\phi}$, where $I_m^{(\alpha)}$ represents the intensity of the $m$th-order quantum coherences in the eigenbasis of $\hat{O}_\alpha$. $I_0^{(\alpha)}$ incorporates both the zero-quantum coherences and populations (diagonal elements). 

In this protocol, we first evolve $\hat{\rho}_\alpha$ with the many-body Hamiltonian $\hat{H}$ for a time duration $t$. Then, we apply a spin rotation with angle $\phi$ given by $\exp(-\text{i}\hat{O}_\alpha \phi)$. This is followed by another evolution under the Hamiltonian $-\hat{H}$ for the same time duration $t$. Afterward, we measure the expectation value $\langle\hat{O}_\alpha\rangle$. Similar to the measurement of the auto-correlation function, this protocol allows us to measure the OTO correlator $F(\phi,t)$, since 
\begin{equation}
\begin{aligned}
\langle\hat{O}_\alpha\rangle&=\text{Tr}[\text{e}^{\text{i}\hat{H}t/\hbar} \text{e}^{-\text{i}\hat{O}_\alpha \phi} \text{e}^{-\text{i}\hat{H}t/\hbar} \hat{\rho}_\alpha \text{e}^{\text{i}\hat{H}t/\hbar} \text{e}^{\text{i}\hat{O}_\alpha \phi} \text{e}^{-\text{i}\hat{H}t/\hbar} \hat{O}_\alpha]\\
&=\text{Tr}[\text{e}^{-\text{i}\hat{O}_\alpha \phi} \hat{\rho}_\alpha(t) \text{e}^{\text{i}\hat{O}_\alpha \phi}  \hat{O}_\alpha(t)]\propto F(\phi, t).
\end{aligned}
\end{equation}
 Then, by varying the rotation angle $\phi$ and time duration $t$, and subsequently applying a Fourier transform with respect to $\phi$, the MQC spectrum $\{I_m^{(\alpha)}(t)\}$ can be obtained. It is worth noting that $\partial^2 F(\phi, t)/\partial\phi^2|_{\phi=0}=-\sum_mI_m^{(\alpha)}(t)m^2=\Tr([\hat{O}_\alpha(t), \hat{O}_\alpha]^2)$, which is the OTO commutator \cite{Garttner2018}. This connection between MQC and OTO commutator allows us to characterize information scrambling in the system \cite{Garttner2018}.

In Fig.~\ref{MQC}, we present the results of $I_m^{(\alpha)}(t)$ for two cases with $\alpha=\hat{z}$. Fig.~\ref{MQC}a depicts the case with $(\xi_x,\xi_y,\xi_z)=(-0.125,-0.025,0.15)$, where $W_z=-0.01875<0$. In this scenario, we observe a monotonic decay of $I_0^{(z)}$, with its weight gradually spreading into higher-order quantum coherences. Fig.~\ref{MQC}b illustrates the case with $(\xi_x,\xi_y,\xi_z)=(-0.1,0.1,0)$ and $W_z=0.06>0$. In this situation, clear oscillations are observed in both $I_0^{(z)}(t)$ and $I_{\pm2}^{(z)}(t)$. Besides, it seems that $I_0^{(z)}(t)$ and $I_{\pm2}^{(z)}(t)$ oscillate with a frequency roughly double that of $\mathcal{C}_z(t)$, as indicated by the time points when $I_0^{(z)}(t)$, $I_{\pm2}^{(z)}(t)$ and $\mathcal{C}_z(t)$ reach their initial trough or peak: $\bar{J}t_{\mathrm{dip}}^{(I)}/2\pi\approx1.29$, $\bar{J}t_{\mathrm{peak}}^{(I)}/2\pi\approx1.22$, and $\bar{J}t_{\mathrm{dip}}^{(\mathcal{C})}/2\pi\approx 2.37$. This is reasonable considering that the OTO commutator $\Tr([\hat{O}_z(t), \hat{O}_z]^2)$ involves a square of $\hat{O}_z(t)\hat{O}_z$. This observation is further verified by varying the Hamiltonian configurations (see Supplementary Information \cite{supple}).

To conclude, we experimentally study the far-from-equilibrium quench dynamics in randomly interacting spin models using solid-state NMR systems. The mean strength of random interaction is the only energy scale in the Hamiltonian governing the dynamics. Hence, this problem is intrinsically a strongly interacting many-body problem that lacks small perturbation parameters, and such a system at non-equilibrium belongs to the most challenging problems for developing physical understandings. The numerical method, like exact diagonalization, is limited to a system size much smaller than the actual physical system and does not provide insightful physical intuitions. The approximation scheme, such as large-M theory, does provide helpful intuition but involves uncontrolled errors. In light of these challenges, quantitative comparison between theory and experiment becomes particularly valuable. To this end, accurate calibration of Hamiltonian parameters and high-quality data on quantum dynamics with insignificant decoherence effects are required. Here, by reaching a consistency between experiment, approximated theory, and numerical diagonalization, we reveal a few universal parameters and uncover dominating interacting processes for this quench dynamics, which can be generalized to similar non-equilibrium dynamics in cold atoms, NV centers, and other systems. 

~

\textbf{Methods}

\textit{Hamiltonian Engineering.} By periodically applying the radio-frequency (RF) pulse sequence to the natural dipolar Hamiltonian, we can engineer the desired form of the anisotropic Heisenberg models \eqref{Heisenberg} as an effective time-independent Hamiltonian by the average Hamiltonian theory \cite{Haeberlen1968}. The basic building block of the RF pulse train is an 8-pulse sequence initially introduced for studying multiple quantum coherences \cite{Suter1987}. 
%More recently, this sequence has been employed to investigate the intricate dynamics of spin chains. 
%This exploration has encompassed phenomena like localization \cite{Wei2018}, prethermalization \cite{Wei2019, Peng2021}, and hydrodynamics \cite{Peng2023}. 
% such sentence should not be written in Method. It is not " method". If needed, it should be cited in maintext. 
Explicitly, the 8-pulse sequence is represented as follows:

\begin{equation*}
    \begin{aligned}(\tau_z,\mathbf{x},\tau_y,\mathbf{y},2\tau_x,\mathbf{y},\tau_y,\mathbf{x},2\tau_z,\mathbf{x},\tau_y,\mathbf{y},2\tau_x,\mathbf{y},\tau_y,\mathbf{x},\tau_z),
    \end{aligned}
\end{equation*}

where $\mathbf{x}$ and $\mathbf{y}$ denote the RF pulses that induce collective $\pi/2$ rotations along the $\hat{x}$ and $\hat{y}$ directions, respectively.
By adjusting the pulse intervals $\tau_\alpha$ such that $\tau_\alpha=\left[1+\xi_\alpha\right]\tau$, we can realize different configurations of the anisotropic parameters $(\xi_x,\ \xi_y,\ \xi_z)$ to the leading order of the Magnus expansion \cite{Magnus1954,supple}.

\textit{Hamiltonian Calibration.} We calibrate the actually realized anisotropic parameters $(\xi_x',\ \xi_y',\ \xi_z')$ and the weight of residual terms other than  $\hat{S}^\alpha_{ia}\hat{S}^\alpha_{jb}$ in the effective Floquet Hamiltonian $\hat{H}_{\mathrm{F}}$. 
%The original sentence is misleading, it is not clear whether $(\xi_x',\ \xi_y',\ \xi_z')$ denotes the realized anisotropic parameters or the deviation from the target values. Please check the modified sentence. 
The idea is to employ Floquet prethermalization hypothesis \cite{Mori2016,Abanin2017}. It assumes that the system attains a quasi-stationary state $\hat{\rho}_{\mathrm{pre}}$, approximately characterized by a canonical ensemble associated with the effective Hamiltonian $\hat{H}_{\mathrm{F}}$, before heated to the infinite temperature. We have
\begin{equation}
    \hat{\rho}_{\mathrm{pre}}\approx\frac{e^{-\beta_{\text{eff}}\hat{H}_{\mathrm{F}}}}{\Tr\small(e^{-\beta_{\text{eff}}\hat{H}_{\mathrm{F}}}\small)}\propto \mathbbm{1}-\beta_{\text{eff}}\hat{H}_{\mathrm{F}},
\end{equation} 
where $\beta_{\text{eff}}$ represents an effective inverse temperature, determined by the initial state $\hat{\rho}_0$ through the energy conservation $\Tr(\hat{\rho}_0\hat{H}_{\text{F}})=\Tr(\hat{\rho}_\text{pre}\hat{H}_{\text{F}})$. The Floquet prethermalization has been experimentally demonstrated in spin chains with dipolar interactions \cite{NMR_prethermal2019,NMR_prethermal2021}. 

As elaborated in the Supplementary Information \cite{supple}, we initially prepare states with finite inverse spin temperatures, and then allow them to prethermalize under the $\hat{H}$ Eq.~\eqref{Heisenberg} with various anisotropic configurations for a time period of $\bar{J} t\ge 14\pi$. In addition, we prepare dipolar-ordered states \cite{Jeener1967, Cho2003} as a reference state using the Jeener-Broekaert method \cite{Jeener1967}. The traceless components of the density matrix of these states are given by
\begin{equation}
  \begin{aligned}
      \delta\hat{\rho}_x^\mathcal{D}&\propto \sum\limits_{i\neq j,ab}J_{ij}(-\hat{S}^y_{ia}\hat{S}^y_{jb}-\hat{S}^z_{ia}\hat{S}^z_{jb}+2\hat{S}^x_{ia}\hat{S}^x_{jb}),
  \end{aligned}
 \end{equation}
and the other two dipolar-ordered states $\delta\hat{\rho}_y^\mathcal{D},\ \delta\hat{\rho}_z^\mathcal{D}$ can be determined through cyclic permutations.
In experiment, we measure the inner products between the prethermal state and each of the dipolar-ordered states. These inner products are proportional to the anisotropic parameters
\begin{equation}
\Tr\left(\hat{\rho}_{\mathrm{pre}}\delta\hat{\rho}_x^\mathcal{D}\right)\propto(2\xi_x'-\xi_y'-\xi_z')\sum\limits_{i\neq j,ab}J_{ij}^2\propto\xi_x'.
\end{equation}
Similarly, we find $\Tr\small(\hat{\rho}_{\mathrm{pre}}\delta\hat{\rho}_y^\mathcal{D}\small)\propto\xi_y'$ and $\Tr\small(\hat{\rho}_{\mathrm{pre}}\delta\hat{\rho}_z^\mathcal{D}\small)\propto\xi_z'$. This determines the actual anisotropic parameters $(\xi_x',\ \xi_y',\ \xi_z')$. The discrepancies between these parameters and their target values $(\xi_x,\ \xi_y,\ \xi_z)$ are quantified by $\Delta\equiv\left|\bm{\xi'}-\bm{\xi}\right|/\left|\bm{\xi}\right|$. Throughout all the realized configurations, the values of $\Delta$ are calibrated to be within $3\%$ \cite{supple}.
% What is the meaning of all configurations and majority of realizations, do they contradict ?
% In the last version, I wanted to say that for the most times of calibration, Delta<2%. In this new version, I make Delta<3%, which means that for all 10 times of calibrations, Delta<3%.

The weight of residual terms denoted as $\cdots$ in the overall effective Hamiltonian Eq.~\eqref{Heisenberg} is defined by $\varepsilon\equiv\sqrt{\Tr\small(\cdots^2\small)/\Tr\small(\hat{H}^2\small)}$. It primarily leverages the orthogonal relationships between the residual term in the prethermal states $\hat{\rho}_\mathrm{pre}$ and the dipolar-ordered states $\delta\hat{\rho}^\mathcal{D}_\alpha$, and incorporates more inner product measurements. The values of $\varepsilon$ are determined to be less than $20\%$ across all the realized configurations \cite{supple}.

\textit{Exact Diagonalization.} In the exact diagonalization calculation, we are restricted to a simplified model consisting of a single spin-1/2 on each molecule with system size up to $N=8$. The Hamiltonian is 
\begin{equation}
\hat{H}_{\text{ED}}=\sum\limits_{i<j}J_{ij}(\xi_x\hat{S}^x_{i}\hat{S}^x_{j}+\xi_y\hat{S}^y_{i}\hat{S}^y_{j}+\xi_z\hat{S}^z_{i}\hat{S}^z_{j}), \label{ED}
\end{equation} 
where $J_{ij}$ is modelled as a random variable obeying a normal distribution $J_{ij}\sim\mathcal{N}[0,(2J/\sqrt{N})^2]$. For each disorder realization of $J_{ij}$, we prepare the initial state as a thermal density matrix, denoted as $\hat{\rho} \propto \exp(-\beta(\hat{H}_{\text{ED}}+\hat{\delta H}))$, where $\beta=\hbar/(k_BT)$ denotes the inverse temperature and an external polarization field is introduced as $\hat{\delta H}=-g \sum_i \hat{S}^x_{i}$. We fix $\beta J=0.2$ and $g/J=2$, as explained later in \textit{Parameters in Numerical Simulations}. The system is then evolved under the Hamiltonian Eq.~\eqref{ED}, and the result is averaged over $10^3$ random realizations. 
%{\color{red}In the Supplementary Information \cite{supple}, we further demonstrate the validity of our results for moderate changes in parameters.} 

\textit{Large-$M$ Expansion.} We can transform the randomly interacting spin model Eq.~\eqref{Heisenberg} into a theory of randomly interacting fermions by adopting the Abrikosov fermion representation. In this representation, the spin operators are expressed as $\hat{S}^\alpha_{ia}=\frac{1}{2}\sum_{ss'}\hat{c}^\dagger_{ia,s}(\sigma^\alpha)_{s s^\prime} \hat{c}_{ia,s^\prime}$ ($s,s'=\uparrow,\downarrow$), limited to the single occupation subspace. Our main interest lies in the spin polarization dynamics, which can be expressed as $\langle \hat{O}_x (t) \rangle=-i N (G^{\gtrless}_{\uparrow\downarrow}(t,t)+G^{\gtrless}_{\downarrow\uparrow}(t,t))/2$. Here the real-time Green's functions are defined as $G^>_{ss'}(t_1,t_2)\equiv-i  \langle c_{ia,s}(t_1)c_{ia,s'}^\dagger(t_2)\rangle$ and $G^<_{ss'}(t_1,t_2)\equiv i  \langle c_{ia,s'}^\dagger(t_2)c_{ia,s}(t_1)\rangle$. The evolution of these Green's functions can be described by a set of classical equations, commonly known as the Kadanoff-Baym equation
\begin{equation}
  \begin{aligned}\label{eq:KBeq}
    i\partial_{t_1}&G^\gtrless =\Sigma^R\circ G^\gtrless+ \Sigma^\gtrless\circ G^A, \\
    -i\partial_{t_2}&G^\gtrless =G^R\circ\Sigma^\gtrless+ G^\gtrless\circ \Sigma^A,
  \end{aligned}
\end{equation}
where $G^{R/A}$ is the retarded/advanced Green's function. $\Sigma^\gtrless$ and $\Sigma^{R/A}$ represent real-time self-energies, which satisfy $\Sigma^{R/A}=\pm \Theta\left(\pm t_{12}\right)\left(\Sigma^>-\Sigma^<\right)$. To make further theoretical advancements, an SU(M)$\times$SU(2) generalization has been introduced, similar to the approach used in \cite{large-M,Zhou}. By taking both the large-$N$ and the large-$M$ limit, melon diagrams play a dominant role in the self-energies, similar as in the SYK model. This leads to
\begin{equation}\label{eq:selfreal}
  \begin{aligned}
    \Sigma^{\gtrless}(t_1,t_2)=\frac{J^2}{4}& \sum_{\alpha,\alpha^\prime} \xi_\alpha \xi_{\alpha^\prime} \sigma^{\alpha^\prime} G^{\gtrless}(t_1,t_2)\sigma^\alpha  \\&\text{Tr}\left[\sigma^{\alpha^\prime} G^{\gtrless}(t_1,t_2)\sigma^\alpha  G^{\lessgtr}(t_2,t_1)\right].
  \end{aligned}
\end{equation}
Numerically, we prepare the system in an initial state described by a thermal ensemble at $\beta J=0.2$ with a polarization field $g/J=2$. The corresponding initial Green's functions are obtained through iterations. Subsequently, we evolve $G^\gtrless$ by combining Eq.~\eqref{eq:KBeq} and Eq.~\eqref{eq:selfreal} to determine $\langle \hat{O}_x (t) \rangle$. Besides, by conducting a quasinormal mode analysis, one
can analytically derive the long-time spin relaxation dynamics Eq.~\eqref{eqn:universalform} within the large-$M$ approximation \cite{Zhou}. This calculation is elaborated in the Supplementary Information \cite{supple}.

\textit{Mean-field Theory.} Another theoretical scheme for analyzing randomly interacting spin models is the mean-field theory. Here, we introduce the average polarization on each molecule as $\hat{M}_{i}^\alpha=\frac{1}{N_a}\sum_{a}\hat{S}_{ia}^\alpha$. Due to the statistical averaging, we expect the fluctuation of $\hat{M}_{i}^\alpha$ to be small, allowing us to approximate it as a classical vector $M_{i}^\alpha$. The Heisenberg equation for $\hat{M}$ then becomes:
\begin{equation}\label{eq:mf}
    \frac{dM_{i}^\alpha}{dt}=N_a \sum_{j,\beta\gamma}J_{ij}\epsilon^{\alpha\beta\gamma} \xi_\beta M_j^\beta M_i^\gamma.
\end{equation}
In the numerical simulation, we investigate a system with $2\times 10^3$ molecules. The initial configuration of $M_{i}^\alpha$ is randomly generated using an independent Gaussian distribution, with mean value $\overline{\bm{M}_{i}}=(\frac{\beta g}{4},0,0)$, $\beta g=0.4$ and variance $\overline{(\delta M_{i}^\alpha)^2}=1/(4 N_a)$. Subsequently, we evolve $M_{i}^\alpha$ according to Eq.~\eqref{eq:mf} for each random realization and compute $\langle \hat{O}^\alpha \rangle= N_a \sum_{i}M_{i}^\alpha$. The final result is then averaged over 20 independent simulations.

\textit{Parameters in Numerical Simulations.}
The NMR experiment is conducted at room temperature, necessitating the conditions $\beta J \ll 1$ and $\beta g \ll 1$. Additionally, the external magnetic field strongly polarizes the state, allowing the initial state to be approximated by $\hat{\rho} \propto \mathrm{e}^{- \beta\hat{H}_{\mathrm{dip}} + \beta g\sum_{ia} \hat{S}_{ia}^\alpha} \approx \hat{\mathbbm{1}} + \beta g \sum_{ia} \hat{S}_{ia}^\alpha$. This requires that the magnitude of the external field must be significantly larger than the characteristic strength of the dipolar interaction $\hat{H}_{\mathrm{dip}}$ [see Eq.~(3)], i.e., $g/J\gg 1$. For numerical simulations, all parameters satisfy these conditions. 
In the Supplementary Information \cite{supple}, we demonstrate that a moderate change of parameters yield qualitatively similar results. In particular, the oscillation frequencies and decay rates remain independent of $\beta$ and $g$ \cite{supple}.

%In summary 
%\begin{itemize}
    %\item $\beta J |\bm{\xi}| \ll 1$, $\beta g  \ll 1$ and $g/(J|\bm{\xi}|) \gg 1$:\ \ for large-$M$ and exact-diagonalization method. 
    %\item $\beta g\ll 1$:\ \ for mean-field method.
%\end{itemize}

%\textbf{Data availability}

%Source data are available at \cite{code}. Further data are available from the corresponding authors upon reasonable request. 

%\textbf{Code availability}

%All codes generated during this study are available at \cite{code}.

~

\end{document}

\title{Supplementary Information}

% \date{\today}

\maketitle
\tableofcontents
\section{Notations}
\begin{table}[htbp]
    \centering
    \renewcommand{\arraystretch}{2}
    \begin{tabular}{p{2cm}<{\centering}|p{6cm}<{\centering}}
        \hline
        \hline
        Notations & Operators \\
        \hline
        $\hat{O}_x$ & $\sum_{ia} \hat{S}^x_{ia}$\\
        \hline
        $\hat{O}_y$ & $\sum_{ia} \hat{S}^y_{ia}$\\
        \hline
        $\hat{O}_z$ & $\sum_{ia} \hat{S}^z_{ia}$\\
        \hline
        $\hat{R}_x(\phi)$ & $\exp(-\mathrm{i}\hat{O}_x\phi)$\\
        \hline
        $\hat{R}_y(\phi)$ & $\exp(-\mathrm{i}\hat{O}_y\phi)$\\
        \hline
        $\hat{R}_z(\phi)$ & $\exp(-\mathrm{i}\hat{O}_z\phi)$\\
        \hline
        $\hat{D}_x$ & $\frac{\hbar}{2}\sum\limits_{i\ne j, ab} J_{ij}\left(-\hat{S}^y_{ia}\hat{S}^y_{jb}-\hat{S}^z_{ia}\hat{S}^z_{jb}+2\hat{S}^x_{ia}\hat{S}^x_{jb}\right)$\\
        \hline
        $\hat{D}_y$ & $\frac{\hbar}{2}\sum\limits_{i\ne j, ab}J_{ij}\left(-\hat{S}^z_{ia}\hat{S}^z_{jb}-\hat{S}^x_{ia}\hat{S}^x_{jb}+2\hat{S}^y_{ia}\hat{S}^y_{jb}\right)$\\
        \hline
        $\hat{D}_z$ & $\frac{\hbar}{2}\sum\limits_{i\ne j, ab} J_{ij}\left(-\hat{S}^x_{ia}\hat{S}^x_{jb}-\hat{S}^y_{ia}\hat{S}^y_{jb}+2\hat{S}^z_{ia}\hat{S}^z_{jb}\right)$\\
        \hline
        \hline
    \end{tabular}
    \caption{\textbf{Notations for operators used in this paper}. $\hat{S}_{ia}^\alpha$ ($\alpha=x,\ y,\ z$) is the spin-1/2 operator for each $^1\mathrm{H}$. $i$ and $j$ label molecules while the indices $a, b = 1, . . . , 16$ label $^1\mathrm{H}$ within each molecule.  }
    \label{<label>}
\end{table}

\section{Experimental system} \label{sec2}
The experimental system we utilize is a sample of adamantane (C\textsubscript{10}H\textsubscript{16}) in its powdered form. Adamantane undergoes a first-order phase transition around 208 K~\cite{chang1960,mccall1960}. At room temperature, it crystallizes into a face-centered cubic structure with the space group Fm3m~\cite{nordman1965}.
In this phase, the nearly-spherical molecules undergo rapid rotations around their centers due to thermal motion, with a characteristic time scale of $10^{-11}$ s \cite{resing1969a}. Therefore, as discussed in the main text, the intra-molecular dipolar interaction is eliminated. Meanwhile, the inter-molecular dipolar interaction primarily depends on molecule indices \cite{resing1969a,mccall1960,smith1961a}.

The focus of our investigation is on the 1/2-spin \textsuperscript{1}H nuclei that comprise the interaction network of interest. The spin-lattice relaxation time is determined to be $ T_1 = 1.3~\text{s} $. Besides, the most abundant carbon isotope (99\% \textsuperscript{12}C) does not possess any spin. During the experiments, the sample is placed at room temperature in a homogeneous magnetic field of $ B_0=9.4 $ T, generated by a superconducting magnet. The Hamiltonian for the \textsuperscript{1}H system is given by:
\begin{equation}
    \hat{H}_0=\hat{H}_{\rm Z}+\hat{H}_{\mathrm{CS}}+\hat{H}_{\mathrm{dip}}=-\hbar\omega_{\rm H}\hat{O}_z+2\pi\hbar\sum_j\nu_{j}\hat{S}^z_j+\hat{H}_{\rm dip}.
    \label{LabSysHam}
\end{equation}
Here the first term $ \hat{H}_{\rm Z} $ is the Zeeman interaction with a uniform Larmor frequency $ \omega_{\rm H}=\gamma_{\rm H}B_0=400.15\times 2\pi$ MHz ($\gamma_{\rm H}$ is the proton's gyromagnetic ratio). The second term represents a combined distribution of the chemical shift and the magnetic field inhomogeneity. According to the liquid-state nuclear magnetic resonance (NMR) data of adamantane \cite{SDBSWeb}, the chemical shift, $\abs{\nu_j}$, varies within a few tens of Hertz. The line-width broadening caused by the magnetic field inhomogeneity is determined to be less than 12 Hz, as estimated from the linewidth of the water sample $1/T_2^*=12$ Hz. This term can thus be discarded since the time period of evolution being studied is only several milliseconds. The third term corresponds to the natural nuclear magnetic dipole-dipole interaction\cite{abragam1961}, which only incorporates the inter-molecular contribution:
\begin{equation}
    \hat{H}_{\mathrm{dip}}=\sum\limits_{i<j,ab}\frac{\mu_0}{4\pi}\frac{\hbar^{2}\gamma_{\mathrm{H}}^{2}}{R_{ij}^3}\left[\hat{\mathbf{S}}_{ia} \cdot \hat{\mathbf{S}}_{jb}-\frac{3\left(\hat{\mathbf{S}}_{ia} \cdot \mathbf{R}_{ij}\right)\left(\hat{\mathbf{S}}_{jb} \cdot \mathbf{R}_{ij }\right)}{R_{ij}^2}\right]. \label{dipolar}
\end{equation}
where $\hat{\mathbf{S}}_{ia}=(\hat{S}_{ia}^x,\hat{S}_{ia}^y,\hat{S}_{ia}^z)$ are spin-1/2 operators for each $^1\text{H}$.  $i$ and $j$ label molecules while the indices $a, b = 1, . . . , 16$ label $^1\text{H}$ within each molecule. $\mu_0$ is the vacuum magnetic permeability. $\mathbf{R}_{ij}$ denotes the displacement between centers of two molecules and $R_{ij}=|\mathbf{R}_{ij}|$.

\begin{figure}[htbp]
\begin{center}
\includegraphics[scale=1]{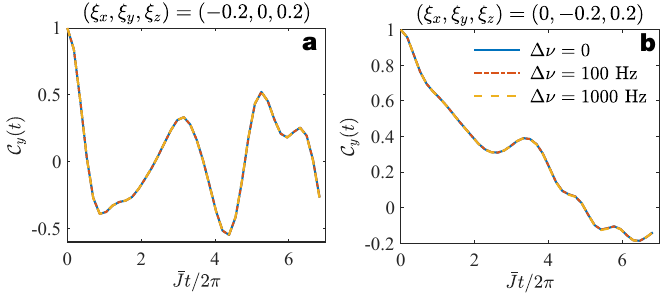}
    \caption{\textbf{Numerical simulations of the effect of static magnetic field inhomogeneities on the auto-correlation function $\mathcal{C}_y(t)$ via exact diagonalization.} In the simulations, the system Hamiltonian is $  \hat{H}=\hbar J\sum_{i<j\le N}(-\hat{S}_i^x\hat{S}_j^x-\hat{S}_i^y\hat{S}_j^y+2\hat{S}_i^z\hat{S}_j^z)$, and a 16-pulse sequence is designed to effectively engineer the target Hamiltonian $\hat{H}_{\mathrm{tar}}=\hbar J\sum_{i<j\le N}(\xi_x\hat{S}_i^x\hat{S}_j^x+\xi_y\hat{S}_i^y\hat{S}_j^y+\xi_z\hat{S}_i^z\hat{S}_j^z)$ with the same parameters as the experiments. The static magnetic field inhomogeneity is modeled as $2\pi\hbar\sum_j^N\nu_j\hat{S}_j^z$ with a random variable $\nu_j$, obeying a normal distribution $\nu_j\sim\mathcal{N} [0,\ (\Delta\nu)^2]$. By varying the standard deviation $\Delta\nu$, we demonstrate how it affects the pulse-engineered evolution of the auto-correlation function $\mathcal{C}_y(t)$ with the system size $N=8$, the constant coupling strength $J= (2\pi)\ 1460$ Hz, and the target Hamiltonian configurations (\textbf{a}) $\bm{\xi}=(-0.2,0,0.2)$ and (\textbf{b}) $\bm{\xi}=(0,-0.2,0.2)$ for instance. The results are averaged over 20 times of random realizations of $\nu_j$ for each $\Delta\nu$. }
\label{fig:simu_inhomogeneity}
\end{center}
\end{figure}

As the readout and the radio-frequency (RF) pulses in NMR are both implemented in a reference frequency $\omega_{\mathrm{r}}\approx\omega_{\mathrm{H}}$ (resonance condition), we consider the system Hamiltonian in the rotating reference frame, which is expressed as $\hat{H}_{\mathrm{s}}=e^{\mathrm{i}\hat{H}_{\mathrm{Z}}t/\hbar}\hat{H}_{0}e^{-\mathrm{i}\hat{H}_{\mathrm{Z}}t/\hbar}-\hat{H}_{\mathrm{Z}}$. The off-resonance effect, originating from uncertainties in calibrating the central transmission frequency, coupled with the chemical shift and magnetic field inhomogeneity, is effectively refocused within our multi-pulse sequences, as confirmed by the numerical simulation in \ref{fig:simu_inhomogeneity}. Considering that the Zeeman interaction $ \hat{H}_{\rm Z} $ is much stronger than the dipolar interaction $ \hat{H}_{\rm dip} $, the dipolar term only preserves the energy-conserving components after applying the secular approximation~\cite{abragam1961}. Consequently, the system Hamiltonian reduces to
\begin{align}
    \hat{H}_{\mathrm{s}}=\hat{H}_{\mathrm{dip}}^{\mathrm{sec}}=\hbar\sum_{i<j,ab}J_{ij}(-\hat{S}_{ia}^x\hat{S}^x_{jb}-\hat{S}_{ia}^y\hat{S}^y_{jb}+2\hat{S}^z_{ia}\hat{S}^z_{jb})=\hat{D}_z.
    \label{sysHam}
\end{align}
Here $ J_{ij}=\left(\frac{\mu_0}{4\pi}\right)\frac{\hbar\gamma_{\mathrm{H}}^{2}}{2R_{ij}^3}\left(1-3\cos^2\theta_{ij}\right) $. $ \theta_{ij} $ is the angle between $\mathbf{R}_{ij}$ and the magnetic field $\hat{z}$ direction.

Calculation of $J_{ij}$ based on the lattice structure shows that $J_{ij}$ has a zero mean ($\overline{J_{ij}}=0$), and its variance is related to the effective dipolar coupling strength $J$ defined by
\begin{equation}
    \begin{aligned}
        J^2&\equiv\frac{N}{4}\overline{\left\langle(J_{ij}-\overline{J_{ij}})^2\right\rangle}=\frac{N}{4}\overline{\langle J_{ij}^2\rangle},
    \end{aligned}
\end{equation}
where the overline $\overline{\cdots}$ denotes the average over molecule indexes $i$ and $j$, and the angle bracket $\left\langle\cdots\right\rangle$ means taking average over random orientations for a powder sample. Here, $N=N_\text{m}N_\text{a}$ is the total number of spins, in which $N_\text{m}$ is the number of molecules and $N_\text{a}=16$ represents the number of $\mathrm{^1H}$. By utilizing the translational symmetry of the lattice, we have 
\begin{equation}
    \begin{aligned}
        J^2\approx\frac{N_\text{a}}{4}\sum\limits_{j,j\ne i}^{N_\text{m}}\langle J_{ij}^2\rangle=\sum\limits_{j,j\ne i}^{N_\text{m}}\left[\left(\frac{\mu_0}{4\pi}\right)\frac{\hbar\gamma_{\mathrm{H}}^{2}}{R_{ij}^3}\right]^2\left\langle(1-3\cos^2\theta_{ij})^2\right\rangle.
    \end{aligned}
\end{equation}
Since
\begin{equation}
    \begin{aligned}
       \left\langle (1-3\cos^2\theta_{ij})^2\right\rangle&=\frac{1}{2}\int_{0}^{\pi}\dd\theta_{ij}\left(1-3\cos^2\theta_{ij}\right)^2\sin\theta_{ij}=\frac{4}{5},
    \end{aligned}
\end{equation}
we get
\begin{equation}
    \begin{aligned}
        J&=\frac{\mu_0}{2\pi}\frac{\hbar\gamma_{\mathrm{H}}^2}{\sqrt{5}}\sqrt{\sum\limits_{j,j\ne i}^{N_\text{m}}\frac{1}{R_{ij}^6}}.
    \end{aligned}
\end{equation}
\begin{figure}[htbp]
\begin{center}
\includegraphics[scale=1]{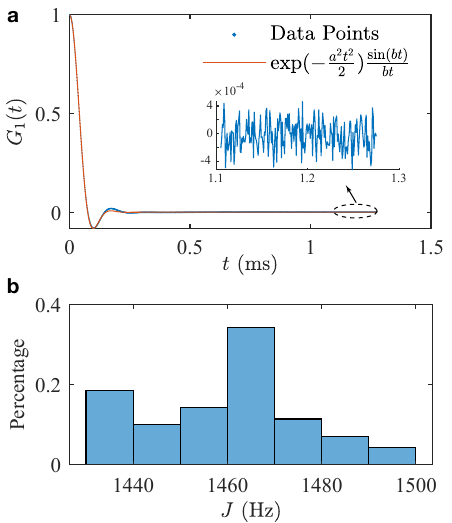}
\caption{\textbf{Measuring the free induction decay $G_1(t)$ and extracting the effective dipolar coupling strengh $J$.} \textbf{a}, $G_1(t)$ is expressed by equation \eqref{Def_FID}, and we measure it using the solid-echo method \cite{kimmich2012} to avoid the transient effects of the dead time at the beginning of sampling. Then we fit the data points with a function \eqref{FID_fit} raised by A. Abragam \cite{abragam1961}, to obtain an estimate of the effective dipolar coupling strength $J$ through the relation \eqref{J_relation_FID}. In this illustration, the delay time of the inversion pulse in the echo sequence is set to be $d_e=60\ \mu\mathrm{s}$, and the fitting results are $a/2\pi=2154\pm 18$ Hz, and $b/2\pi=6631\pm 14$ Hz, giving an estimate of $J/2\pi$ to be $1464\pm3$ Hz. The standard deviation of the noise, quantified as $\sim10^{-4}$, is determined from the end range of the FID, and subsequently incorporated into the error bar analysis. \textbf{b}, The uncertainty of the estimate of $J$ mainly comes from the fluctuation due to varying the delay time $d_e$ in a range of $[50,\ 80]\ \mu\mathrm{s}$. We find that $1432~\text{Hz}\le J/2\pi \le 1502~\text{Hz}$, and we take the average value $\bar{J}/2\pi=1460$ Hz.}
\label{FID}
\end{center}
\end{figure}
The value of $J$ can be experimentally determined by measuring and fitting the curve of the free induction decay (FID). Here, the FID monitors the $\hat{O}_x$ magnetization of a state $\hat{\rho}_x\propto\hat{\mathbbm{1}}+\epsilon_0\hat{O}_x$, which freely evolves under the secular dipolar Hamiltonian $\hat{H}_{\text{dip}}^{\text{sec}}$ \eqref{sysHam}. The expression of the this FID normalized to its initial-time value reduces to an auto-correlation function:
\begin{equation}
    G_1(t)\equiv\frac{\tr(\mathrm{e}^{-\frac{\mathrm{i}}{\hbar}\hat{H}_{\mathrm{dip}}^{\mathrm{sec}}t}\hat{\rho}_x\mathrm{e}^{\frac{\mathrm{i}}{\hbar}\hat{H}_{\mathrm{dip}}^{\mathrm{sec}}t}\hat{O}_x)}{\tr(\hat{\rho}_x\hat{O}_x)}=\frac{\tr(\mathrm{e}^{-\frac{\mathrm{i}}{\hbar}\hat{H}_{\mathrm{dip}}^{\mathrm{sec}}t}\hat{O}_x\mathrm{e}^{\frac{\mathrm{i}}{\hbar}\hat{H}_{\mathrm{dip}}^{\mathrm{sec}}t}\hat{O}_x)}{\tr(\hat{O}_x^2)}.
    \label{Def_FID}
\end{equation}
$G_1(t)$ can be expanded in a power series as
\begin{align}
    G_1(t)=\sum_n(-1)^n\frac{M_{2n}t^{2n}}{(2n)!},
\end{align}
where 
\begin{equation}
    M_{2n}\equiv(-1)^n\left(\frac{\dd^{2n}G_1(t)}{\dd t^{2n}}\right)_{t=0}.
\end{equation}
In particular, the second moment reads
\begin{equation}
    \begin{aligned}
            M_2\equiv&-\left(\frac{\dd^{2}G_1(t)}{\dd t^{2}}\right)_{t=0}
            =-\frac{\tr([\hat{H}_{\mathrm{dip}}^{\mathrm{sec}},~\hat{O}_x]^2)}{\hbar^2\tr(\hat{O}_x^2)}.
    \end{aligned}
\end{equation}
The derivation for the explicit expression of the second moment is illustrated in Ref. \cite{vanvleck1948,abragam1961}, and the final expression is 
\begin{equation}
    M_2=\left(\frac{\mu_0}{2\pi}\right)^2\frac{9\hbar^2\gamma_{\mathrm{H}}^4}{5}\sum\limits_{j,j\ne i}^{N_\text{m}}\frac{1}{R_{ij}^6}=9J^2.
    \label{m2}
\end{equation}
Then the question comes to the estimation of the second moment from the FID \eqref{Def_FID}. In reference \cite{abragam1961}, A. Abragam raised that the experimental curves of $G_1(t)$ bear a strong resemblance to the analytical expression
\begin{equation}
    g(t)\equiv\exp(-\frac{a^2t^2}{2})\frac{\sin bt}{bt},
    \label{FID_fit}
\end{equation}
which has a series expansion
\begin{equation}
    g(t)=1-\frac{t^2}{2}\left(a^2+\frac{b^2}{3}\right)+\frac{t^4}{4!}\left(3a^4+2a^2b^2+\frac{b^4}{5}\right)+\cdots.
\end{equation}
Therefore, by fitting the FID with the function \eqref{FID_fit} we can obtain an estimate of the second moment as 
\begin{equation}
    M_2=a^2+\frac{b^2}{3}.
\end{equation}
Combing with equation \eqref{m2}, an estimate of $J$ reads
\begin{equation}
    J=\frac{\sqrt{M_2}}{3}=\frac{\sqrt{a^2+b^2/3}}{3}.
    \label{J_relation_FID}
\end{equation}
In experiments, we measure the FID using the solid-echo method\cite{kimmich2012} to avoid the transient effects of the dead time at the beginning of sampling. Therefore, the error bar includes both the fitting error and the fluctuation due to varying the delay time of the inversion pulse in the echo sequence. We find that $1432~\mathrm{Hz}\le J/2\pi \le 1502~\mathrm{Hz}$ and we take the average value $\bar{J}/2\pi=1460$ Hz. 

\section{Hamiltonian engineering}
By periodically applying RF pulse train to the internal Hamiltonian of the system $\hat{H}_{\mathrm{s}}$ \eqref{sysHam}, we engineer the target anisotropic Heisenberg model stroboscopically utilizing the average Hamiltonian theory \cite{haeberlen1968}. The anisotropic Heisenberg model reads
\begin{equation}
    \hat{H}\equiv\hbar\sum\limits_{i<j,ab}J_{ij}\left(\xi_x\hat{S}^x_{ia}\hat{S}^x_{jb}+\xi_y\hat{S}^y_{ia}\hat{S}^y_{jb}+\xi_z\hat{S}^z_{ia}\hat{S}^z_{jb}\right).
    \label{HeiModel}
\end{equation}
In a general Floquet engineering protocol, the total time-dependent Hamiltonian is $\hat{H}(t)=\hat{H}_{\mathrm{s}}+\hat{H}_{\mathrm{rf}}(t)$, where $\hat{H}_{\mathrm{rf}}(t)=2\pi\hbar[h_x(t)\hat{O}_x+h_y(t)\hat{O}_y]$ is the periodic Hamiltonian induced by the RF pulses, with $h_x(t)=h_x(t+T)$ and $h_y(t)=h_y(t+T)$. By observing the system stroboscopically at multiples of the periods $T$, the evolution dynamics can be described effectively by a static Hamiltonian denoted as $\hat{H}_{\mathrm{F}}$. This Hamiltonian is obtained from the Floquet-Magnus (FM) expansion truncated to some specific order $n_*$ \cite{ magnus1954a,blanes2009,bukov2015,mori2016,kuwahara2016,abanin2017}, and can be expressed as $\hat{H}_{\mathrm{F}}\equiv\sum_{n=0}^{n_*}\hat{H}_{\mathrm{F}}^{(n)}$. The dominant terms of $\hat{H}_{\mathrm{F}}$ are given by \cite{blanes2009,bukov2015}:
\begin{equation}
    \begin{aligned}
        \hat{H}_{\mathrm{F}}^{(0)}\equiv&\frac{1}{T}\int_0^T\dd t\hat{H}(t),\\
        \hat{H}_{\mathrm{F}}^{(1)}\equiv&\frac{-\mathrm{i}}{2 T\hbar} \int_0^{T} \dd t_1 \int_0^{t_1} \dd t_2\left[\hat{H}(t_1), \hat{H}(t_2)\right],\\
        \hat{H}_{\mathrm{F}}^{(2)}\equiv&-\frac{1}{6 T\hbar^2} \int_0^T \dd t_1 \int_0^{t_1} \dd t_2 \int_0^{t_2} \dd t_3\cdot\left\{\left[\hat{H}(t_1),[\hat{H}(t_2), \hat{H}(t_3)]\right]+\left[\hat{H}(t_3),[\hat{H}(t_2), \hat{H}(t_1)]\right]\right\}.
    \end{aligned}
    \label{FM expansion}
\end{equation}
A successful engineering protocol should realize an effective $\hat{H}_\mathrm{F}$ close enough to the target Hamiltonian.

In this work, the basic building block of the RF pulse train is an 8-pulse sequence firstly raised by Suter et al. \cite{suter1987}:
\begin{equation}
    \begin{aligned}
    \mathrm{P}(\tau_z,\ \mathbf{x},\ \tau_y,\ \mathbf{y},\ 2\tau_x,\ \mathbf{y},\ \tau_y,\ \mathbf{x},\ 2\tau_z,\ \mathbf{x},\ \tau_y,\ \mathbf{y},\ 2\tau_x,\ \mathbf{y},\ \tau_y,\ \mathbf{x},\ \tau_z),
    \end{aligned}
    \label{sequence}
\end{equation}
where $\mathbf{x}$ and $\mathbf{y}$ represent RF pulses with length $\tau_{\mathrm{p}}=2\ \mathrm{\mu s}$ in our experiments, realizing global $\pi/2$ rotations along the $\hat{x}$ and $\hat{y}$ directions, respectively. To further reduce errors caused by the inhomogeneity of the RF field, we add a copy of the above 8-pulse sequence with opposite pulse phases, leading to a 16-pulse sequence that has a zero net rotation (\ref{pulse}a). This 16-pulse sequence has been employed in the study of localization and thermalization in spin chains \cite{wei2018,wei2019,peng2021,peng2023}. The pulse intervals (between midpoints of two adjacent pulses) are given by
\begin{equation}
    \begin{aligned}
        \tau_\alpha&=\left[1+\xi_\alpha^{(0)}\right]\tau,
    \end{aligned}
    \label{PulseIntervals}
\end{equation}
where the average pulse interval $\tau=(\tau_x+\tau_y+\tau_z)/3$, leading to a cycle time $T=24\tau$. Note that correspondingly, the pulse intervals calculated between the edges of two consecutive pulses are give by the following equations:
\begin{equation}
    \begin{aligned}
        \tau_x'&=\tau_x-\frac{\tau_\mathrm{p}}{2},\\
        \tau_y'&=\tau_y-\tau_\mathrm{p},\\
        \tau_z'&=\tau_z-\frac{\tau_\mathrm{p}}{2}.
      \end{aligned}
    \label{eq:Interval between edges}
\end{equation}
Here $\tau_\mathrm{p}=2~\mu\mathrm{s}$ is the $\pi/2$ pulse width. In experiment, $\tau'_\alpha>0$ should be guaranteed. This pulse sequence \ref{pulse}a realizss the zeroth-order term of the FM expansion as 
\begin{equation}
    \hat{H}_{\mathrm{F}}^{(0)}=\hbar\sum\limits_{i<j,ab}J_{ij}\left(\xi_x^{(0)}\hat{S}^x_{ia}\hat{S}^x_{jb}+\xi_y^{(0)}\hat{S}^y_{ia}\hat{S}^y_{jb}+\xi_z^{(0)}\hat{S}^z_{ia}\hat{S}^z_{jb}\right).
\end{equation}
where the anisotropic parameters satisfy $\sum_\alpha\xi_\alpha^{(0)}=0$, inherited from the secular dipolar Hamiltonian \eqref{sysHam} and conserved by the global rotations. Furthermore, $\hat{H}_{\mathrm{F}}^{(1)}$ is zero due to the symmetry of this pulse sequence. Therefore, the total effective Hamiltonian reads
\begin{equation}
    \begin{aligned}
    \hat{H}_{\mathrm{F}}&=\hat{H}_{\mathrm{F}}^{(0)}+\hat{H}_{\mathrm{F}}^{(2)}+\mathcal{O}[(JT)^3]\\
        &=\hbar\sum\limits_{i<j,ab}J_{ij}\left(\xi_x'\hat{S}^x_{ia}\hat{S}^x_{jb}+\xi_y'\hat{S}^y_{ia}\hat{S}^y_{jb}+\xi_z'\hat{S}^z_{ia}\hat{S}^z_{jb}\right)+\hat{H}_\epsilon,
    \end{aligned}
    \label{FloHam}
\end{equation}
where the synthetic configuration $(\xi_x',\ \xi_y',\ \xi_z')=\bm{\xi}'=\bm{\xi}^{(0)}+\bm{\xi}^{(2)}+\cdots$, preserving the relation $\sum_\alpha\xi_\alpha'=0$. During experiments, we set the zeroth-order configuration $\bm{\xi}^{(0)}$ to match the target configuration $\bm{\xi}$. We then optimize the value of the average pulse interval $\tau$ to be $5\ \mathrm{\mu s}$ (\ref{Opt of Para}f), which is neither too long to significantly incorporate contributions from the higher-order terms of the FM expansion, nor too short to introduce errors caused by pulse transients. Within $\hat{H}_{\mathrm{F}}$, the first term holds a primary influence, with discrepancies of $\bm{\xi}'$ from the target values $\bm{\xi}$ in \eqref{HeiModel}, quantified by $\Delta\equiv\abs*{\bm{\xi}'-\bm{\xi}}/\abs*{\bm{\xi}}$, calibrated to remain within $3\%$ across all the configurations (\ref{tomo result}b). The second term, $\hat{H}_\epsilon$, incorporates the residue terms other than $\hat{S}_{ia}^\alpha\hat{S}_{jb}^\alpha$. The weight of $\hat{H}_\epsilon$ in the total effective Hamiltonian $\hat{H}_\mathrm{F}$, defined as $\varepsilon\equiv\sqrt{\tr(\hat{H}_\epsilon^2)/\tr(\hat{H}_\mathrm{F}^2)}$ is calibrated to be less than $20\%$ across all the configurations (\ref{tomo result}c).

\section{Hamiltonian calibration}
In this section we present the protocol and results of Hamiltonian calibration. \ref{sec4a} introduces the formulation of the principle and subsequently clarifies the pulse sequence in detail. \ref{sec4b} presents the optimization of the experimental parameters involved in the protocol. Finally, \ref{sec4c} presents the calibration results.

\subsection{Protocol} \label{sec4a}
The central idea is to employ Floquet prethermalization hypothesis\cite{mori2016,abanin2017}. That is, under Floquet driving, the effective Hamiltonian $\hat{H}_{\mathrm{F}}$ acts as a quasiconserved quantity and dominates the system dynamics for a relatively long time, making the system reach a quasistationary state before it is heated to infintie temperature. The quasistationary state, as referred to the  prethermal state $\hat{\rho}_{\mathrm{pre}}$, can be approximated as the canonical ensemble of the effective Hamiltonian $\hat{H}_{\mathrm{F}}$, written as 
\begin{equation}
    \hat{\rho}_{\mathrm{pre}}\approx\frac{e^{-\beta_{\text{eff}}\hat{H}_{\mathrm{F}}}}{\Tr\small(e^{-\beta_{\text{eff}}\hat{H}_{\mathrm{F}}}\small)}\approx \frac{\hat{\mathbbm{1}}-\beta_{\text{eff}}\hat{H}_{\mathrm{F}}}{\Tr(\hat{\mathbbm{1}})},
    \label{PreTherState}
\end{equation} 
where $\beta_{\text{eff}}$ is the inverse effective temperature determined by the initial state $\hat{\rho}_0$ and the energy conservation $\Tr(\hat{\rho}_0\hat{H}_{\text{F}})=\Tr(\hat{\rho}_\text{pre}\hat{H}_{\text{F}})$. A finite inverse effective temperature $\beta_{\text{eff}}\ne 0$ is required for the prethermal state $\hat{\rho}_{\text{pre}}$ to retain information of the effective Hamiltonian $\hat{H}_\text{F}$. The phenomenon of Floquet prethermalization has been experimentally investigated in spin chain systems with dipolar interactions \cite{wei2019,peng2021,yin2021}. The occurrence of Floquet prethermalization is closely linked to the chaotic nature of the dipolar Hamiltonian \cite{fine2004long,jyoti2017dipolarly,Deutsch2018}.

The prethermal state \eqref{PreTherState} is a fingerprint of the Floquet Hamiltonian. Consequently, the question arises as to how to perform tomography on $\hat{\rho}_\text{pre}$. Full tomography is infeasible due to the absence of universal quantum control over the system. Instead, several methods for partial tomography exist. One approach involves expanding the state in basis of irreducible spherical tensor operators (ISTOs) \cite{suter1988a,vanbeek2005}. Another approach employs multiple quantum coherences (MQCs)\cite{Ernst1976,Ernst1977,Drobny1978,Bodenhausen1980,yen1983,Pines1985}. In our work, we simplify the ``basis'' by focusing on the traceless components of the dipolar-ordered states\cite{jeener1967,cho2003}
\begin{equation}
\begin{aligned}
    \delta\hat{\rho}_x^\mathcal{D}\propto\hat{D}_x&\equiv\hbar\sum\limits_{i<j, ab} J_{ij}\left(-\hat{S}^y_{ia}\hat{S}^y_{jb}-\hat{S}^z_{ia}\hat{S}^z_{jb}+2\hat{S}^x_{ia}\hat{S}^x_{jb}\right),\\
    \delta\hat{\rho}_y^\mathcal{D}\propto\hat{D}_y&\equiv\hbar\sum\limits_{i< j, ab} J_{ij}\left(-\hat{S}^z_{ia}\hat{S}^z_{jb}-\hat{S}^x_{ia}\hat{S}^x_{jb}+2\hat{S}^y_{ia}\hat{S}^y_{jb}\right),\\
    \delta\hat{\rho}_z^\mathcal{D}\propto\hat{D}_z&\equiv\hbar\sum\limits_{i< j, ab} J_{ij}\left(-\hat{S}^x_{ia}\hat{S}^x_{jb}-\hat{S}^y_{ia}\hat{S}^y_{jb}+2\hat{S}^z_{ia}\hat{S}^z_{jb}\right).
\end{aligned}
    \label{DO_states_pro}
\end{equation}
The dipolar-ordered states, serving as reference states, are adequate for achieving the calibration of our Hamiltonian. 

Firstly, we formulate the procedure for calibrating the anisotropic parameters $\{\xi_x',\xi_y',\xi_z'\}$ in the Floquet Hamiltonian $\hat{H}_\text{F}$ \eqref{FloHam}. What we need to do is measure the inner products between the prethermal state $\hat{\rho}_{\mathrm{pre}}$ \eqref{PreTherState} and the dipolar-ordered states $\delta\hat{\rho}_\alpha^\mathcal{D}$ such as:
\begin{equation}
    \begin{aligned}
        y_1&=\tr(\hat{\rho}_{\mathrm{pre}}\delta\hat{\rho}_x^\mathcal{D})\propto\beta_{\text{eff}}\tr(\hat{H}_{\mathrm{F}}\hat{D}_x),
        \label{y1}
    \end{aligned}
\end{equation}
Using the explicit expression of $\hat{H}_\text{F}$ \eqref{FloHam} and $\hat{D}_x$ \eqref{DO_states_pro}, we find
\begin{equation}
    \begin{aligned}
        y_1\propto(2\xi_x'-\xi_y'-\xi_z')\sum_{i\ne j,\ ab}J_{ij}^2\propto\xi_x',
    \end{aligned}
    \label{y123}
\end{equation}
where we have used the relation $\sum_\alpha\xi_\alpha'=0$. Analogously, we have $y_2\propto\tr(\hat{H}_{\mathrm{F}}\hat{D}_y)\propto\xi_y'$ and $y_3\propto\tr(\hat{H}_{\mathrm{F}}\hat{D}_z)\propto\xi_z'$. 

Notice that in principle we can only obtain the ratio $\xi_x':\xi_y':\xi_z'=y_1:y_2:y_3$. Considering that across all the configurations we set the target value $\xi_z$ to be a constant value of $0.2$, in the next we investigate the deviation of the real value $\xi_z'$ from its target value $\xi_z$, described by a relation $\xi_z'=c\xi_z$. In order to calibrate $c$ we make a time-scale rescaling between the evolution dynamics governed by the secular dipolar Hamiltonian $\hat{H}_{\mathrm{dip}}^{\mathrm{sec}}=\hat{D}_z$ \eqref{sysHam} and the one produced by a to-be-calibrated effective Hamiltonian $\hat{H}_{\mathrm{F}}$ \eqref{FloHam}. For $\hat{H}_{\mathrm{dip}}^{\mathrm{sec}}$ the configuration is known as $\bm{\xi}^{\mathrm{dip}}=(-1,\ -1,\ 2)$, while for $\hat{H}_{\mathrm{F}}$ we set the target configuration to be $\bm{\xi}=(-0.1,\ -0.1,\ 0.2)=\bm{\xi}^{\mathrm{dip}}/10$. If the real configuration $\bm{\xi}'$ is very close to $\bm{\xi}$, we expect the dynamics of $\hat{H}_{\mathrm{dip}}$ and $\hat{H}_{\mathrm{F}}$ to collapse after rescaling the time scale by a factor of 10.

\begin{figure}[htbp]
    \begin{center}
    \includegraphics[scale=1]{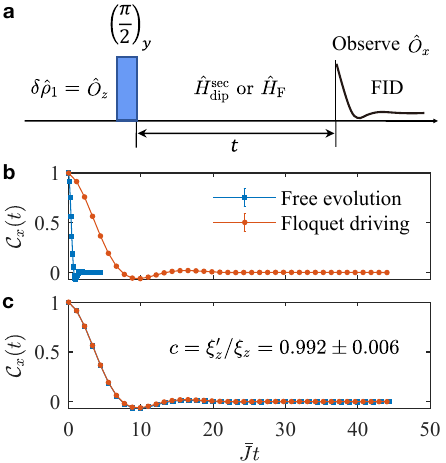}
    \caption{\textbf{Calibration of the scaling factor $c$.} \textbf{a}, The pulse sequence simply starts from the thermal equilibrium state, polarized along the $z$ direction, and then followed by a $(\pi/2)_y$ pulse to turn it into the $\hat{x}$ polarization. Afterwards, the state evolves for a varying duration $t$, under the secular dipolar Hamiltonian $\hat{H}^\mathrm{sec}_\mathrm{dip}$ (free evolution) or the effective Hamiltonian $\hat{H}_\mathrm{F}$ (Floquet driving, using periodically applied pulse sequence in \ref{pulse}a). For $\hat{H}_{\mathrm{dip}}^{\mathrm{sec}}$ the configuration is known as $\bm{\xi}^{\mathrm{dip}}=(-1,\ -1,\ 2)$, while for $\hat{H}_{\mathrm{F}}$ we set the target configuration to be $\bm{\xi}=(-0.1,\ -0.1,\ 0.2)=\bm{\xi}^{\mathrm{dip}}/10$. Finally we measure the polarization $\hat{O}_x$. \textbf{b}, The blue data points represent the evolution dynamics for  $\hat{H}_\mathrm{dip}^{\mathrm{sec}}$, denoted by equation \eqref{Sx1}, while the red data points correspond to the dynamics governed by $\hat{H}_\mathrm{F}$, described by equation \eqref{Sx2}. Each data point is read out from its corresponding free induction decay, with an error bar (incorporated within the markers) representing the noise amplitude, as shown in \ref{FID}a. \textbf{c}, The two curves collapse after rescaling $\bar{J}t$ of the blue points by a factor of $c'=\xi_z^{(\mathrm{dip})}/\xi_z'=10.084\pm 0.063$, thus we can determine that $\xi_z'=0.198\pm 0.001$ and $c=\xi_z'/\xi_z=10/c'=0.092\pm 0.006$, with confidence intervals estimated from the least squares method.}
    \label{EvolutionScaling}
    \end{center}
\end{figure}

Using the pulse sequence displayed by \ref{EvolutionScaling}a, we measure the evolution dynamics of the total spin operator $\hat{O}_x=\sum_i\hat{S}_i^x$, under the two Hamiltonians, respectively:
\begin{align}
    \mathcal{C}_x^1(t)\equiv&\frac{\tr(\mathrm{e}^{-\mathrm{i}\hat{H}_{\mathrm{dip}}^{\mathrm{sec}}t/\hbar}\hat{O}_x\mathrm{e}^{\mathrm{i}\hat{H}_{\mathrm{dip}}^{\mathrm{sec}}t/\hbar}\hat{O}_x)}{\tr(\hat{O}_x^2)},
        \label{Sx1}\\
        \mathcal{C}_x^2(t)\equiv&\frac{\tr(\mathrm{e}^{-\mathrm{i}\hat{H}_{\mathrm{F}}t/\hbar}\hat{O}_x\mathrm{e}^{\mathrm{i}\hat{H}_{\mathrm{F}}t/\hbar}\hat{O}_x)}{\tr(\hat{O}_x^2)}.
        \label{Sx2}
\end{align}           
The two curves in \ref{EvolutionScaling}b collapse in \ref{EvolutionScaling}c after rescaling $\bar{J}t$ of the blue points by a factor of $c'=\xi_z^{(\mathrm{dip})}/\xi_z'=10.084\pm 0.063$, thus we can determine that $\xi_z'=0.198\pm 0.001$ and $c=\xi_z'/\xi_z=10/c'=0.992\pm 0.006$, with confidence intervals estimated from the least squares method. Although we are not able to make such a rescaling analysis for other configurations that are not in the same ratio $\xi_x:\xi_y:\xi_z$ as in the secular dipolar Hamiltonian $\hat{H}_\mathrm{dip}^\mathrm{sec}$, it is reasonable to assume that $\xi_z'=0.198\pm 0.001$ and $c=\xi_z'/\xi_z=0.992\pm 0.006$ across the configurations we investigated, considering that we maintain a constant target value of $\xi_z=0.2$ and realize them using the same pulse protocol with the same Floquet driving period $T$. Combining the information of $\xi_z'$ and the ratio $\xi_x':\xi_y':\xi_z'$ fulfills the calibration of their absolute values.

Next, we will introduce the principle of calibrating the proportion of the residue terms $\hat{H}_\epsilon$ in the total effective Hamiltonian $\hat{H}_\mathrm{F}$ \eqref{FloHam}. The proportion is defined by
\begin{equation}
\varepsilon\equiv\sqrt{\frac{\tr\small(\hat{H}_\epsilon^2\small)}{\tr\small(\hat{H}_{\mathrm{F}}^2\small)}}.
    \label{varepsilon}
\end{equation}
The traceless component of the prethermal state \eqref{PreTherState} is proportional to the effective Hamiltonian $\hat{H}_\text{F}$, and can be decomposed as
\begin{equation}
\delta\hat{\rho}_{\mathrm{pre}}=u\hat{H}_{\mathrm{F}}=a (\delta\hat{\rho}_z^{\mathcal{D}})+b(\delta\hat{\rho}_x^{\mathcal{D}})+u\hat{H}_\epsilon,
\end{equation}
where $a,\ b$ are the coefficients to be determined, and $(\delta\hat{\rho}_z^{\mathcal{D}}),\ (\delta\hat{\rho}_x^{\mathcal{D}})$ are the dipolar-ordered states, as shown by equation \eqref{DO_states_pro}.
We can determine the proportion $\varepsilon$ by measuring the following five quantities:

1. The inner products among the dipolar-ordered states: 
\begin{align}
    C_1\equiv&\tr\left(\delta\hat{\rho}_z^\mathcal{D}\delta\hat{\rho}_z^\mathcal{D}\right)=\tr\left(\delta\hat{\rho}_x^\mathcal{D}\delta\hat{\rho}_x^\mathcal{D}\right),
    \label{quantity1}\\
        C_2\equiv&\tr\left(\delta\hat{\rho}_z^\mathcal{D}\delta\hat{\rho}_x^\mathcal{D}\right).
\end{align}

2. The inner products between $\delta\hat{\rho}_{\mathrm{pre}}$ and $\delta\hat{\rho}_z^\mathcal{D},\ \delta\hat{\rho}_x^\mathcal{D}$, which have the same form as in equation \eqref{y1}:
\begin{align}
    C_3\equiv\tr\left(\delta\hat{\rho}_{\mathrm{pre}}\delta\hat{\rho}_z^\mathcal{D}\right)=y_3=&aC_1+bC_2,\\
    C_4\equiv\tr\left(\delta\hat{\rho}_{\mathrm{pre}}\delta\hat{\rho}_x^\mathcal{D}\right)=y_1=&bC_1+aC_2.\label{quantity4}
\end{align}

3. The auto inner product of $\delta\hat{\rho}_{\mathrm{pre}}$:
\begin{equation}
    \begin{aligned}
        C_5\equiv\tr[(\delta\hat{\rho}_{\mathrm{pre}})^2]=u^2\tr(\hat{H}_\mathrm{F}^2)=(a^2+b^2)C_1+2ab C_2+u^2\tr\small(\hat{H}_\epsilon^2\small).
    \end{aligned}
    \label{quantity5}
\end{equation}
Measurements of $C_1$ to $C_4$ give the values of $a$ and $b$. Subsequently, we substitute $a,\ b$ into the formula for $C_5$, yielding the expression for the proportion $\varepsilon$:
\begin{equation}
    \begin{aligned}
        \varepsilon^2=&\frac{\tr\small(\hat{H}_\epsilon^2\small)}{\tr\small(\hat{H}_{\mathrm{F}}^2\small)}=\frac{u^2\tr\small(\hat{H}_\epsilon^2\small)}{\tr[(\delta\hat{\rho}_{\mathrm{pre}})^2]}\\
                =&1-\frac{(a^2+b^2)C_1+2ab C_2}{C_5}.
    \end{aligned}
\end{equation}

\begin{figure*}[htbp]
    \begin{center}
    \includegraphics[scale=1]{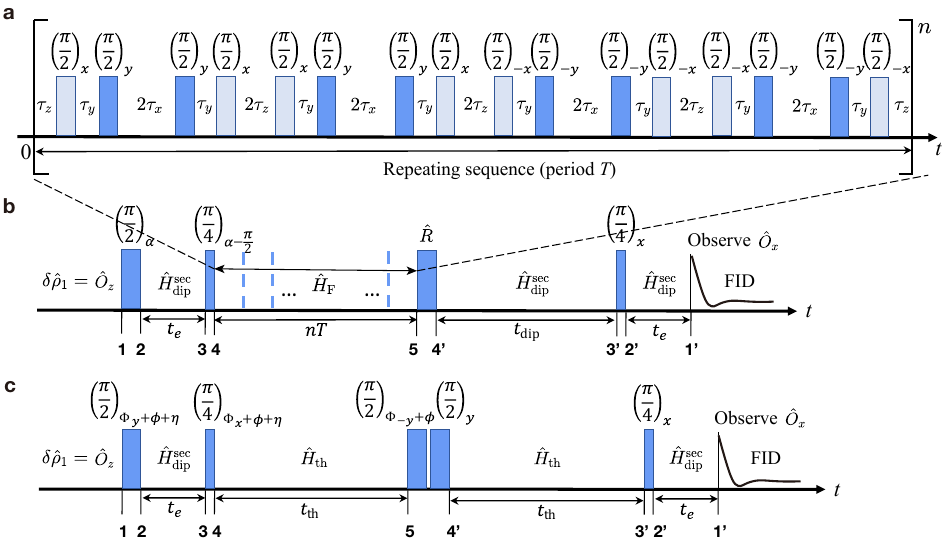}
    \caption{\textbf{The pulse sequences for Hamiltonian engineering and calibration.} \textbf{a}, One period of the pulse train for the Hamiltonian engineering. The $(\pi/2)_x$ pulse with length $\tau_{\mathrm{p}}=2\ \mathrm{\mu s}$ induces a $\pi/2$ rotation along the $\hat{x}$ direction, and similarly for the other pulses. The pulse intervals, which are the time intervals between the midpoints of two adjacent pulses, are set to be $\tau_\alpha=[1+\xi_\alpha^{(0)}]\tau$. Here, the average pulse interval is $\tau=(\tau_x+\tau_y+\tau_z)/3=5\ \mu \mathrm{s}$, the period is $T=24\tau=120\ \mathrm{\mu s}$, and the Hamiltonian anisotropic parameters satisfy $\sum\xi_\alpha^{(0)}=0$. The zeroth order of the Floquet-Magnus expansion of this sequence is $\hat{H}_{\mathrm{F}}^{(0)}=\frac{\hbar}{2}\sum_{i\ne j,ab}J_{ij}\left(\xi_x^{(0)}\hat{S}^x_{ia}\hat{S}^x_{jb}+\xi_y^{(0)}\hat{S}^y_{ia}\hat{S}^y_{jb}+\xi_z^{(0)}\hat{S}^z_{ia}\hat{S}^z_{jb}\right)$, which is dominant in the total effective Hamiltonian $\hat{H}_{\mathrm{F}}$ \eqref{FloHam}. \textbf{b}, The pulse sequence for the Hamiltonian calibration. The first half of the sequence is used to prepare the prethermal state $\delta\hat{\rho}_5=\delta\hat{\rho}_{\mathrm{pre}}\propto\hat{H}_{\mathrm{F}}$ after $n$ cycles of Floquet driving. The second half is used to prepare the dipolar-ordered state $\hat{O}_{4'}=\delta\hat{\rho}_z^{\mathcal{D}}\propto \hat{H}_{\mathrm{dip}}^{\mathrm{sec}}=\hat{D}_z$ in the Heisenberg picture. After implementing an operation $\hat{R}\in\{\hat{R}_y(\pi/2),\ \hat{R}_x(\pi/2),\ \hat{\mathbbm{1}}\}$, we measure the inner products between the prethermal states and the dipolar-ordered states. \textbf{c}, The sequence referred from Ref. \cite{cho2003} for assessing the dipolar-ordered state and optimizing the experimental parameters involved in \textbf{a} and \textbf{b}. The difference between this sequence and that in \textbf{b} lies in, firstly, that the azimuthal angles of the pulses in the initial half are shifted by combinations of $\phi$ and $\eta$, such as $(\pi/2)_{\Phi_y+\phi+\eta}$, which induces an operation $\hat{R}_z(\phi+\eta)\hat{R}_{y}(\pi/2)\hat{R}_z^\dagger(\phi+\eta)$. The overall effect is to realize successive rotations $\hat{R}_x(\phi)\hat{R}_z(\eta)$ in the middle of the sequence, instead of the $\hat{R}$ operator in \textbf{b}. Secondly, for both periods of thermalization, we choose the same Hamiltonian $\hat{H}_{\mathrm{th}}$. Specifically, this Hamiltonian can be either the secular dipolar Hamiltonian $\hat{H}_{\mathrm{dip}}^{\mathrm{sec}}=\hat{D}_z$ \eqref{sysHam} or the effective Hamiltonian $\hat{H}_{\mathrm{F}}$ \eqref{FloHam}. We denote the states at each step by the numbers below.}
    \label{pulse}
    \end{center}
\end{figure*}

In the following, we will clarify the pulse sequence used to implement the aforementioned principles. The calibration of the anisotropic parameters is achieved using the sequence shown in \ref{pulse}b, where we prepare the prethermal states $\hat{\rho}_{\text{pre}}$ \eqref{PreTherState} and measure their inner products with the dipolar-ordered states $\delta\hat{\rho}_\alpha^\mathcal{D}$. The first half of the sequence is dedicated to preparing the prethermal states, and the procedures are as follows:

1: We start from the thermal equilibrium state of the system without RF irradiation, which is the steady state of the spin-lattice relaxation: 
\begin{equation}
    \begin{aligned}
        \hat{\rho}_1=&\frac{\mathrm{e}^{-\beta_0\hat{H}_0}}{\tr\small(\mathrm{e}^{-\beta_0\hat{H}_0}\small)}\approx\frac{\mathrm{e}^{\epsilon_0\hat{O}_z}}{\tr\small(\mathrm{e}^{\epsilon_0\hat{O}_z}\small)}\approx\frac{\hat{\mathbbm{1}}+\epsilon_0 \hat{O}_z}{2^N}=\frac{\hat{\mathbbm{1}}+\epsilon_0(\delta\hat{\rho}_1)}{2^N},
    \end{aligned}
    \label{rho1}
\end{equation}
where the first approximation is due to the fact that the Zeeman interaction is dominant, and in the second approximation we use the high-temperature expansion, discarding the higher-order terms $\mathcal{O}(\epsilon_0^2)$ since the polarization rate $\epsilon_0=\beta_0\hbar\gamma_{\mathrm{H}}B_0\sim 10^{-5}$ at room temperature ($\beta_0=1/k_BT,\ T\approx 300$ K). As the identity matrix doesn't contribute to the observable signal, we focus on the component
\begin{equation}
    \delta\hat{\rho}_1=\hat{O}_z, 
    \label{partialrho1}
\end{equation}

\smallskip
2: By applying the $(\pi/2)_y$ pulse (set the azimuthal angle $\alpha$ in the $\hat{x}-\hat{y}$ plane equal to that of the $\hat{y}$ direction, denoted by $\Phi_{y}$, as an example), which induces a $\pi/2$ rotation along the $\hat{y}$ direction, the density matrix becomes:
\begin{equation}
    \delta\hat{\rho}_2=\hat{R}_y(\pi/2)\delta\hat{\rho}_1\hat{R}_y^\dagger(\pi/2)=\hat{O}_x.
\end{equation}

\smallskip
3: The state evolves under the secular dipolar Hamiltonian $\hat{H}_{\mathrm{dip}}^{\rm{sec}}=\hat{D}_z$ for time $t_e$, which gives
\begin{equation}
    \begin{aligned}
        \delta\hat{\rho}_3=&\mathrm{e}^{-\frac{\mathrm{i}}{\hbar}\hat{H}_{\mathrm{dip}}^{\rm{sec}}t_e}\delta\hat{\rho}_2\mathrm{e}^{\frac{\mathrm{i}}{\hbar}\hat{H}_{\mathrm{dip}}^{\rm{sec}}t_e}
        =\hat{O}_x-\frac{\mathrm{i}}{\hbar}[\hat{D}_z,\ \hat{O}_x]t_e+\cdots\\
        =&\hat{O}_x+\frac{3}{2}t_e\sum_{i\ne j, ab}J_{ij}\left(\hat{S}^z_{ia}\hat{S}^y_{jb}+\hat{S}^y_{ia}\hat{S}^z_{jb}\right)+\cdots.
    \end{aligned}
    \label{rho3}
\end{equation}

\smallskip
4: The successive $\left(\pi/4\right)_x$ pulse (set the azimuth angle $\alpha-\pi/2$ equal to that of the $\hat{x}$ direction, denoted by $\Phi_{x}$, as an example) makes up a Jeener-Broekaert pulse pair\cite{jeener1967}, producing 
\begin{equation}
    \begin{aligned}
        \delta\hat{\rho}_4=&\hat{R}_x(\pi/4)\delta\hat{\rho}_3\hat{R}_x^\dagger(\pi/4)\\
        =&\hat{O}_x+\frac{3}{2}t_e\sum_{i\ne j, ab}J_{ij}\left(\hat{S}^z_{ia}\hat{S}^z_{jb}-\hat{S}^y_{ia}\hat{S}^y_{jb}\right)+\cdots.
    \end{aligned}
    \label{rho4}
\end{equation}
as the initial state of the prethermalization process. The delay time $t_e$ can be optimized to give the maximal experimental signals, i.e., as high inverse effective temperature $\beta_\mathrm{eff}$ as possible (\ref{Opt of Para}d).

\smallskip
5: Following the prethermalization hypothesis, after enough Floquet driving cycles $n$ the effective Hamiltonian $\hat{H}_\text{F}$ thermalizes $\hat{\rho}_4$ into the so-called prethermal state $\hat{\rho}_{\mathrm{pre}}$ \eqref{PreTherState}. Since $\hat{\rho}_{\mathrm{pre}}$ is still a high-temperature state, we can approximate it as 
\begin{equation}
    \hat{\rho}_{\mathrm{pre}}\approx \frac{\hat{\mathbbm{1}}-\beta_{\text{eff}}\hat{H}_{\mathrm{F}}}{2^N}=\frac{\hat{\mathbbm{1}}+\epsilon_0(\delta\hat{\rho}_{\mathrm{pre}})}{2^N},
    \label{rho5}
\end{equation}
with inverse effective temperature 
\begin{equation}
    \beta_{\text{eff}}=-\epsilon_0\tr\small(\delta\hat{\rho}_4\hat{H}_{\mathrm{F}}\small)/\tr\small(\hat{H}_{\mathrm{F}}^2\small).
    \label{SpinTem}
\end{equation} 
Therefore, we get
\begin{equation}
    \delta\hat{\rho}_5=e^{-\frac{\mathrm{i}}{\hbar}\hat{H}_{\mathrm{F}}nT}\delta\hat{\rho}_4e^{\frac{\mathrm{i}}{\hbar}\hat{H}_{\mathrm{F}}nT}=\delta\hat{\rho}_{\mathrm{pre}}=u\hat{H}_{\mathrm{F}},
    \label{PartialRho5}
\end{equation}
where $u=-\beta_{\text{eff}}/\epsilon_0$. The number of cycles $n$ should be experimentally optimized to ensure that a quasi-stationary state has been reached (\ref{Opt of Para}e).

\smallskip
\textbf{Notes}: We need to calibrate the relative errors of thirteen different configurations of the anisotropic parameters $\bm{\xi}'$ in the effective Hamiltonian $\hat{H}_{\mathrm{F}}$ \eqref{FloHam}, with respect to the target values $\bm{\xi}$. We fix $\xi_z=0.2$, which leads to $\xi_x+\xi_y=-0.2$, and independently tune $\xi_x$ from $-0.4$ to $0.2$ in a step of $0.05$. To ensure a proper inverse effective temperature $\beta_{\text{eff}}$ \eqref{SpinTem} for a given configuration of $\bm{\xi}$, it is necessary to select the azimuthal angles ($\alpha$ and $\alpha-\pi/2$) of the Jeener-Broekaert pulse pair in step 2-4 appropriately. Taking $\bm{\xi}=(-0.4,\ 0.2,\ 0.2)$ as example, if we set $\alpha$ equal to the azimuthal angle of the $\hat{y}$ axis $\Phi_y$, $\delta\hat{\rho}_4$ will be the form of \eqref{rho4}, and the corresponding inverse effective temperature \eqref{SpinTem} will be approximately zero. Therefore, for $\xi_x\in\left[-0.4,-0.1\right]$, we set $\alpha=\Phi_{x}$ or $\Phi_{-x}$, and for $\xi_x\in\left[-0.1,0.2\right]$, set $\alpha=\Phi_{y}$ or $\Phi_{-y}$. 

\smallskip
The goal of the next half of the sequence is to prepare the dipolar-ordered states $\delta\hat{\rho}_\alpha^\mathcal{D}$ \eqref{DO_states_pro} as reference states. The preparation process can be analyzed by the Heisenberg evolution of the observable:

1': The final observable is 
\begin{equation}
    \hat{O}_{1'}=\hat{O}_x.
\end{equation}

\smallskip
2': Free evolution for the same time $t_e$ as in step 3 under the secular dipolar Hamiltonian $\hat{H}_{\mathrm{dip}}^{\mathrm{sec}}=\hat{D}_z$ \eqref{sysHam} gives 
\begin{equation}
    \begin{aligned}
        \hat{O}_{2'}=&\mathrm{e}^{\frac{\mathrm{i}}{\hbar}\hat{H}_{\mathrm{dip}}^{\mathrm{sec}}t_e}\hat{O}_{1'}\mathrm{e}^{-\frac{\mathrm{i}}{\hbar}\hat{H}_{\mathrm{dip}}^{\mathrm{sec}}t_e}
        =\hat{O}_x+\frac{\mathrm{i}}{\hbar}[\hat{D}_z,\hat{O}_x]t_e+\cdots\\
        =&\hat{O}_x-\frac{3}{2}t_e\sum_{i\ne j,\ ab}J_{ij}\left(\hat{S}^z_{ia}\hat{S}^y_{jb}+\hat{S}^y_{ia}\hat{S}^z_{jb}\right)+\cdots.
    \end{aligned}
\end{equation}

\smallskip
3': The subsequent $(\pi/4)_x$ pulse makes 
\begin{equation}
    \begin{aligned}
        \hat{O}_{3'}=&\hat{R}^\dagger_x(\pi/4)\hat{O}_{2'}\hat{R}_x(\pi/4)\\
        =&\hat{O}_x+\frac{3}{2}t_e\sum_{i\ne j,\ ab}J_{ij}\left(\hat{S}^z_{ia}\hat{S}^z_{jb}-\hat{S}^y_{ia}\hat{S}^y_{jb}\right)+\cdots.
    \end{aligned}
\end{equation}

\smallskip
4': Similar to the step $5$, the operator $\hat{O}_{3'}$ thermalizes to the dipolar-ordered state \cite{jeener1967,cho2003} \eqref{DO_states_pro}, after a period of free evolution under the secular dipolar Hamiltonian  $\hat{H}_{\mathrm{dip}}^{\mathrm{sec}}=\hat{D}_z$ \eqref{sysHam}:
\begin{equation}
    \hat{O}_{4'}=\mathrm{e}^{\frac{\mathrm{i}}{\hbar}\hat{H}_{\mathrm{dip}}^{\mathrm{sec}}t_{\mathrm{dip}}}\hat{O}_{3'}\mathrm{e}^{-\frac{\mathrm{i}}{\hbar}\hat{H}_{\mathrm{dip}}^{\mathrm{sec}}t_{\mathrm{dip}}}
    =u'\hat{D}_z=\delta\hat{\rho}_z^\mathcal{D},
    \label{O4'}
\end{equation}
where $u'=\tr\small(\hat{O}_{3'}\hat{D}_z\small)/\tr\small(\hat{D}_z^2\small)$, and the evolution time $t_{\mathrm{dip}}$ should be experimentally optimized to ensure that a quasi-stationary state has been reached (\ref{Opt of Para}c).

\smallskip
By implementing an operation $\hat{R}\in\{\hat{R}_y(\pi/2),\ \hat{R}_x(\pi/2)\}$, we can realize the dipolar-ordered states along the $\hat{x},\ \hat{y}$ directions, respectively:
\begin{equation}
    \begin{aligned}
        \delta\hat{\rho}_x^\mathcal{D}=&\hat{R}^\dagger_y(\pi/2) \hat{O}_{4'}\hat{R}_y(\pi/2)=u'\hat{D}_x,\\
        \delta\hat{\rho}_y^\mathcal{D}=&\hat{R}^\dagger_x(\pi/2) \hat{O}_{4'}\hat{R}_x(\pi/2)=u'\hat{D}_y.
    \end{aligned}
    \label{DOstate}
\end{equation}
The complete experimental sequence then effectively measures the inner products between the prethermal state $\delta\hat{\rho}_5=\delta\hat{\rho}_{\mathrm{pre}}$ \eqref{rho5} and the dipolar-ordered states $\delta\hat{\rho}_\alpha^\mathcal{D}$, which is exactly proportional to the to-be-calibrated parameter $\xi_\alpha'$ (see equation \eqref{y123} for derivation):
\begin{equation}
    \begin{aligned}
        y_1&=\tr(\delta\hat{\rho}_{\mathrm{pre}}\delta\hat{\rho}_x^\mathcal{D})=uu'\tr(\hat{H}_{\mathrm{F}}\hat{D}_x)\propto \xi_x',\\
        y_2&=\tr(\delta\hat{\rho}_{\mathrm{pre}}\delta\hat{\rho}_y^\mathcal{D})=uu'\tr(\hat{H}_{\mathrm{F}}\hat{D}_y)\propto \xi_z',\\
        y_3&=\tr(\delta\hat{\rho}_{\mathrm{pre}}\delta\hat{\rho}_z^\mathcal{D})=uu'\tr(\hat{H}_{\mathrm{F}}\hat{D}_z)\propto \xi_z'.
    \end{aligned}
\end{equation}
\begin{table}[htbp]
    \centering
    \begin{tabular}{|c|c|c|c|}
    \hline
      ~ & $\hat{H}_{\mathrm{th}} \ (4\to 5)$  & $\hat{R}$ & $\hat{H}_{\mathrm{th}} \ (3'\to 4')$ \\ \hline
    $C_1$ & $\hat{H}_{\mathrm{dip}}^{\mathrm{sec}}$ &  $\hat{\mathbbm{1}}$ & $\hat{H}_{\mathrm{dip}}^{\mathrm{sec}}$ \\ \hline
    $C_2$ & $\hat{H}_{\mathrm{dip}}^{\mathrm{sec}}$ &  $\hat{R}_y(\frac{\pi}{2})$ & $\hat{H}_{\mathrm{dip}}^{\mathrm{sec}}$ \\ \hline
    $C_3$ & $\hat{H}_{\mathrm{F}}$ &  $\hat{\mathbbm{1}}$ & $\hat{H}_{\mathrm{dip}}^{\mathrm{sec}}$ \\ \hline   
    $C_4$ & $\hat{H}_{\mathrm{F}}$ &  $\hat{R}_y(\frac{\pi}{2})$ & $\hat{H}_{\mathrm{dip}}^{\mathrm{sec}}$ \\ \hline
    $C_5$ & $\hat{H}_{\mathrm{F}}$ &  $\hat{\mathbbm{1}}$ & $\hat{H}_{\mathrm{F}}$ \\ \hline
    \end{tabular}
    \caption{\textbf{Guidelines for measuring $C_1$ to $C_5$ in equations \eqref{quantity1}-\eqref{quantity5}.} Utilizing the same pulse scheme depicted in \ref{pulse}b, it is feasible to measure all the five quantities $C_1$ through $C_5$. This process necessitates only the modification of the (pre)thermalization Hamiltonian during the transition of step $4\to5$ and $3'\to4'$, as well as the operation $\hat{R}$.} 
    \label{tab:measure C_i}
\end{table}

As for the calibration of the proportion of the residue terms $\hat{H}_\epsilon$, we need to measure the five quantities $C_1$ to $C_5$ as demonstrated in equations \eqref{quantity1}-\eqref{quantity5}. The values of $C_3=y_3$ and $C_4=y_1$ have already been determined. The measurement procedures for $C_1$, $C_2$, and $C_5$ bear a significant resemblance to those of $C_3$ and $C_4$, necessitating only the substitution of the (pre)thermalization process in the pulse sequence depicted in \ref{pulse}b. These guidelines are succinctly encapsulated in \ref{tab:measure C_i}.

\subsection{Assessment of the dipolar-ordered state and optimization of the experimental parameters} \label{sec4b}
The quality of the preparation of the dipolar-ordered states\cite{jeener1967,cho2003} and the sufficiency of the Floquet prethermalization are the crucial ingredients in our calibration protocol. In the following discussions, we will first introduce how to assess the dipolar-ordered states and then present the optimized experimental parameters for preparing these states and achieving prethermalization.

The verification of the dipolar-ordered states is via measuring the multiple quantum coherences (MQCs) in two non-commuting bases~ \cite{cho2003}. In the $\hat{z}$ basis, MQCs for a Hermitian operator $\hat{V}$ reads
\begin{equation}
    \begin{aligned}
        \tr\left(\hat{V} e^{-\mathrm{i}\hat{O}_z\phi}\hat{V} e^{\mathrm{i}\hat{O}_z\phi}\right)&=\sum_{p,q}\abs{\bra{p}\hat{V}\ket{q}}^2e^{-\mathrm{i}(Z_{p}-Z_{q})\phi}\\
        &=\sum_m\left(\sum_{Z_{p}-Z_{q}=m}\abs{\bra{p}\hat{V}\ket{q}}^2\right)e^{-\mathrm{i}m\phi}\\
        &=\sum_mI_m^{(z)} e^{-\mathrm{i}m\phi},
    \end{aligned}
\end{equation}
Here, $\ket*{p}$, $\ket*{q}$ symbolize the eigenvectors of the global spin operator $\hat{O}_z$, and $Z_p$, $Z_q$ represent their eigenvalues. $I_m^{(z)}\equiv \sum_{Z_{p}-Z_{q}=m}\abs{\bra{p}\hat{V}\ket{q}}^2$ is defined as the MQC intensity of order $m$. Similarly we could define MQC with respect to other operators such as $\hat{O}_x$. For a dipolar-ordered state such as  $\delta\hat{\rho}_{z}^\mathcal{D}\propto\hat{D}_z\equiv\frac{\hbar}{2}\sum_{i\ne j,\ ab}J_{ij}(-\hat{S}^x_{ia}\hat{S}^x_{jb}-\hat{S}^y_{ia}\hat{S}^y_{jb}+2\hat{S}^z_{ia}\hat{S}^z_{jb})$, we can calculate its MQCs in the eigenbases of $\hat{O}_z$ and $\hat{O}_x$, respectively\cite{ramanathan2003,cho2003}. In the $\hat{z}$ basis,
\begin{equation}
    \delta\hat{\rho}_{z}^\mathcal{D}\propto\sum_{i<j,\ ab}J_{ij}\left\{-\frac{1}{2}\left[\hat{S}^{+(z)}_{ia}\hat{S}_{jb}^{-(z)}+\hat{S}_{ia}^{-(z)}\hat{S}_{jb}^{+(z)}\right]+2\hat{S}^z_{ia}\hat{S}^z_{jb}\right\},
\end{equation}
where $\hat{S}_{ia}^{+(z)}=\hat{S}^x_{ia}+\mathrm{i}\hat{S}^y_{ia},\ S_{ia}^{-(z)}=\hat{S}^x_{ia}-\mathrm{i}\hat{S}^y_{ia}$. Thus, only zeroth-order quantum coherence $I_0^{(z)}$ presents. In the $\hat{x}$ basis,
\begin{equation}
\begin{aligned}
    \delta\hat{\rho}_{z}^\mathcal{D}\propto&\sum_{i<j,\ ab}J_{ij}\left\{-\hat{S}^x_{ia}\hat{S}^x_{jb}+\frac{1}{4}\left[\hat{S}_{ia}^{+(x)}\hat{S}_{jb}^{-(x)}+\hat{S}_{ia}^{-(x)}\hat{S}_{jb}^{+(x)}\right]\right\}-\sum_{i<j,\ ab}J_{ij}\frac{3}{4}\left[\hat{S}_{ia}^{+(x)}\hat{S}_{jb}^{+(x)}+\hat{S}_{ia}^{-(x)}\hat{S}_{jb}^{-(x)}\right],
\end{aligned}
\end{equation}
where $\hat{S}^{+(x)}_{ia}=\hat{S}^y_{ia}+\mathrm{i}\hat{S}_{ia}^z,\ \hat{S}^{-(x)}_{ia}=\hat{S}^y_{ia}-\mathrm{i}\hat{S}^z_{ia}$. There are both the second- and zeroth-order quantum coherences, denoted by the first and second terms, respectively. \textbf{The ratio between the intensities of the second- and zeroth-order quantum coherences is calculated to be $\bm{r=I_2^{(x)}/I_0^{(x)}=1.5}$}, which value can be experimentally measured to verify the dipolar-ordered state $\delta\hat{\rho}_{z}^\mathcal{D}$. 
\begin{figure}[htbp]
    \begin{center}
    \includegraphics[scale=0.9]{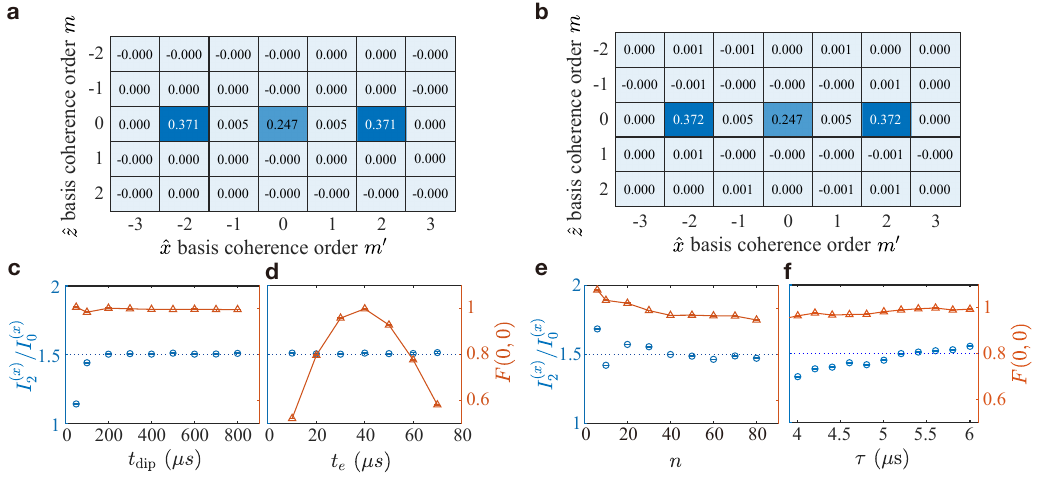}
    \caption{\textbf{Assessment of the dipolar-ordered states and optimization of experimental parameters.} \textbf{a},\textbf{b}, Using the pulse sequence referred from Cho et al. \cite{cho2003} (\ref{pulse}c), we present the two-dimensional multiple quantum coherence (MQC) spectrum of the dipolar-ordered states, thermalized by either (\textbf{a}) the secular dipolar Hamiltonian $\hat{H}_{\mathrm{dip}}^{\mathrm{sec}}=\hat{D}_z$ \eqref{sysHam}, or \textbf{b} the effective Hamiltonian under Floquet driving, with configuration set to be $\bm{\xi}^{(0)}=(-0.1,\ -0.1,\ 0.2)$, leading $\hat{H}_{\mathrm{F}}\approx 0.1\hat{D}_z$. In the $\hat{z}$ basis, only the zeroth-order coherence $I_0^{(z)}$ is significantly present; while in the $\hat{x}$ basis, the ratio between the second- and zeroth-order coherence intensities is $r=I_2^{(x)}/I_0^{(x)}=1.5038(12)$ in \textbf{a} and $r=1.5068(6)$ in \textbf{b}, indicating that the system has reached close to the dipolar-ordered state, which has an exact ratio of $r=1.5$. \textbf{c}-\textbf{f}, We illustrate how the ratio $r=I_2^{(x)}/I_0^{(x)}$ and the signal amplitude $F(0,\ 0)$ defined by equation \eqref{F} depend on the four experimental parameters: $t_{\mathrm{dip}}$ \eqref{O4'}, $t_e$ \eqref{rho3}, $n$ \eqref{PartialRho5}, and $\tau$ \eqref{PulseIntervals}, respectively. In this context, we simply measure the one-dimensional MQC spectrum $\left\{I_m\right\}$ in the $\hat{x}$ basis. In \textbf{c}, with $t_e=40\ \mu\mathrm{s}$ fixed, we determine that the ratio $r$ becomes stable at approximately $1.5$ after $t_{\mathrm{dip}}\ge 200\ \mu\mathrm{s}$. In \textbf{d}, we fix $t_{\mathrm{dip}}=600\ \mu\mathrm{s}$, leading to $t_e=40\ \mu\mathrm{s}$ as the optimal value as it generates the maximum signal $F(0,\ 0)$. In \textbf{e}, setting $t_{e}=40\ \mu\mathrm{s}$, $\tau=5\ \mu\mathrm{s}$, and assigning $\bm{\xi}^{(0)}=(-0.05,\ -0.05,\ 0.1)$ (creating $\hat{H}_{\mathrm{F}}\approx0.05\hat{D}_z$), we find that the ratio stabilizes at roughly $1.5$ after $n\ge 40$. Lastly, in \textbf{f}, keeping $t_e=40\ \mu\mathrm{s}$, $\bm{\xi}^{(0)}=(-0.1,\ -0.1,\ 0.2)$ (resulting in $\hat{H}_{\mathrm{F}}\approx0.1\hat{D}_z$), and $\tau\cdot n=200$, it is shown that the ratio increases with $\tau$. We choose $\tau=5\ \mu\mathrm{s}$, since it is neither too long to significantly incorporate contributions from the higher-order terms of the Floquet-Magnus expansion, nor too short to introduce errors caused by pulse transients. According to the average Hamiltonian theory, it is expected the ratio $r$ should converge to a plateau when $\tau$ becomes smaller. However, this is absent in our parameter range. We attribute this to the limitation of the value range of $\tau$ set by pulse transient effects \cite{tabuchi2010total,wittmann2015compensating}, with analysis further given in \ref{fig:Calibration of tau}. All values of $F(0,\ 0)$ are normalized with respect to the value of the fourth point in \textbf{d}, where $t_e=40\ \mathrm{\mu s}$ and $t_{\mathrm{dip}}=600\ \mathrm{\mu s}$. The error bars in \textbf{a},\textbf{b}, estimated as $\sim 10^{-5}$, are not depicted. All error bars are derived considering the noise amplitude in the free induction decay.}
    \label{Opt of Para}
    \end{center}
\end{figure}

Using the same pulse sequence (\ref{pulse}c) as used by Cho et al.  ~\cite{cho2003}, we can encode the MQCs in both the $\hat{z}$ and $\hat{x}$ basis simultaneously. The difference between this sequence and that in \ref{pulse}b lies in, firstly, that the azimuthal angles of the pulses in the initial half sequence are shifted by combinations of $\phi$ and $\eta$. For example, a pulse $(\pi/2)_{\Phi_y+\phi+\eta}$ induces an operation $\hat{R}_z(\phi+\eta)\hat{R}_{y}(\pi/2)\hat{R}_z^\dagger(\phi+\eta)$. The overall effect of these shifts is to realize successive rotations $\hat{R}_x(\phi)\hat{R}_z(\eta)$ in the middle of the sequence, instead of the $\hat{R}$ operator introduced in \ref{pulse}b. Secondly, for both periods of thermalization, we choose the same Hamiltonian $\hat{H}_{\mathrm{th}}$. Specifically, this Hamiltonian can be either the secular dipolar Hamiltonian: $\hat{H}_{\mathrm{dip}}^{\mathrm{sec}}=\hat{D}_z$ \eqref{sysHam}; or the effective Hamiltonian \eqref{FloHam} under Floquet driving: $\hat{H}_{\mathrm{F}}\approx 0.1\hat{D}_z$ with configuration chosen to be $\bm{\xi}^{(0)}=(-0.1,\ -0.1,\ 0.2)$. After a sufficiently long period of evolution $t_\text{th}$, at step $5$ and step $4'$ in the pulse sequence (\ref{pulse}c) we have prepared two dipolar-ordered states as
\begin{equation}
\delta\hat{\rho}_{5}=\hat{O}_{4'}=\delta\hat{\rho}_{z}^\mathcal{D}\propto\hat{D}_z, 
\end{equation}
then through the combined rotation $\hat{R}_x(\phi)\hat{R}_z(\eta)$, we measure the two-dimensional encoding function below:
\begin{equation}
    \begin{aligned}
        F(\phi,\eta)\equiv&\tr\left[\hat{R}_{x}(\phi)\hat{R}_{z}(\eta)\delta\hat{\rho}_{z}^\mathcal{D}\hat{R}_{z}^\dagger(\eta) \hat{R}_{x}^\dagger(\phi)\delta\hat{\rho}_{z}^\mathcal{D}\right]\\
        =&\sum_{i,j,p,q}e^{-\mathrm{i}(Z_{p}-Z_{q})\eta}e^{\mathrm{i}(X_{i}-X_{j})\phi}\bra{p}\delta\hat{\rho}_{z}^\mathcal{D}\ket{q}\bra{i}\delta\hat{\rho}_{z}^\mathcal{D}\ket{j}\braket*{q}{i}\braket*{j}{p}.
        \label{F}
    \end{aligned}
\end{equation}
Here, $\ket*{p}$, $\ket*{q}$, $Z_p$ and $Z_q$ symbolize the eigenvectors and eigenvalues of the global spin operator $\hat{O}_z$, while $\ket*{i}$, $\ket*{j}$, $X_i$ and $X_j$ represent those of $\hat{O}_x$. $\phi$ and $\eta$ are independently incremented from $0$ to $2\pi$, with a step $\Delta\phi=\Delta\eta=\pi/4$. Subsequently, we conduct a two-dimensional Fourier transform on $F(\phi,~\eta)$ with respect to $\phi$ and $\eta$. Consequently, the MQC spectrum in the $\hat{x}-\hat{z}$ plane is obtained, which is expressed as
\begin{equation}
    I(m,m')=\sum_{Z_{q}-Z_{p}=m}\sum_{X_{i}-X_{j}=m'}\bra{p}\delta\hat{\rho}_{z}^\mathcal{D}\ket{q}\bra{i}\delta\hat{\rho}_{z}^\mathcal{D}\ket{j}\braket*{q}{i}\braket*{j}{p},
\end{equation}
where $m$ and $m'$ denote the $\hat{z}$ and $\hat{x}$ coherence order, respectively. To obtain the one-dimensional MQC spectrum in the $\hat{x}$ or $\hat{z}$ basis, it suffices to keep either $\eta$ or $\phi$ at zero.

In \ref{Opt of Para}a,b, we present the experimentally measured two-dimensional MQC spectrum under the secular dipolar Hamiltonian $\hat{H}_{\mathrm{dip}}^{\mathrm{sec}}=\hat{D}_z$ \eqref{sysHam} and the effecive Hamiltonian $\hat{H}_{\mathrm{F}}\approx 0.1\hat{D}_z$, respectively. In both cases, only the zeroth-order coherence $I_0^{(z)}$ is significantly present in the $\hat{z}$ basis. Meanwhile, in the $\hat{x}$ basis, the ratios between the intensities of the second- and zeroth-order coherences, are $\bm{r=I_2^{(x)}/I_0^{(x)}=1.5038(12)}$ and $\bm{1.5068(6)}$, respectively, suggesting that we have prepared the dipolar-ordered states with high quality. These outcomes also indicate the validity of the (pre)thermalization hypothesis.

We further optimize the experimental parameters involved in the (pre)thermalization process. These parameters include the duration of the thermalization under the secular dipolar Hamiltonian $\hat{H}_{\mathrm{dip}}^{\mathrm{sec}}$, represented by $t_\mathrm{dip}$ in equation \eqref{O4'}, the Jeener-Broekaert pulse pair spacing time denoted as $t_{e}$ in equation \eqref{rho3}, the number of Floquet driving cycles depicted as $n$ in equation \eqref{PartialRho5}, and the average pulse interval in the 16-pulse sequence provided in the Hamiltonian engineering section \eqref{PulseIntervals}, demonstrated as $\tau$. \ref{Opt of Para}c-f illustrate how the ratio $r=I_2^{(x)}/I_0^{(x)}$ and the signal amplitude $F(\phi=0,\ \eta=0)$ (as provided in equation \eqref{F}) depend on the four experimental parameters: $t_{\mathrm{dip}}$, $t_e$, $n$, and $\tau$. In this context, we maintain a fixed value of $\eta=0$, and increment the angle $\phi$ from $0$ to $2\pi$ with a step $\Delta\phi=\pi/4$ or $\pi/8$, leading to the encoding of the MQCs in the $\hat{x}$ basis only and resulting a one-dimensional MQC spectrum $\left\{I_m^{(x)}\right\}$. 

In \ref{Opt of Para}c, with $t_e=40\ \mu\mathrm{s}$ fixed, we determine that the ratio $r$ becomes stable at approximately $1.5$ after the thermalization time $t_{\mathrm{dip}}\ge 200\ \mu\mathrm{s}$. In \ref{Opt of Para}d, we fix $t_{\mathrm{dip}}=600\ \mu\mathrm{s}$, leading to $t_e=40\ \mu\mathrm{s}$ as the optimal value since it generates the maximum signal $F(0,\ 0)$. 

\begin{figure}[htbp]
    \begin{center}
        \includegraphics[scale=1]{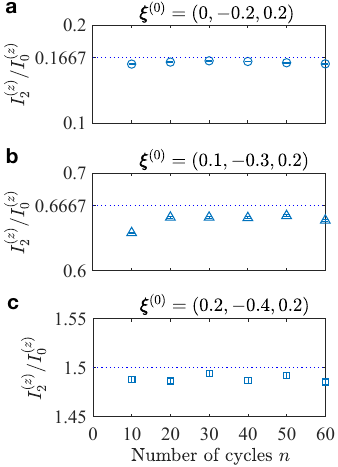}
        \caption{\textbf{Verification of the prethermalization process for several Hamiltonian configurations other than the dipolar configuration.} We measure the MQC ratio $\hat{I}_2^{(z)}/I_0^{(z)}$ for the three configurations in the $\hat{z}$ basis, instead of in the $\hat{x}$ basis as in \ref{Opt of Para}e. The dashed lines denote the expected ratios for the target configurations.}
        \label{fig:Prethermal_VaryLoops}
    \end{center}
\end{figure}

In \ref{Opt of Para}e, setting $t_{e}=40\ \mu\mathrm{s}$, $\tau=5\ \mu\mathrm{s}$, and assigning $\bm{\xi}^{(0)}=(-0.05,\ -0.05,\ 0.1)$ (creating the effective Hamiltonian $\hat{H}_{\mathrm{F}}\approx0.05\hat{D}_z$), we find that the ratio $r$ stabilizes at roughly $1.5$ after the number of cycles $n\ge 40$. Equivalently, for $\bm{\xi}^{(0)}=(-0.1,-0.1,0.2)$ it is sufficient to set $n\ge 20$. Among the collection of configurations we explored: $\bm{\xi}^{(0)}=(\xi_x,-0.2-\xi_x,0.2)$ and $\xi_x\in[-0.4,0.2]$, $\bm{\xi}^{(0)}=(-0.1,-0.1,0.2)$ has the smallest interaction strength. Therefore, it is expected that a value of $n\ge 20$ is also adequate for the other configurations in this collection. This is indeed consistent with our experimental observation shown in \ref{fig:Prethermal_VaryLoops}, where the prethermalization process for three configurations\textemdash$\bm{\xi}^{(0)}=(\xi_x,-0.2-\xi_x,0.2)$, $\xi_x\in\{0,0.1,0.2\}$\textemdash are verified. Here we measure the MQC ratio $\hat{I}_2^{(z)}/I_0^{(z)}$ in the $\hat{z}$ basis, instead of in the $\hat{x}$ basis when calibrating the dipolar Hamiltonian (\ref{Opt of Para}e). For all the three configurations explored, the ratios reach plateaus close to their target values after a number of evolution cycles $n\ge 20$. During our implementation of the calibration protocol, we set $n=40$. 

Lastly, in \ref{Opt of Para}f, in order to calibrate the average pulse interval $\tau$ we vary $\tau\in[4,6]~\mu\mathrm{s}$ while fixing $n*\tau=200$ and $t_e=40\ \mu\mathrm{s}$. The configuration is $\bm{\xi}^{(0)}=(-0.1,-0.1,0.2)$ (resulting in $\hat{H}_{\mathrm{F}}\approx0.1\hat{D}_z$). It is shown that the ratio increases with $\tau$. According to the average Hamiltonian theory, it is expected the ratio $r$ should converge to a plateau of $r=1.5$ when $\tau$ becomes smaller. However, this is absent in our parameter range of $\tau$. We attribute this to the limitation of the value range of $\tau$ set by pulse transient effects \cite{tabuchi2010total,wittmann2015compensating}. We choose $\tau=5\ \mu\mathrm{s}$, since it is neither too long to significantly incorporate contributions from the higher-order terms of the Floquet-Magnus expansion, nor too short to introduce errors caused by pulse transients. 

\begin{figure}[htbp]
    \begin{center}
        \includegraphics[scale=1]{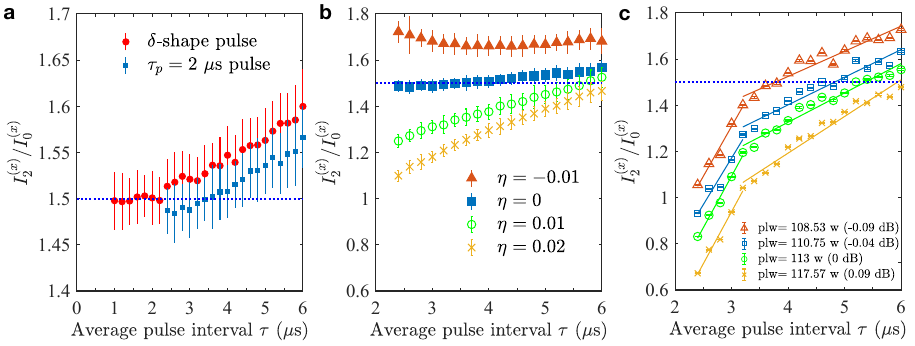}
        \caption{\textbf{Detailed investigation of how the ratio $r=I_2^{(x)}/I_0^{(x)}$ relies on the average pulse interval $\tau$.} \textbf{a}, Via numerical simulation it is shown that for the $\delta$-shape pulse case (red dots), $r$ converges to a plateau of $1.5$ when $\tau\le 2.2~\mu\mathrm{s}$. While for the finite-pulse case ($\pi/2$-pulse width $\tau_\mathrm{p}=2~\mu\mathrm{s}$) denoted by the blue squares, $r$ is decreased compared with the $\delta$-pulse case. Besides, it seems that $r$ reaches a plateau when $\tau\le 3~\mu\mathrm{s}$. \textbf{b}, We further simulate how the flip angle error $\eta=(\phi-\phi_{\mathrm{tar}})/\phi_{\mathrm{tar}}$ affects the behavior of $r$ versus $\tau$. The ED calculation in \textbf{a,b} is conducted for a system size $N=10$, averaged over $50$ times of random realizations. \textbf{c}, We experimentally varied the RF pulse power from the calibrated value $\mathrm{plw}=113~\mathrm{w}$ to provide a variation of $\eta$. At the point $\tau\approx 3.2~\mu\mathrm{s}$ there seem to be sudden changes of slopes, implying that the phase transients begin to play a role.}
        \label{fig:Calibration of tau}
    \end{center}
\end{figure}

In the following we give our analysis of the lack of a plateau in \ref{Opt of Para}f, combining both the numerical and experimental investigations. Via exact diagonalization (ED) calculation, we firstly simulate the $\delta$-pulse case, for which the ratio $r$ converges to a plateau of $1.5$ when $\tau\le 2.2~\mu\mathrm{s}$ (red dots in \ref{fig:Calibration of tau}a), consistent with the expectation from the average Hamiltonian theory. Secondly, we simulate the finite-pulse effect by introducing a $\pi/2$-pulse width $\tau_\mathrm{p}=2~\mu\mathrm{s}$. Accordingly, the specific pulse intervals $\tau_\alpha'$ in the sequence \ref{pulse}a, which are calculated between the edges of two consecutive pulses, are set according to \eqref{eq:Interval between edges} (reproduced below):
\begin{equation*}
    \begin{aligned}
        \tau_x'&=(1+\xi_x^{(0)})\tau-\frac{\tau_\mathrm{p}}{2}=0.9\tau-\frac{\tau_\mathrm{p}}{2},\\
        \tau_y'&=(1+\xi_y^{(0)})-\tau_\mathrm{p}=0.9\tau-\tau_\mathrm{p},\\
        \tau_z'&=(1+\xi_z^{(0)})-\frac{\tau_\mathrm{p}}{2}=1.1\tau-\frac{\tau_\mathrm{p}}{2}.
    \end{aligned}
\end{equation*}
The condition $\tau'_\alpha>0$ sets a lower bound $\tau>2.22~\mu\mathrm{s}$. It is shown that the finite-pulse effect causes a decrease of the ratio $r$ (blue squares in \ref{fig:Calibration of tau}a), compared with the $\delta$-pulse case. Besides, it seems that $r$ reaches a plateau when $\tau\le 3~\mu\mathrm{s}$. In \ref{fig:Calibration of tau}b we further simulate how the flip angle error $\eta=(\phi-\phi_{\mathrm{tar}})/\phi_{\mathrm{tar}}$ affects the behavior of $r$ versus $\tau$. Basically, a positive $\eta$ results in a decrease of $r$, while a negative $\eta$ causes an increase of $r$. This is consistent with the experimental observation, where the RF pulse power is varied from the calibrated value $\mathrm{plw}=113~\mathrm{w}$ to provide a variation of $\eta$ (\ref{fig:Calibration of tau}c). On the other hand, there are discrepancies between the behaviors in \ref{fig:Calibration of tau}b and \ref{fig:Calibration of tau}c, especially when $\tau\le 3.2~\mu\mathrm{s}$. We attribute this to the pulse transient effects. At the point $\tau\approx 3.2~\mu\mathrm{s}$ there seem to be sudden changes of slopes, indicating that $0.88~\mu\mathrm{s}$\textemdash the smallest value of $\tau_\alpha'$ when $\tau=3.2~\mu\mathrm{s}$\textemdash is possibly a critical bound, below which the pulse transients start playing a significant role. This is the basic reason for the absence of a plateau of the ratio $r$ in experiment. 

\subsection{Calibration results of the anisotropic parameters and the residue Hamiltonian} \label{sec4c}
Upon optimization of the experimental parameters involved in the Hamiltonian engineering and calibration protocol, it is achievable to prepare the prethermal states $\hat{\rho}_\text{pre}$ \eqref{PreTherState} and the dipolar-ordered states $\delta\hat{\rho}_\alpha^{\mathcal{D}}$ \eqref{DO_states_pro} in high quality. These two kinds of states are crucial ingredients in the calibration protocol described in \ref{sec4a}. In this subsection we present the calibration results of $\bm{\xi}'$ and the residue terms $\hat{H}_{\epsilon}$ in the effective Hamiltonian $\hat{H}_\text{F}$ \eqref{FloHam}.

\begin{figure}[htbp]
    \begin{center}
        \includegraphics[scale=1]{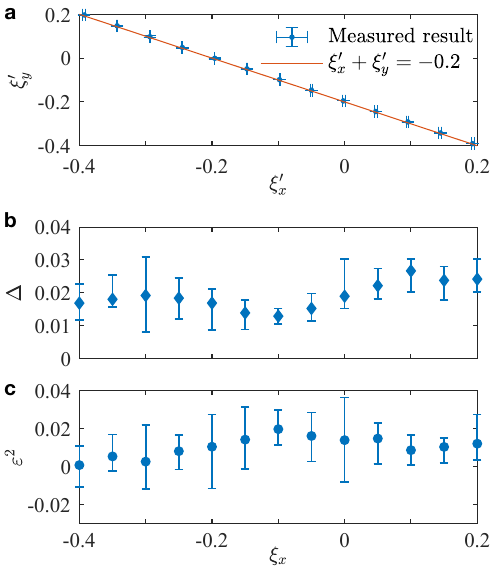}
        \caption{\textbf{Calibration results of the effective Hamiltonian.} \textbf{a}, Following the protocol described in \ref{sec4a}, we exhibit the calibration results of the realized configurations $\{\xi_x',\ \xi_y',\ \xi_z'\}$, where $\xi_z'$ is already calibrated to be $0.198\pm 0.001$ when introducing the protocol. The target line  $\xi_x'+\xi_y'=-0.2$ is depicted as comparison. \textbf{b}, We show the deviation $\Delta\equiv\abs*{\bm{\xi}'-\bm{\xi}}/\abs*{\bm{\xi}}< 3\%$ across all the configurations. \textbf{c}, We depict the square of the proportion $\varepsilon$ of the residual terms $\hat{H}_\epsilon$ in the total effective Hamiltonian $\hat{H}_\mathrm{F}$, defined by $\varepsilon\equiv\sqrt{\tr(\hat{H}_\epsilon^2)/\tr(\hat{H}_\mathrm{F}^2)}$. $\varepsilon^2<4\%$ so that $\varepsilon<20\%$ across all the configurations. The error bars denote the lower and upper bounds in 10 repetitive calibrations.}
        \label{tomo result}
    \end{center}
\end{figure}

In \ref{tomo result}a, we exhibit the calibration results of the realized configurations $\{\xi_x',\ \xi_y',\ \xi_z'\}$, where $\xi_z'$ is already calibrated to be $0.198\pm 0.001$ when introducing the protocol. The target line  $\xi_x'+\xi_y'=-0.2$ is depicted as comparison. We define the deviation from the target configuration as
\begin{equation}
    \Delta\equiv\frac{\abs*{\bm{\xi}'-\bm{\xi}}}{\abs*{\bm{\xi}}}=\frac{\sqrt{\sum_\alpha(\xi_\alpha'-\xi_\alpha)^2}}{\abs*{\bm{\xi}}}.
    \label{Delta}
\end{equation}
In \ref{tomo result}b, we show the deviation $\Delta$ for each target value of $\xi_{x}$, which remains within $3\%$. We conduct $10$ times of repetitive calibrations.

Next, we calibrate the proportion of the residue terms $\hat{H}_\epsilon$ in the total effective Hamiltonian $\hat{H}_{\mathrm{F}}$ \eqref{FloHam}, reproduced as below:
\begin{equation*}
    \hat{H}_{\mathrm{F}}
        =\frac{\hbar}{2}\sum\limits_{i\ne j,ab}J_{ij}\left(\xi_x'\hat{S}^x_{ia}\hat{S}^x_{jb}+\xi_y'\hat{S}^y_{ia}\hat{S}^y_{jb}+\xi_z'\hat{S}^z_{ia}\hat{S}^z_{jb}\right)+\hat{H}_\epsilon.
\end{equation*}
The proportion is defined by equation \eqref{varepsilon} reproduced as below:
\begin{equation*}
\varepsilon\equiv\sqrt{\frac{\tr\small(\hat{H}_\epsilon^2\small)}{\tr\small(\hat{H}_{\mathrm{F}}^2\small)}}.
\end{equation*}
In \ref{tomo result}c we present the calibration results of $\varepsilon^2$, leading to the conclusion that $\varepsilon<20\%$ across all the configurations. We conduct $10$ times of repetitive calibrations.

\begin{figure}[htbp]
    \begin{center}
        \includegraphics[scale=1]{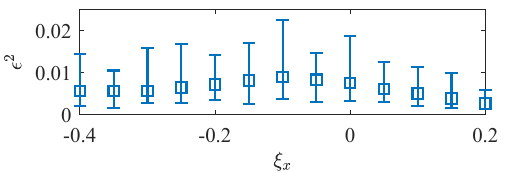}
        \caption{\textbf{Numerical simulation of the weight of the residue terms.} We display the square of the weight $\varepsilon^2$ for different values of $\xi_x$, thus making a quantitative comparison with the experimental calibration results in \ref{tomo result}c. In order to obtain $\varepsilon^2$, we calculate the Floquet-Magnus expansion up to the second order \eqref{FM expansion} of the effective Hamiltonian for the pulse sequence \eqref{sequence}, from which the residue terms are extracted. The system size is limited to $N=8$, and the system Hamiltonian is denoted by equation \eqref{Hami_nume}. In numerical calculations, we set the average pulse interval to $\tau=5\ \mathrm{\mu\mathrm{s}}$, which is the same as in the experiment. The squares and the errorbars denote the mean values, and the lower and upper bounds, respectively, in $10^2$ disorder realizations.}
        \label{Nume_epsilon}
    \end{center}
\end{figure}
In order to explain the source of the experimental calibration results of $\varepsilon$, we numerically calculate the FM expansion of the effective Hamiltonian up to the second order \eqref{FM expansion} for the pulse sequence \eqref{sequence}. Due to computational limitations, the calculation is limited to a small system size of $N=8$, with a simplified dipolar Hamiltonian:
\begin{equation}
    \hat{H}_{\mathrm{dip}}=\hbar\sum\limits_{i<j}^NJ_{ij}(-\hat{S}^x_{i}\hat{S}^x_{j}-\hat{S}^y_{i}\hat{S}^y_{j}+2\hat{S}^z_{i}\hat{S}^z_{j}).
    \label{Hami_nume}
\end{equation}
Here, $J_{ij}$ is randomly sampled from a Gaussian distribution, i.e., $J_{ij}\sim \mathcal{N}(0,\sigma^2)$, with $\sigma=2J/\sqrt{N}$, and $J/2\pi=1460$ Hz calibrated from experiment (see \ref{sec2}). In each disorder realization of the couplings $J_{ij}$, we extract the residue terms $\hat{H}_\epsilon$ from the second-order term of the FM expansion, and then their total weight $\varepsilon$ is calculated. We conduct $10^2$ disorder realizations and summarize the results in \ref{Nume_epsilon}, providing a quantitative comparison with the experimental calibration results displayed in \ref{tomo result}c.

\section{Comparison between dynamics of the global auto-correlation function and the multiple quantum coherences}

%\begin{figure}[htbp] 
    %\begin{center}  
    %\includegraphics[scale=1]{supp_figure/FigS9.pdf}
    %\caption{\textbf{Comparison of dynamics between the auto-correlation function $\mathcal{C}_z$ and the MQCs.} We depict the dynamics of the auto-correlation functions, in addition to the MQC dynamics displayed in Fig.~4 in the main text. The auto-correlation function and the MQCs were measured under the same effective Floquet Hamiltonian $\hat{H}_\mathrm{F}$, of which $(\xi_x,\xi_y,\xi_z)=(-0.125,-0.025,0.15$) for the monotonic case (\textbf{a}) or $(-0.1,0.1,0$) for the oscillatory case (\textbf{b}). The error bars ($\sim 10^{-4}$) are incorporated within the markers of the data points. $I_0^{(z)}(t)$ reaches its initial trough at $\bar{J}t_{\mathrm{dip}}^{(I)}/2\pi\approx1.28$, and $I_{\pm 2}^{(z)}(t)$ reaches its initial peak at $\bar{J}t_{\mathrm{peak}}^{(I)}/2\pi\approx1.22$, while $\mathcal{C}_z(t)$ reaches its initial trough at $\bar{J}t_{\mathrm{dip}}^{(\mathcal{C})}/2\pi\approx 2.39$.} 
   % \label{fig:Compare_MQC_autocorrelation}
    %\end{center}
%\end{figure}

\begin{figure}[htbp] 
    \begin{center}  
    \includegraphics[scale=1]{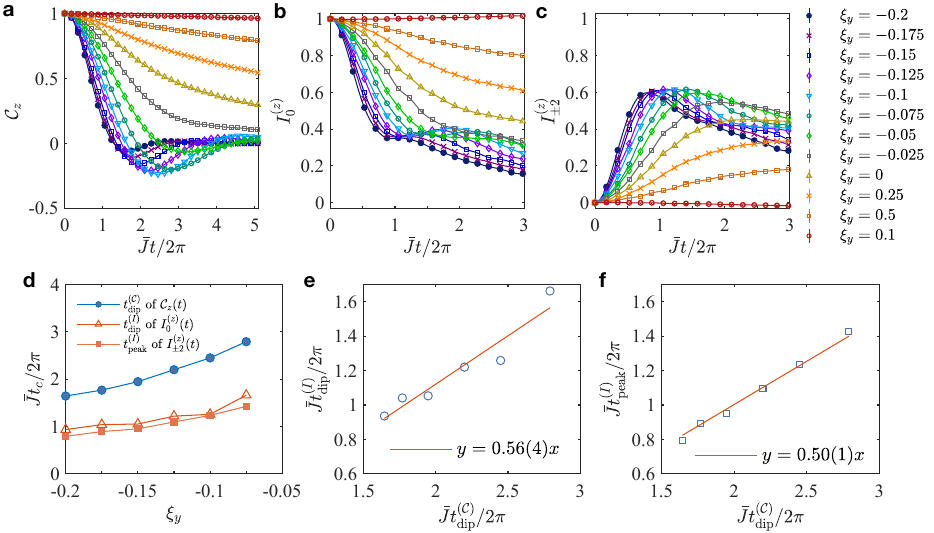}
    \caption{\textbf{The relations among $t_{\mathrm{dip}}^{(\mathcal{C})}$, $t_{\mathrm{peak}}^{(I)}$ and $t_{\mathrm{dip}}^{(I)}$.} \textbf{a-c}, We experimentally measured the dynamics of (\textbf{a}) the auto-correlation function $\mathcal{C}_z(t)$, (\textbf{b}) the zero quantum coherence $I_0^{(z)}(t)$, and (\textbf{c}) the double quantum coherence $I_{\pm 2}^{(z)}(t)$, for various Hamiltonian configurations. Across all the configurations we vary $\xi_y\in[-0.2,0.1]$ while fixing $\xi_x=0.1$. In this setup, $\xi_y=0$ is expected to be the oscillatory-monotonic conversion point. The solid lines denote the cubic spline interpolation. \textbf{d}, For the obvious oscillatory cases $\xi_y\in[-0.2,-0.075]$ we extract the three time points $t_{\mathrm{dip}}^{(\mathcal{C})}$, $t_{\mathrm{dip}}^{(I)}$ and $t_{\mathrm{peak}}^{(I)}$. \textbf{e,f}, We performed a linear fit of $t_{\mathrm{dip}}^{(I)}$ versus $t_{\mathrm{dip}}^{(\mathcal{C})}$ in \textbf{e} and $t_{\mathrm{peak}}^{(I)}$ versus $t_{\mathrm{dip}}^{(\mathcal{C})}$ in \textbf{f}, both revealing an approximate doubling relationship. } 
    \label{fig:Tpeak_Tdip_fit}
    \end{center}
\end{figure}

In Fig.~4 of the main text, we simultaneously depict the dynamics of the global auto-correlation functions and the MQCs, indicating that for the oscillatory case, the zero and double quantum coherences possibly oscillate with a frequency double that of the auto-correlation function. To be clear, we reproduce the definition of the auto-correlation function
\begin{equation}
    \mathcal{C}_z(t)\equiv\frac{1}{c_O}\Tr[\hat{O}_z(t)\hat{O}_z(0)],
\end{equation}
where $\hat{O}_z(t)=\mathrm{e}^{-\mathrm{i}\hat{H}_\mathrm{F}t}\hat{O}_z\mathrm{e}^{\mathrm{i}\hat{H}_\mathrm{F}t}$ and $c_O$ is a normalization constant such that $\mathcal{C}_z(0)=1$. The MQC intensity is defined as
\begin{equation}
   I_m^{(z)}(t)\equiv\sum_{\substack{p,q~\text{s.t.}\\\bra{p}\hat{O}_z\ket{p}-\bra{q}\hat{O}_z\ket{q}=m}}\frac{1}{c_O}\left|\bra{p}\hat{O}_z(t)\ket{q}\right|^2.
\end{equation}
In order to further verify this observation, we experimentally measured the dynamics of both the MQCs and the global auto-correlation function $\mathcal{C}_z$ under various Hamiltonian configurations. Across all the configurations we vary $\xi_y\in[-0.2,0.1]$ while fixing $\xi_x=0.1$. In this setup, $\xi_y=0$ is expected to be the oscillatory-monotonic conversion point. The results are depicted in \ref{fig:Tpeak_Tdip_fit}, where subfigures \textbf{a,b,c} correspond to the dynamics of the auto-correlation function $\mathcal{C}_z$, the zero quantum coherence $I_0^{(z)}$, and the double quantum coherence $I_{\pm 2}^{(z)}$, respectively. For the cases $\xi_y\in[-0.2,-0.075]$ where the oscillation is obvious, we extracted the three time points $t_{\mathrm{dip}}^{(\mathcal{C})}$, $t_{\mathrm{dip}}^{(I)}$ and $t_{\mathrm{peak}}^{(I)}$ via cubic spline interpolation (\ref{fig:Tpeak_Tdip_fit}d). Subsequently, We performed a linear fit of $t_{\mathrm{dip}}^{(I)}$ versus $t_{\mathrm{dip}}^{(\mathcal{C})}$ in \ref{fig:Tpeak_Tdip_fit}e, and $t_{\mathrm{peak}}^{(I)}$ versus $t_{\mathrm{dip}}^{(\mathcal{C})}$ in \ref{fig:Tpeak_Tdip_fit}f, both revealing an approximate doubling relationship. 

\section{Theoretical analysis}
In this section, we first provide an explanation of the theoretical analysis presented in the main text. The analysis comprises three key components: exact diagonalization for small subsystem sizes \ref{ED}, along with two widely employed approximation methods—large-$M$ expansion \ref{LargeM} and mean-field theory \ref{MF}. The code can be found on the website\cite{code}. Secondly, we will discuss the contributions of the four diagrams illustrated in the main text and provide a detailed derivation in \ref{diagrams}.

\subsection{Exact diagonalization} \label{ED}
In the exact diagonalization simulation, due to computational limitations, we are constrained to investigate a relatively small system size with $N=8$. As a result, we introduce a simplified model that consists of a single spin-1/2 on each molecule:
\begin{equation}
\hat{H}_{\text{ED}}=\sum\limits_{i<j}J_{ij}(\xi_x\hat{S}^x_{i}\hat{S}^x_{j}+\xi_y\hat{S}^y_{i}\hat{S}^y_{j}+\xi_z\hat{S}^z_{i}\hat{S}^z_{j}), \label{EDH}
\end{equation} 
where $J_{ij}$ is modelled as a random variable following a normal distribution $J_{ij}\sim \mathcal{N}[0,(2J/\sqrt{N})^2]$. This simplified model represents an essential building block of the larger system and provides valuable insights into its dynamics. In each disorder realization of the couplings $J_{ij}$, we prepare the initial state as a thermal density matrix denoted as $\hat{\rho} \propto \exp(-\beta(\hat{H}_{\text{ED}}+\delta \hat{H}))$, where an external polarization field is introduced as $\delta \hat{H}=-g \sum_i \hat{S}^x_{i}$. In the main text, the parameters $\beta J=0.2$ and $g/J=2$ are fixed for our simulations. The system is then evolved under the Hamiltonian \eqref{EDH}, and we compute the dynamics of spin depolarization. The results are averaged over $10^3$ disorder realizations, providing a robust estimate of the system's behavior under various disorder configurations.

\subsection{Large-$M$ expansion}   \label{LargeM}
The Abrikosov fermion representation allows us to reformulate the original spin model as a fermionic system, opening up access to various field-theory approaches. In this representation, spin operators are expressed as $\hat{S}^\alpha_{ia}=\frac{1}{2}\sum_{ss'}\hat{c}^\dagger_{ia,s}(\sigma^\alpha)_{s s^\prime} \hat{c}_{ia,s^\prime}$ ($s,s'=\uparrow,\downarrow$), with a projection into the single occupation subspace. The spin polarization can be expressed as $\langle \hat{O}_x (t) \rangle=-i N (G^{\gtrless}_{\uparrow\downarrow}(t,t)+G^{\gtrless}_{\downarrow\uparrow}(t,t))/2$, in which the real-time Green's functions are defined as $G^>_{ss'}(t_1,t_2)\equiv-i  \langle c_{ia,s}(t_1)c_{ia,s'}^\dagger(t_2)\rangle$ and $G^<_{ss'}(t_1,t_2)\equiv i  \langle c_{ia,s'}^\dagger(t_2)c_{ia,s}(t_1)\rangle$. The evolution of these Green's functions can be described by the Schwinger-Dyson equation on the Keldysh contour \cite{kamenev2023field}, commonly known as the Kadanoff-Baym equation
\begin{equation}
  \begin{aligned}\label{eq:KBeq}
    i\partial_{t_1}&G^\gtrless =\Sigma^R\circ G^\gtrless+ \Sigma^\gtrless\circ G^A, \\
    -i\partial_{t_2}&G^\gtrless =G^R\circ\Sigma^\gtrless+ G^\gtrless\circ \Sigma^A,
  \end{aligned}
\end{equation}
where $G^{R/A}=\pm \Theta\left(\pm t_{12}\right)\left(G^>-G^<\right)$ is the retarded/advanced Green's function. $\Sigma^\gtrless$ and $\Sigma^{R/A}$ represent real-time self-energies, which satisfy a similar relation as Green's functions $\Sigma^{R/A}=\pm \Theta\left(\pm t_{12}\right)\left(\Sigma^>-\Sigma^<\right)$. To proceed, we need to express self-energies as a functional of Green's functions. In references \cite{PhysRevResearch2021, Tian-GangZhou2023}, this relation is established by introducing an SU(M)$\times$SU(2) generalization. Taking the large-$N$ and the large-$M$ limit assists in selecting a set of diagrams when computing the self-energy, known as SYK-like melon diagrams. Following the discussions in Ref. \cite{Tian-GangZhou2023}, the SU(M)$\times$SU(2) generalization of our random spin model reads
\begin{equation}
	\hat{H}= \frac{1}{\sqrt{M}} \sum_{i_1<i_2,ab,\alpha u} J_{i_1i_2}\xi_\alpha\hat{S}^{\alpha,u}_{i_1a}\hat{S}^{\alpha,u}_{i_2b}.
\end{equation}
We have introduced SU(M)$\times$SU(2) spins
\begin{equation}
\hat{S}^{\alpha,u}_{ia}=\frac{1}{2}\sum_{s_1,s_2,m_1,m_2}\hat{c}^{\dagger}_{ia,s_1,m_1}(\sigma^\alpha)_{s_1s_2}(T^u)_{m_1m_2}\hat{c}^{}_{ia,s_2,m_2}.
\end{equation}
Here, $m_k\in \{1,2,...M\}$ and $u\in \{1,2,...,M^2-1\}$ labels generators $T^u$ in the SU(M) group, which satisfy the completeness relation 
\begin{equation}
\sum_\gamma T^\gamma_{m_1m_2}T^\gamma_{m_3m_4}=\delta_{m_1m_4}\delta_{m_2m_3}-\frac{1}{M}\delta_{m_1m_2}\delta_{m_3m_4}\approx \delta_{m_1m_4}\delta_{m_2m_3}.
\end{equation}
Using this relation, the Hamiltonian can be rewritten as 
\begin{equation}
	\hat{H}= \frac{1}{4\sqrt{M}} \sum_{i_1<i_2,ab,\alpha m_1m_2} J_{i_1i_2}\xi_\alpha(\sigma^\alpha)_{s_1s_2}(\sigma^\alpha)_{s_3s_4} \hat{c}^{\dagger}_{i_1a,s_1,m_1}\hat{c}^{\dagger}_{i_2b,s_3,m_2}\hat{c}^{}_{i_1a,s_2,m_2}\hat{c}^{}_{i_2b,s_4,m_1}.
\end{equation}
For a fixed $M$, fermion operators $\hat{c}^{\dagger}_{i_1a,s_1,m_1}$ and $\hat{c}^{}_{i_1a,s_2,m_2}$ operate on the same mode with a probability of $\sim 1/M$, as $m_k$ can take $M$ different values. Therefore, in the large-$M$ limit, we can safely assume that the four fermion operators in each term are independent, as in the Hamiltonian of the complex SYK model \cite{Complex:SYK}. This leads to
\begin{equation}\label{eq:selfreal}
  \begin{aligned}
    \Sigma^{\gtrless}(t_1,t_2)=\frac{J^2}{4}& \sum_{\alpha,\alpha^\prime} \xi_\alpha \xi_{\alpha^\prime} \sigma^{\alpha^\prime} G^{\gtrless}(t_1,t_2)\sigma^\alpha \text{Tr}\left[\sigma^{\alpha^\prime} G^{\gtrless}(t_1,t_2)\sigma^\alpha  G^{\lessgtr}(t_2,t_1)\right].
  \end{aligned}
\end{equation}
In our numerical simulations at main text, we initialize the system in a thermal ensemble at $\beta J=0.2$ with a polarization field $g/J=2$, following a procedure similar to the one used in the exact diagonalization simulation. The initial Green's functions corresponding to this state are obtained through iterative methods, as detailed in \cite{PhysRevResearch2021,Tian-GangZhou2023}. Subsequently, we evolve $G^\gtrless$ using a combination of \eqref{eq:KBeq} and \eqref{eq:selfreal}, enabling us to determine the time evolution of the spin polarization $\langle \hat{O}_x (t) \rangle$.

\subsubsection{Quasi-normal mode }
The quasi-normal mode physics plays a central role in the NMR experiment, which leads to the universal form of the two-point correlation function. With the large-$M$ framework, we are able to predict the analytical form of quasi-normal mode in terms of parameters in Hamiltonian.

The universal long-time region requires that the off-diagonal term $s \neq s'$ in any real-time Green's function $G_{ss'}^{R/A/K}$ or $G_{ss'}^{\gtrless}$ should be smaller than the diagonal term $s=s'$. Therefore, by treating the off-diagonal term as a small perturbation, we can perform a linearized analysis of the Schwinger-Dyson equation to reveal the mechanism for the oscillation and decay. Without loss of generality, we assume the initial polarization is at $x$, and the total magnetization can be expressed as its off-diagonal component in the Keldysh Green's function:
\begin{equation}\label{eq:mtGK}
m(t)=-iG^K_{\uparrow\downarrow}(t,t)/2,
\end{equation}
where $G^K_{ss'}(t_1,t_2) = G^>_{ss'}(t_1,t_2)+G^<_{ss'}(t_1,t_2)$ and $G^{\gtrless}(t_1,t_2)$ is defined in \ref{LargeM}. The full Schwinger-Dyson equation in real time after Kelydsh read
\begin{equation}\label{eq:SD_Keldysh_full}
    \begin{pmatrix}
        (G_0^R)^{-1} -\Sigma^R & (G_0^{-1})^K -\Sigma^K\\
        0 & (G_0^{-1})^A -\Sigma^A\\
    \end{pmatrix}
    \circ
    \begin{pmatrix}
        G^R & G^K\\
        0 & G^A\\
    \end{pmatrix}
    = \mathbb{I},
\end{equation}
which immediately leads to the equation that involves our linearized calculation
\begin{equation}\label{eq:SD_GK_Keldysh}
  G^K=G^R \circ \Sigma^K \circ G^A,\ \ \ \ \ \text{with}\ \Sigma^K=\Sigma^>+\Sigma^<.
\end{equation}

 We expand all types of the Green's function around its equilibrium value by taking $G^{a}_{s  s'}(t_1,t_2)=G^{a,\beta_f}_{s  s'}(t_{12})+\delta G^{a}_{s  s'}(t_1,t_2)$, where $G^{a,\beta_f}(t)$ is the equilibrium Green's function on the final state. Lengthy calculation shows the off-diagonal element of Eq~.\eqref{eq:SD_GK_Keldysh} reads 
\begin{equation}\label{eq:SD_GK_Keldyshlinear}
\begin{aligned}
  \delta G^K_{\uparrow \downarrow}&=G^{R,\beta_f}_{\uparrow \uparrow} \circ \delta \Sigma^K_{\uparrow \downarrow}\circ G^{A,\beta_f}_{\uparrow \uparrow},\\ \delta \Sigma^K_{\uparrow \downarrow}&= \frac{1}{4} J^2 W_x \left((G^{>,\beta_f}_{\uparrow\uparrow})^2  + (G^{<,\beta_f}_{\uparrow\uparrow})^2 \right) \delta G^{K}_{\uparrow\downarrow} + O\left( (\delta G^{K}_{\uparrow\downarrow})^3 \right).
  \end{aligned}
\end{equation} 
where the polynomial $W_x=-\xi_1^2+\xi_2^2-4\xi_2\xi_3+\xi_3^2$ is the same as we proposed in the main text. To obtain the linearized equation Eq.~\eqref{eq:SD_GK_Keldyshlinear} from Eq~.\eqref{eq:SD_GK_Keldysh}, we use the following argument
\begin{itemize}
    \item Since $G^{K,\beta_f}_{\uparrow \downarrow}=0$, Eq. \eqref{eq:mtGK} is equivalent to $m(t)=-i\delta G^K_{\uparrow \downarrow}(t,t)/2$.
    \item To simplify the self-energy and obtain a certain combination of anisotropic parameters, we apply two symmetries to Green's function of fermion. The first can be regarded as $\pi$ rotation of axis $x$.
    \begin{equation}\label{eq:Sym1}
	\begin{split}
		&\hat{c}_{s_1} \to \sum_{s'} \left(\iu \sigma^x \right)_{s_1 s'}\hat{c}_{s'},\ \ \hat{c}_{s_1}^{\dagger} \to \sum_{s'} \hat{c}_{s'}^{\dagger} \left(-\iu \sigma^x \right)_{s' s_1} \\
        &G^>_{s_1 s_2}(t_1,t_2) \rightarrow \sum_{s' s''} \sigma^x_{s_1 s'} G^>_{s' s''}(t_1,t_2) \sigma^x_{s'' s_2}=\begin{pmatrix}
			G^>_{\downarrow \downarrow}(t_1,t_2) & G^>_{\downarrow \uparrow}(t_1,t_2) \\
			G^>_{\uparrow \downarrow}(t_1,t_2) & G^>_{\uparrow \uparrow}(t_1,t_2) \\
		\end{pmatrix}_{s_1 s_2} \\
	\end{split}
    \end{equation}
    which actually maps $\left\{\hat{S}^{x},\hat{S}^{y},\hat{S}^{z}\right\}$ to $\left\{\hat{S}^{x},-\hat{S}^{y},-\hat{S}^{z}\right\}$, and therefore keeps the Hamiltonian invariant.

    The second is a combination of particle-hole symmetry and rotation, which reads as
    \begin{equation}\label{eq:Sym2}
	\begin{split}
	&	\hat{c}_{s_1} \to \sum_{s'} \left(\iu \sigma^y \right)_{s_1 s'}\hat{c}_{s'}^{\dagger},\ \  \hat{c}_{s_1}^{\dagger} \to \sum_{s'} \hat{c}_{s'} \left(-\iu \sigma^y \right)_{s' s_1}\\
    &G^>_{s_1 s_2}(t_1,t_2) \rightarrow -\sum_{s' s''} \sigma^y_{s_1 s'} G^<_{s'' s'}(t_2,t_1) \sigma^y_{s'' s_2}
    =\begin{pmatrix}
			-G^<_{\uparrow \uparrow}(t_2,t_1) & G^<_{\uparrow \downarrow}(t_2,t_1) \\
			G^<_{\uparrow \downarrow}(t_2,t_1) & -G^<_{\uparrow \uparrow}(t_2,t_1) \\
		\end{pmatrix}_{s_1 s_2}
	\end{split}
    \end{equation}
    which maps $\left\{\hat{S}^{x},\hat{S}^{y},\hat{S}^{z}\right\}$ to $\left\{\hat{S}^{x},\hat{S}^{y},-\hat{S}^{z}\right\}$ and also keeps the Hamiltonian invariant. Finally Eq.~\eqref{eq:selfreal} can be simplified as
    \begin{equation}\label{eqS:full_self_energy}
	\Sigma^{\gtrless}(t_1,t_2)
	= \frac{1}{2} J^2 \begin{pmatrix}
		-\Gamma G^{\gtrless}_{\uparrow\uparrow}(t_1,t_2)^3 + W_x G^{\gtrless}_{\uparrow\uparrow}(t_1,t_2) G^{\gtrless}_{\uparrow\downarrow}(t_1,t_2)^2& 
		W_x G^{\gtrless}_{\uparrow\uparrow}(t_1,t_2)^2 G^{\gtrless}_{\uparrow\downarrow}(t_1,t_2) + \Gamma G^{\gtrless}_{\uparrow\downarrow}(t_1,t_2)^3 \\
		W_x G^{\gtrless}_{\uparrow\uparrow}(t_1,t_2)^2 G^{\gtrless}_{\uparrow\downarrow}(t_1,t_2) + \Gamma G^{\gtrless}_{\uparrow\downarrow}(t_1,t_2)^3& 
		-\Gamma G^{\gtrless}_{\uparrow\uparrow}(t_1,t_2)^3 + W_x G^{\gtrless}_{\uparrow\uparrow}(t_1,t_2) G^{\gtrless}_{\uparrow\downarrow}(t_1,t_2)^2 \\
	\end{pmatrix},
\end{equation}
where the polynomials are $\Gamma=\xi_1^2+\xi_2^2+\xi_3^2$ and $W_x=-\xi_1^2+\xi_2^2-4\xi_2\xi_3+\xi_3^2$.
\item With perturbation on the off-diagonal term on the Eq.~\eqref{eq:SD_GK_Keldysh}, in principle we have
\begin{equation}\label{eq:linearization}
     \delta G^K_{\uparrow \downarrow}=\sum_{s_1, s_2} \underbrace{ G^{R,\beta_f}_{\uparrow s_1} \circ \delta \Sigma^K_{s_1 s_2}\circ G^{A,\beta_f}_{s_2 \downarrow} }_{\text{Eq.~\eqref{eq:linearization}(a)}} 
  + \underbrace{\delta G^{R}_{\uparrow s_1} \circ  \Sigma^{K,\beta_f}_{s_1 s_2}\circ G^{A,\beta_f}_{s_2 \downarrow} }_{\text{Eq.~\eqref{eq:linearization}(b)}}
  + \underbrace{G^{R,\beta_f}_{\uparrow s_1} \circ  \Sigma^{K,\beta_f}_{s_1 s_2}\circ \delta G^{A}_{s_2 \downarrow} }_{\text{Eq.~\eqref{eq:linearization}(c)}}.
\end{equation}
For Eq.~\eqref{eq:linearization}(a), all off-diagonal components of $G^{R/A,\beta_f}$ are zero, and consequently we only take $s_1 = \uparrow$ and $s_2=\downarrow$. For Eq.~\eqref{eq:linearization}(b), the vanishing off-diagonal component of $G^{A,\beta_f}, \Sigma^{K,\beta_f}$ requires $s_2=\downarrow$ and then $s_1 = \downarrow$. However, in the infinite high-temperature region we have $\Sigma^{K,\beta_f}_{\downarrow \downarrow}=0$ as a result of the fluctuation-dissipation theorem\cite{kamenev2023field}. In the following content, we can also check this with the specific ansatz for equilibrium Green's function. Therefore Eq.~\eqref{eq:linearization}(b) vanishes, and we can show Eq.~\eqref{eq:linearization}(c) vanishes with similar arguments. Combining all the argument, only the off-diagonal term in Eq.~\eqref{eq:linearization}(a) survives and finally leads to Eq.~\eqref{eq:SD_GK_Keldyshlinear}.
% \item Due to the fluctuation-dissipation theorem $G^K_{s s'}(\omega) = \tanh{\frac{\beta \omega}{2}} (G^R(\omega)-G^A(\omega))_{s s'}$, in the high temperature phase we can estimate that $\delta G^K_{\uparrow \downarrow}(t) \Theta(t) = \frac{\beta}{2}\iu \partial_t \delta G^R_{\uparrow \downarrow}(t)$. The term $\delta G^{R,\beta_f}_{\uparrow \downarrow} \circ  \Sigma^K_{\downarrow \downarrow}\circ G^{A,\beta_f}_{\downarrow \downarrow}$ still contributes to the quasinormal mode equation. Therefore the full equation will be
% \begin{equation}\label{eq:SD_GK_Keldyshlinear}
% \begin{aligned}
%   \delta G^K_{\uparrow \downarrow}&=G^{R,\beta_f}_{\uparrow \uparrow} \circ \delta \Sigma^K_{\uparrow \downarrow}\circ G^{A,\beta_f}_{\uparrow \uparrow} 
%   + \delta G^{R}_{\uparrow \downarrow} \circ  \Sigma^{K,\beta_f}_{\downarrow \downarrow}\circ G^{A,\beta_f}_{\downarrow \downarrow} 
%   + G^{R,\beta_f}_{\uparrow \uparrow} \circ  \Sigma^{K,\beta_f}_{\uparrow \uparrow}\circ \delta G^{A}_{\uparrow \downarrow} \\
%   &=G^{R,\beta_f}_{\uparrow \uparrow} \circ \delta \Sigma^K_{\uparrow \downarrow}\circ G^{A,\beta_f}_{\uparrow \uparrow} 
%   -\frac{2\iu}{\beta} \left(\int \delta G^{K}_{\uparrow \downarrow}\Theta(t)\right) \circ  \Sigma^{K,\beta_f}_{\downarrow \downarrow}\circ G^{A,\beta_f}_{\downarrow \downarrow} 
%   -\frac{2\iu}{\beta} G^{R,\beta_f}_{\uparrow \uparrow} \circ  \Sigma^{K,\beta_f}_{\uparrow \uparrow}\circ \left(\int \delta G^{K}_{\uparrow \downarrow}\Theta(-t)\right) \\
%   \end{aligned}
% \end{equation} 
\end{itemize}

Starting from the linearized solution Eq.\eqref{eq:SD_GK_Keldyshlinear}, we can further make assumption about the equilibrium solution. In thermal equilibrium with $h=0$, the self-energies \eqref{eq:selfreal} can be simplified as $\Sigma^{\gtrless,\beta_f}_{ss'}(t)=-\frac{J^2\Gamma}{2}G^{\gtrless,\beta_f}_{ss}(t)^3\delta_{ss'}$, which matches the self-energy of the Majorana SYK$_4$ model but with effective coupling constant $J\sqrt{\Gamma}/\sqrt{2}$. It is known that at high temperatures $\beta J\ll1$, the SYK model can be described by weakly interacting quasi-particles \cite{zhangObstacleSubAdSHolography2021}.
\begin{equation}\label{eq:highT_GF}
	\begin{aligned}
		G_{\uparrow\uparrow}^{R/A,\beta_f}(t)&\approx\mp i\Theta(\pm t)e^{-\Gamma |t|/2},\ \ \ \ \ \ 
		G_{\uparrow\uparrow}^{\gtrless,\beta_f}(t)&\approx\mp ie^{-\Gamma |t|/2}/2.
	\end{aligned}
\end{equation}
with quasi-particle decay rate $\Gamma \propto J\sqrt{\Gamma}$. Besides, we can also verify $G^{K, \beta_f} = G^{>, \beta_f} + G^{<, \beta_f}\approx 0$ and $\Sigma^{K, \beta_f} = \Sigma^{>, \beta_f} + \Sigma^{<, \beta_f}\approx 0$, which is consistent with our assumption above.

With the equilibrium Green's function, the linearized equation ~\eqref{eq:SD_GK_Keldyshlinear} becomes
\begin{equation}\label{eq:conv}
	\begin{split}
		\delta G^K_{\uparrow \downarrow}(t_1, t_4) &= \int \diff t_2 \diff t_3 \  G^R_{\uparrow \uparrow}(t_1, t_2)  \left( -\frac{1}{8}J^2 W_x e^{-|t_2-t_3|\Gamma } \delta G^K_{\uparrow \downarrow}(t_2, t_3) \right)  G^A_{\downarrow \downarrow}(t_3, t_4)  \\
		&= -\frac{1}{8}J^2 W_x \int \diff t_2 \diff t_3 
		e^{-\frac{\Gamma}{2}(t_1-t_2)} \Theta(t_1-t_2)
		e^{-|t_2-t_3|\Gamma } \delta G^K_{\uparrow \downarrow}(t_2, t_3)
		e^{\frac{\Gamma}{2}(t_3-t_4)} \Theta(-t_3+t_4).
		\\
	\end{split}
\end{equation}
To simplify Eq.~\eqref{eq:conv}, we multiply $e^{\frac{\Gamma}{2} t_1}$ and take $\partial_{t_1}$ on both side, and then similarly multiply $e^{\frac{\Gamma}{2} t_4}$ and take $\partial_{t_4}$ on both side. The result reads
\begin{equation}\label{eq:SD_GK_diff}
\left(\partial_{t_1}+\frac{\Gamma}{2}\right)\left(\partial_{t_4}+\frac{\Gamma}{2}\right)\delta G^K_{\uparrow \downarrow}=-\frac{W_x}{8}J^2
   e^{-\Gamma|t_{14}|} \delta G^K_{\uparrow \downarrow}.
\end{equation}
The ordinary differential equation of the non-equilibrium Green's function without time translation invariant can be parameterized using the ansatz $\delta G^K_{\uparrow \downarrow}(t_1,t_4)=\text{Re}~e^{-\lambda\frac{t_1+t_4}{2}}\varphi_{\lambda}(t_{14})$, where $\lambda$ is the collective dynamics of quasi-normal mode. It finally arrives at the quasi normal mode equations 
\begin{equation}\label{eq:normal_modes_final}
	\begin{split}
		-\frac{(\Gamma-\lambda)^2}{4}\varphi_{\lambda}(t_{14}) =-\partial_{t_{14}}^2 \varphi_{\lambda}(t_{14}) + \frac{W_x}{8} J^2e^{-\Gamma |t_{14}|}\varphi_{\lambda}(t_{14}).
	\end{split}
\end{equation}
Eq.~\eqref{eq:normal_modes_final} suggests that the quasi-normal mode problem can be described in a Schr\"{o}dinger-like equation, where $-\frac{(\Gamma-\lambda)^2}{4}$ represents the energy $E$ for given state and $\frac{W_x}{8} J^2e^{-\Gamma |t_{12}|}$ is regarded as potential $V$ in the Hamiltonian. With the wisdom of 1D quantum mechanism, we argue that the oscillating regime and the non-oscillating regime actually depends on the sign of potential energy: For $W_x<0$, the potential energy is negative. Since in 1D, any attractive potential exhibits at least one bound state, we can denote the energy of the ground state as $-|E_0|$. Then we can solve the quasi-normal mode $\lambda=\Gamma-2\sqrt{|E_0|}$, which is a real number. Consequently, we expect the magnetization relaxes monotonically. For $W_x>0$, the potential is repulsive. The eigenstates of the \eqref{eq:normal_modes_final} are scattering modes with continuous positive energy $E$. We find $\lambda=\Gamma\pm2i\sqrt{E}$, which is a complex number. We argue this leads to oscillations in the relaxation process. 

We can further determine the typical oscillation frequency $\Omega$ by determining the typical energy $E$ that contributes to the quench dynamics. In principle all the eigenstates $\varphi_{\lambda}(t_{14})$ contributes to the magnetization dynamics, i.e. $\delta G^K_{\uparrow \downarrow}(t_1,t_4)=\sum_{\lambda} c(\lambda)\text{Re}~e^{-\lambda\frac{t_1+t_4}{2}}\varphi_{\lambda}(t_{14})$, where $c(\lambda)$ is the corresponding superposition coefficient. According to Eq. \eqref{eq:mtGK}, the magnetization probes the decay of the wave function $\varphi(t_{14})$ at $t_{14}=0$ and $t_1=t_4=t$, where the typical potential energy is $\sim AJ^2$. For $E\ll W_x J^2$, the eigenstate has exponentially small weight near $t_{14}=0$. As a result, the corresponding contribution to $m(t)$ can be neglected. We can approximate
\begin{equation}
m(t)\sim \Re \int_{E\sim W_x J^2} dE~c(E)e^{-\Gamma t-2i\sqrt{E}t}.
\end{equation}
Here $c(E)$ is some smooth function determined by the initial condition, which is dominant at $E$ around $W_x J^2$. We then expect $\Omega \approx c_0 \sqrt{W_x}J$, with some $O(1)$ constant $c_0$ which does not depend on parameters in the Hamiltonian Eq.~\eqref{HeiModel} and should be extracted using numerics. We have verified this point in \ref{supp:more_param}. Interestingly, the result predicts the oscillation period $2\pi/\Omega$ diverges as we approach $W_x=0$, which can be viewed as an analog of the divergence of the correlation length in traditional phase transition described by order parameters.

% We comment that our results unveil the universality of relaxation dynamics in random spin models. Although the microscopic model in \eqref{eq:random_spin} contains several parameters, the criterion for the different relaxation behaviors, as well as the oscillation frequency, only depends on a specific combination $A$. This is a direct analog of universality in the scattering theory, where for a complicated potential, the low-energy scattering problem can only depend on a specific combination of microscopic parameters, which is the scattering length. 

\subsection{Mean-field theory}  \label{MF}
Another powerful theoretical scheme for analyzing random spin models is mean-field theory. In this approach, we introduce the collective spin polarization on each molecule as $\hat{M}_{i}^\alpha=\frac{1}{N_a}\sum_{a}\hat{S}_{ia}^\alpha$. Due to the statistical averaging over spins in the same molecule, we expect the fluctuation of $\hat{M}_{i}^\alpha$ to be small. We thus approximate the operator $\hat{M}_{i}^\alpha$ as a classical vector $M_{i}^\alpha$, the evolution of which is governed by the Heisenberg equation:
\begin{equation}\label{eq:mf}
    \frac{dM_{i}^\alpha}{dt}=N_a \sum_{j,\beta\gamma}J_{ij}\epsilon^{\alpha\beta\gamma} \xi_\beta M_j^\beta M_i^\gamma.
\end{equation}
Our numerical simulation is based on classical equations \eqref{eq:mf}. We investigate a system with $N_m=2\times 10^3$ molecules. The initial configuration of $M_{i}^\alpha$ is sampled using an independent Gaussian distribution, with mean value $\overline{\bm{M}_{i}}=(\frac{\beta g}{4},0,0)$, $\beta g=0.4$, and variance $\overline{(\delta M_{i}^\alpha)^2}=1/(4 N_a)$. $M_{i}^\alpha$ is then evolved according to \eqref{eq:mf} for each disorder realization of $J_{ij}$ and we compute $\langle \hat{O}^\alpha \rangle= N_a \sum_{i}M_{i}^\alpha$. The final result is averaged over 20 independent simulations.

\subsection{Parameters in Numerical Simulations}\label{supp:more_param}
\begin{figure}[htbp] % 插图的时候使用[H]固定悬浮位置，不然图会乱飘。
    \begin{center}  
    \includegraphics[scale=0.5]{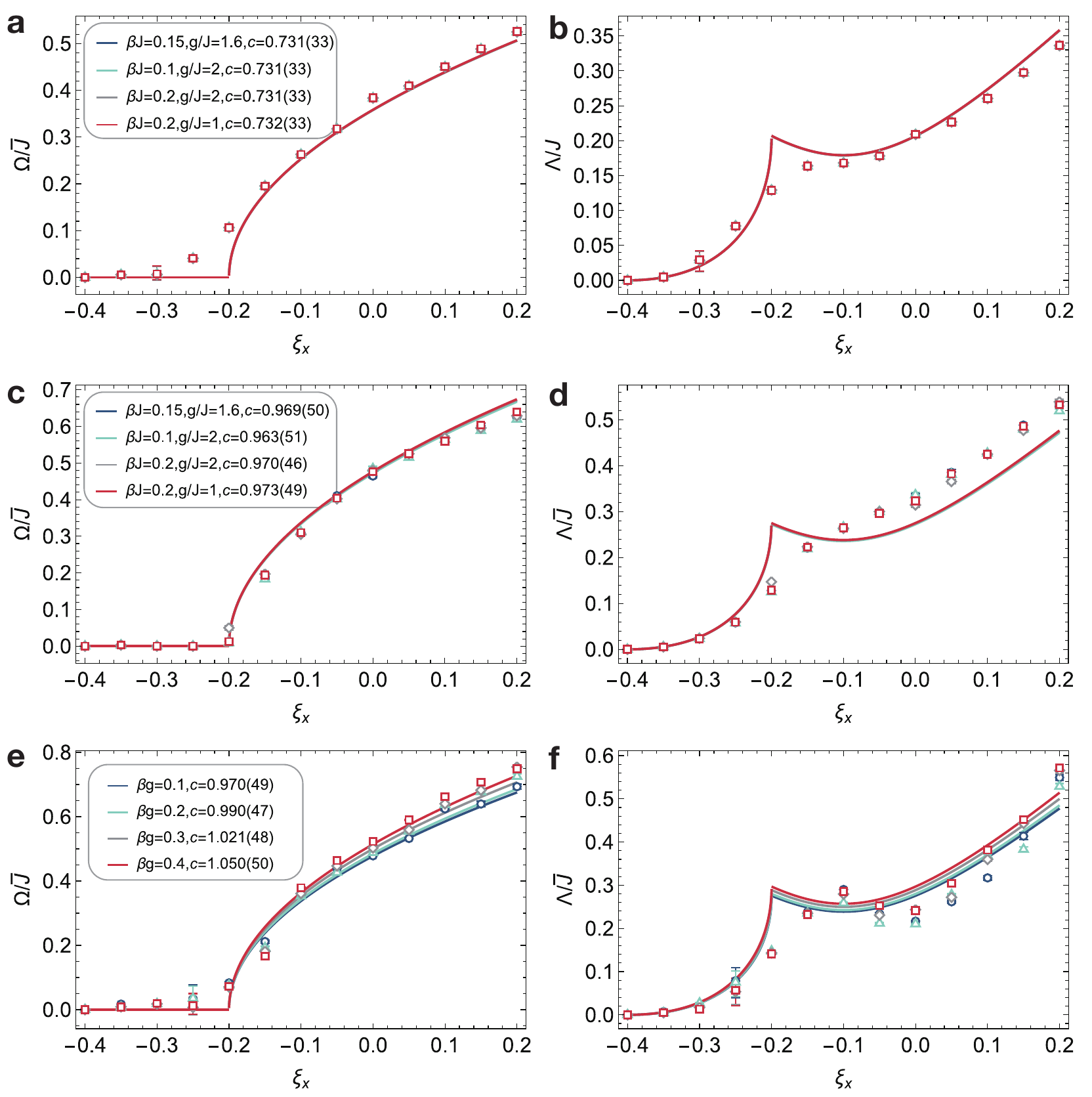}
    \caption{\textbf{The benchmark of three numerical methods in the valid parameter region}. The oscillation frequency and decay rate in different parameters are shown for (\textbf{a,b}) large-$M$ expansion, (\textbf{c,d}) exact diagonalization, with $N=8$ and 200 random realizations, and (\textbf{e,f}) mean-field theory, with $N=1500$ molecules and 20 random realizations. The legends label different parameters, and the overall constant $c$ is introduced in the theoretical prediction for the frequency and decay rate. } 
    \label{fig:supp_theory_moreparam1}
    \end{center}
\end{figure}

The NMR experiment is taken at room temperature. Therefore we require the high-temperature condition $\beta J |\bm{\xi}| \ll 1$ and $\beta g  \ll 1$. Also, the external magnetic field strongly polarized the state so that the initial state can be approximated by $\hat{\rho} \propto \mathrm{e}^{- \beta\hat{H} + \beta g\sum_{ia} \hat{S}_{ia}^\alpha} \approx \hat{\mathbbm{1}} + \beta g \sum_{ia} \hat{S}_{ia}^\alpha$. To satisfy that, the magnitude of the external field must be significantly larger than the characteristic energy of the NMR spin system, i.e. $g/(J|\bm{\xi}|) \gg 1$. For numerics, all the chosen parameter regions satisfy these conditions, except for mean-field numerics where only the composite parameter $\beta g\ll 1$ needs to be considered, as it is the only adjustable parameter. We have verified that different parameter choices yield qualitatively similar results. From \ref{fig:supp_theory_moreparam1}, it can be observed that the fitting frequency and decay rates are independent of the specific values of parameters $\beta$ and $g$, but solely depend on $J$, provided they satisfy the relation
\begin{itemize}
    \item $\beta J |\bm{\xi}| \ll 1$, $\beta g  \ll 1$ and $g/(J|\bm{\xi}|) \gg 1$:\ \ for large-$M$ and exact-diagonalization method. 
    \item $\beta g\ll 1$:\ \ for mean-field method.
\end{itemize}
The quantitative analysis reveals that the fitting parameter $c$ remains consistent within the error bars for each numerical method. This benchmark is also a concrete verification of the universality of the quasi-normal mode in the NMR spin system and the mode introduced in Eqs.~(5), (6) in the main text. 

\subsection{Analysis for the four diagrams}
\label{diagrams}

Without loss of generality, we assume that the initial system is prepared at high temperatures and polarized in the $\hat{x}$-direction. Consequently, the evolution of the magnetization can be simplified to the two-point correlator.
	\begin{equation}\label{eq:mag_evolve}
		\sum_{ia} \langle \hat{S}_{ia}^x(t) \rangle \approx \beta g \sum_{ij,ab} \frac{1}{2^N} \Tr\left(\hat{S}_{ia}^x(t) \hat{S}_{jb}^x(0)\right),
	\end{equation}
	where $N=N_\text{m} N_\text{a}$ is the total number of spins, in which $N_\text{m}$ is the number of molecules and $N_\text{a} = 16$ represents the number of ${}^1$H in each molecule. $ 2^{N}$ is the normalization constant in total Hilbert space. Here the summation is over proton index $a,b$, and molecule index $i,j$. In the experiment, we use normalized magnetization
	\begin{equation}\label{eq:normalized_mag}
	\mathcal{C}^{\alpha}(t) \equiv	\left( \sum_{ia} \langle \hat{S}_{ia}^x(t) \rangle \right) / \left( \sum_{ia} \langle \hat{S}_{ia}^x(0) \rangle \right) = \frac{4}{ N} \sum_{ij,ab} \Tr\left(\hat{S}_{ia}^x(t) \hat{S}_{jb}^x(0)\right).
	\end{equation}
	
	The magnetization decays exponentially and possibly with oscillation. To obtain the information of frequency $\omega$ and decay rate $\lambda$, we take the second derivative of the normalized magnetization
	\begin{equation}\label{eq:normalized_mag_2_deriv}
		\begin{split}
			&\frac{4}{N} \sum_{ij,ab} \frac{\diff^2}{\diff t^2} \Tr\left(\hat{S}_{ia}^x(t) \hat{S}_{jb}^x(0)\right) \Big|_{t=0} \\
			=&\frac{8}{N} \Bigg\{\sum_{i,a} \left[ \Tr\left(\hat{H} \hat{S}_{ia}^x \hat{H} \hat{S}_{ia}^x \right) - \Tr\left(\hat{H}^2   \hat{S}_{ia}^x \hat{S}_{ia}^x \right) \right]  + \sum_{(i,a)\neq (j,b)} \left[ \Tr\left(\hat{H} \hat{S}_{ia}^x \hat{H} \hat{S}_{jb}^x \right) - \Tr\left(\hat{H}^2   \hat{S}_{ia}^x \hat{S}_{jb}^x \right) \right]  \Bigg\}\\
            =& \mathcal{G}_c - \mathcal{G}_d + \mathcal{G}_e - \mathcal{G}_f 
		\end{split}
	\end{equation}

     The four terms without minus signs correspond to four diagrams, as shown in Fig.~3 in the main text.  We summarize the results of each diagram $\mathcal{G}_c,\mathcal{G}_d,\mathcal{G}_e,\mathcal{G}_f$ in the limit of large $N_m$
	\begin{equation}\label{eq:diagram}
		\begin{split}
			\mathcal{G}_c &= \frac{J^2}{2} \left(\xi_x^2 + \xi_y^2 + \xi_z^2\right),\\
			\mathcal{G}_d &=\frac{J^2}{2} \left(\xi_x^2 - \xi_y^2 - \xi_z^2\right), \\
			\mathcal{G}_e &= -J^2 \xi_y \xi_z, \\
			\mathcal{G}_f &= J^2 \xi_y \xi_z. \\
		\end{split}
	\end{equation}
We find that diagram $\mathcal{G}_c$ is proportional to the polynomial $\Gamma \equiv \xi_x^2 + \xi_y^2 + \xi_z^2$, and a linear combination of diagrams (d), (e), (f), i.g. $-\mathcal{G}_d+\mathcal{G}_e-\mathcal{G}_f$ is proportional to polynomial $W_x  \equiv -\xi_x^2 + \xi_y^2 - 4 \xi_y \xi_z + \xi_z^2$. 

This argument in diagrams is more general to the large-$M$ analysis. In low-energy physics, universality arises when a specific set of diagrams becomes the most relevant one and dominates the physical process, such as the (Cooper channel) ladder diagram in superconductivity. Here, by computing the second derivative of the correlator, we aim to identify important diagrams for the spin relaxation process. Under the Schwinger Boson/Fermion representation, the two-point function of spin operators becomes a four-point function of partons. This four-point function satisfies a recursion relation known as the Bethe–Salpeter equation \cite{altland2010condensed}, which sums up a series of two-particle irreducible (2PI) diagrams. This criterion nicely divides all diagrams $\mathcal{G}_c, \mathcal{G}_e, \mathcal{G}_d, \mathcal{G}_f$ into two groups: Diagram c represents a correction of the propagator, which is not 2PI, while diagrams d-f are all 2PI diagrams. This motivates us to separate them out and investigate the contribution of each group of diagrams, which turns out to match the experimental results and numerical simulations. This indicates that spin relaxation is dominated by a ladder of diagrams d-f, in which each propagator is renormalized by diagram c, as in the large-$M$ analysis.
	
    %Based on the experiment result together with the large-$N$ and large-$M$ justification\cite{Tian-GangZhou:2023hou}, we conjecture that the oscillation frequency $\omega_x$ and decay rate $\lambda$ obey the universal function in terms of the anisotropic function\begin{equation}\label{eq:scaling_function}
		%(\omega_x/\bar{J}, \lambda/\bar{J}) =
		%\begin{cases}
			%c\left( \sqrt{W_x}, \sqrt{\Gamma}\right)  & W_x \geq 0 \\
			%c\left(0, \sqrt{\Gamma} - \sqrt{-W_x}\right) & W_x<0 \\
		%\end{cases},
	%\end{equation}
    %where $c$ is the overall constant. 
    %There is also an intuitive but not strict argument about the scaling of oscillation frequency $\omega_x$ and decay rate $\lambda$. For simplicity, we focus on the $W_x \geq 0$ case and take an ansatz of the relaxing dynamics $\mahtcal{C}^x(t) = A \cos(\omega_x t + \phi)e^{-\lambda t}$. The second derivative of $\mahtcal{C}^x(t)$ at $t=0$ leads to the value $2 \lambda  \omega_x  \sin (A)+\cos (A) (\lambda^2 -\omega_x^2 )$. Therefore, it is reasonable that the scaling of $\lambda^2$ and $\omega_x^2$ can be proportional to polynomial $\Gamma$ and $W_x$.

    Finally, we provide the detailed derivation of each diagram. Notice that both the six-spin correlators $\frac{8}{N} \sum_{i,a} \left[ \Tr\left(\hat{H} \hat{S}_{ia}^x \hat{H} \hat{S}_{ia}^x \right) \right]$ and $\frac{8}{N} \sum_{i,a} \left[  \Tr\left(\hat{H}^2   \hat{S}_{ia}^x \hat{S}_{ia}^x \right) \right]$ generally have a disconnected diagram which is composited by three pairs of two-spin correlators disconnected term. However, the disconnected term is not included in the definition of diagram $\mathcal{G}_c$ and $\mathcal{G}_d$ for two reasons. Firstly, the disconnected part contribution will be exactly canceled in the original formula $\frac{4}{N} \sum_{ij,ab} \frac{\diff^2}{\diff t^2} \Tr\left(\hat{S}_{ia}^x(t) \hat{S}_{jb}^x(0)\right) \Big|_{t=0}$. Secondly, the disconnected term is the order of $J^2 O(N)$, which is not in the order of magnitude of oscillation frequency and decay rate, and therefore should be excluded from the diagrammatic contribution.
	\paragraph*{Diagram $\mathcal{G}_c$}
	\begin{equation}\label{eq:termA}
		\begin{split}
		&\frac{8}{N} \sum_{i,a} \Tr\left(\hat{H}^2   \hat{S}_{ia}^x \hat{S}_{ia}^x \right)  \\
			=& \frac{8}{N} \sum_{i,a,\alpha \beta} \xi_{\alpha} \xi_{\beta} \Tr \left( \sum_{k<l,cd}J_{kl} \hat{S}_{kc}^{\alpha} \hat{S}_{ld}^{\alpha} \sum_{k'<l',c'd'}J_{k'l'} \hat{S}_{k'c'}^{\beta} \hat{S}_{l'd'}^{\beta} \hat{S}_{ia}^x \hat{S}_{ia}^x \right) \\
			=& \frac{8}{N} \frac{4J^2}{N} \sum_{i,a,\alpha \beta} \xi_{\alpha} \xi_{\beta} \Tr \left( \sum_{k<l,cd c'd'} \hat{S}_{kc}^{\alpha} \hat{S}_{ld}^{\alpha}  \hat{S}_{kc'}^{\beta} \hat{S}_{ld'}^{\beta} \hat{S}_{ia}^x \hat{S}_{ia}^x \right) \\
			=& \frac{32 J^2}{N^2} \sum_{i,a,\alpha \beta} \xi_{\alpha} \xi_{\beta} \sum_{k\neq l,cd} \Bigg[\Tr \left( \hat{S}_{ld}^{\alpha}  \hat{S}_{ld}^{\beta} \hat{S}_{ia}^x \hat{S}_{ia}^x \right) \Tr \left( \hat{S}_{kc}^{\alpha}  \hat{S}_{kc}^{\beta} \right)+ \Tr \left( \hat{S}_{ld}^{\alpha}  \hat{S}_{ld}^{\beta}  \right) \Tr\left( \hat{S}_{ia}^x  \hat{S}_{ia}^x \right) \Tr \left( \hat{S}_{kc}^{\alpha}  \hat{S}_{kc}^{\beta} \right) \Bigg] \\
			=& \frac{32 J^2}{N^2} \sum_{i,a,\alpha \beta} \xi_{\alpha} \xi_{\beta} \sum_{k\neq l,cd} \delta_{l i} \delta_{d a} \Bigg[\Tr \left( \hat{S}_{ld}^{\alpha} \hat{S}_{ld}^{\beta} \hat{S}_{ia}^x \hat{S}_{ia}^x \right) \Tr \left( \hat{S}_{kc}^{\alpha}  \hat{S}_{kc}^{\beta} \right) \Bigg] + \text{Disconnected} \\
			\approx& \frac{J^2 }{2} \left(\xi_x^2 + \xi_y^2 + \xi_z^2 \right)+ \text{Disconnected}. \\
            \equiv& \mathcal{G}_c + \text{Disconnected}. \\
		\end{split}
	\end{equation}
	
	\paragraph*{Diagram $\mathcal{G}_d$}
	\begin{equation}\label{eq:termb}
		\begin{split}
		&\frac{8}{N} \sum_{i,a} \Tr\left(\hat{H} \hat{S}_{ia}^x \hat{H} \hat{S}_{ia}^x \right)  \\
			=& \frac{8}{N} \sum_{i,a,\alpha \beta} \xi_{\alpha} \xi_{\beta} \Tr \left( \sum_{k<l,cd}J_{kl} \hat{S}_{kc}^{\alpha} \hat{S}_{ld}^{\alpha} \hat{S}_{ia}^x \sum_{k'<l',c'd'}J_{k'l'} \hat{S}_{k'c'}^{\beta} \hat{S}_{l'd'}^{\beta} \hat{S}_{ia}^x \right) \\
			=& \frac{8}{N} \frac{4J^2}{N} \sum_{i,a,\alpha \beta} \xi_{\alpha} \xi_{\beta} \Tr \left( \sum_{k<l,cd c'd'} \hat{S}_{kc}^{\alpha} \hat{S}_{ld}^{\alpha} \hat{S}_{ia}^x  \hat{S}_{kc'}^{\beta} \hat{S}_{ld'}^{\beta} \hat{S}_{ia}^x \right) \\
			=& \frac{32 J^2}{N^2} \sum_{i,a,\alpha \beta} \xi_{\alpha} \xi_{\beta} \sum_{k\neq l,cd} \Bigg[\Tr \left( \hat{S}_{ld}^{\alpha} \hat{S}_{ia}^x  \hat{S}_{ld}^{\beta} \hat{S}_{ia}^x \right) \Tr \left( \hat{S}_{kc}^{\alpha}  \hat{S}_{kc}^{\beta} \right) + \Tr \left( \hat{S}_{ld}^{\alpha}  \hat{S}_{ld}^{\beta}  \right) \Tr\left( \hat{S}_{ia}^x  \hat{S}_{ia}^x \right) \Tr \left( \hat{S}_{ld}^{\alpha}  \hat{S}_{kc}^{\beta} \right) \Bigg] \\
			=& \frac{32 J^2}{N^2} \sum_{i,a,\alpha \beta} \xi_{\alpha} \xi_{\beta} \sum_{k\neq l,cd} \delta_{l i} \delta_{da} \Bigg[\Tr \left( \hat{S}_{ld}^{\alpha} \hat{S}_{ia}^x  \hat{S}_{ld}^{\beta} \hat{S}_{ia}^x \right) \Tr \left( \hat{S}_{kc}^{\alpha}  \hat{S}_{kc}^{\beta} \right) \Bigg] + \text{Disconnected} \\
			\approx& \frac{J^2 }{2} \left(\xi_x^2 - \xi_y^2 - \xi_z^2 \right)+ \text{Disconnected} \\
            \equiv& \mathcal{G}_d + \text{Disconnected}. \\
		\end{split}
	\end{equation}

	\paragraph*{Diagram $\mathcal{G}_e$}
	\begin{equation}\label{eq:termc}
		\begin{split}
			&\frac{8}{N} \sum_{(i,a)\neq (j,b)} \Tr\left(\hat{H} \hat{S}_{ia}^x \hat{H} \hat{S}_{jb}^x \right)  \\
			=& \frac{8}{N} \sum_{(i,a)\neq (j,b)}\sum_{\alpha \beta} \xi_{\alpha} \xi_{\beta} \Tr \left( \sum_{k<l,cd}J_{kl} \hat{S}_{kc}^{\alpha} \hat{S}_{ld}^{\alpha} \hat{S}_{ia}^x \sum_{k'<l',c'd'}J_{k'l'} \hat{S}_{k'c'}^{\beta} \hat{S}_{l'd'}^{\beta} \hat{S}_{jb}^x \right) \\
			=& \frac{8}{N} \frac{4J^2}{N} \sum_{(i,a)\neq (j,b)}\sum_{\alpha \beta} \xi_{\alpha} \xi_{\beta} \Tr \left( \sum_{k<l,cd c'd'} \hat{S}_{kc}^{\alpha} \hat{S}_{ld}^{\alpha} \hat{S}_{ia}^x  \hat{S}_{kc'}^{\beta} \hat{S}_{ld'}^{\beta} \hat{S}_{jb}^x \right) \\
			=& \frac{32 J^2}{N^2} \sum_{(i,a)\neq (j,b)}\sum_{\alpha \beta} \xi_{\alpha} \xi_{\beta} \sum_{k\neq l,cd} \Bigg[\Tr \left( \hat{S}_{kc}^{\alpha} \hat{S}_{ia}^x  \hat{S}_{kc}^{\beta} \right) \Tr \left( \hat{S}_{ld}^{\alpha}  \hat{S}_{ld}^{\beta} \hat{S}_{jb}^x \right) + \Tr \left( \hat{S}_{ld}^{\alpha}  \hat{S}_{ld}^{\beta}  \right) \Tr\left( \hat{S}_{ia}^x  \hat{S}_{jb}^x \right) \Tr \left( \hat{S}_{kc}^{\alpha}  \hat{S}_{kc}^{\beta} \right) \Bigg] \\
			=& \frac{32 J^2}{N^2} \sum_{(i,a)\neq (j,b)}\sum_{\alpha \beta} \xi_{\alpha} \xi_{\beta} \sum_{k\neq l,cd} \delta_{k i} \delta_{lj} \delta_{ac} \delta_{db} \Bigg[ \Tr \left( \hat{S}_{kc}^{\alpha} \hat{S}_{ia}^x  \hat{S}_{kc}^{\beta} \right) \Tr \left( \hat{S}_{ld}^{\alpha}  \hat{S}_{ld}^{\beta} \hat{S}_{jb}^x \right) \Bigg] + 0 \\
			\approx& -J^2 \xi_y \xi_z \\
            \equiv& \mathcal{G}_e. \\
		\end{split}
	\end{equation}
	
	\paragraph*{Diagram $\mathcal{G}_f$}
	\begin{equation}\label{eq:termd}
		\begin{split}
			&\frac{8}{N} \sum_{(i,a)\neq (j,b)} \Tr\left(\hat{H}^2   \hat{S}_{ia}^x \hat{S}_{jb}^x \right)  \\
			=& \frac{8}{N} \sum_{(i,a)\neq (j,b)}\sum_{\alpha \beta} \xi_{\alpha} \xi_{\beta} \Tr \left( \sum_{k<l,cd}J_{kl} \hat{S}_{kc}^{\alpha} \hat{S}_{ld}^{\alpha} \sum_{k'<l',c'd'}J_{k'l'} \hat{S}_{k'c'}^{\beta} \hat{S}_{l'd'}^{\beta} \hat{S}_{ia}^x \hat{S}_{jb}^x \right) \\
			=& \frac{8}{N} \frac{4J^2}{N} \sum_{(i,a)\neq (j,b)}\sum_{\alpha \beta} \xi_{\alpha} \xi_{\beta} \Tr \left( \sum_{k<l,cd c'd'} \hat{S}_{kc}^{\alpha} \hat{S}_{ld}^{\alpha}  \hat{S}_{kc'}^{\beta} \hat{S}_{ld'}^{\beta} \hat{S}_{ia}^x \hat{S}_{jb}^x \right) \\
			=& \frac{32 J^2}{N^2} \sum_{(i,a)\neq (j,b)}\sum_{\alpha \beta} \xi_{\alpha} \xi_{\beta} \sum_{k\neq l,cd} \Bigg[\Tr \left( \hat{S}_{kc}^{\alpha} \hat{S}_{kc}^{\beta}  \hat{S}_{ia}^x \right) \Tr \left( \hat{S}_{ld}^{\alpha}  \hat{S}_{ld}^{\beta} \hat{S}_{jb}^x \right) + \Tr \left( \hat{S}_{ld}^{\alpha}  \hat{S}_{ld}^{\beta}  \right) \Tr\left( \hat{S}_{ia}^x  \hat{S}_{jb}^x \right) \Tr \left( \hat{S}_{kc}^{\alpha}  \hat{S}_{kc}^{\beta} \right) \Bigg] \\
			=& \frac{32 J^2}{N^2} \sum_{(i,a)\neq (j,b)}\sum_{\alpha \beta} \xi_{\alpha} \xi_{\beta} \sum_{k\neq l,cd} \delta_{k i} \delta_{lj}  \delta_{lj} \delta_{ac} \Bigg[ \Tr \left( \hat{S}_{kc}^{\alpha}  \hat{S}_{kc}^{\beta} \hat{S}_{ia}^x\right) \Tr \left( \hat{S}_{ld}^{\alpha}  \hat{S}_{ld}^{\beta} \hat{S}_{jb}^x \right) \Bigg] + 0 \\
			\approx& J^2 \xi_y \xi_z \\
            \equiv& \mathcal{G}_f. \\
		\end{split}
	\end{equation}

%\bibliography{ref}% Produces the bibliography via BibTeX.
%\bibliographystyle{naturemag}